\documentclass[reprint, onecolumn, superscriptaddress]{revtex4-1}

\usepackage{graphicx}
\usepackage{color}
\usepackage{bm}
\usepackage{comment}
\usepackage{amsfonts}

\newcommand{\bnabla}{\boldsymbol{\nabla}}
\newcommand{\corr}[1]{\textcolor{red}{[#1]}}

\begin{document}

\setcounter{page}{1}

\title{Wave turbulence description of interacting particles: Klein-Gordon model with a Mexican-hat potential}
\author{Basile Gallet}
\affiliation{Laboratoire SPHYNX, Service de
Physique de l'\'Etat Condens\'e, DSM, CEA Saclay, CNRS UMR 3680, 91191
Gif-sur-Yvette, France}
\author{Sergey Nazarenko}
\affiliation{Mathematics Institute, University of Warwick, Coventry, CV4 7AL, UK}
\author{B\'ereng\`ere Dubrulle}
\affiliation{Laboratoire SPHYNX, Service de
Physique de l'\'Etat Condens\'e, DSM, CEA Saclay, CNRS UMR 3680, 91191
Gif-sur-Yvette, France}

\begin{abstract}
In field theory, particles are waves or excitations that propagate on the fundamental state. In experiments or cosmological models one typically wants to compute the out-of-equilibrium evolution of a given initial distribution of such waves. Wave Turbulence deals with out-of-equilibrium ensembles of weakly nonlinear waves, and is therefore well-suited to address this problem. As an example, we consider the complex Klein-Gordon equation with a Mexican-hat potential. This simple equation displays two kinds of excitations around the fundamental state: massive particles and massless Goldstone bosons. The former are waves with a nonzero frequency for vanishing wavenumber, whereas the latter obey an acoustic dispersion relation. Using wave turbulence theory, we derive wave kinetic equations that govern the coupled evolution of the spectra of massive and massless waves. We first consider the thermodynamic solutions to these equations and study the wave condensation transition, which is the classical equivalent of Bose-Einstein condensation. We then focus on nonlocal interactions in wavenumber space: we study the decay of an ensemble massive particles into massless ones. Under rather general conditions, these massless particles accumulate at low wavenumber. We study the dynamics of waves coexisting with such a strong condensate, and we compute rigorously a nonlocal Kolmogorov-Zakharov solution, where particles are transferred non-locally to the condensate, while energy cascades towards large wave numbers through local interactions. This nonlocal cascading state constitute the intermediate asymptotics between the initial distribution of waves and the thermodynamic state reached in the long-time limit.
\end{abstract}

\maketitle

\section{Introduction}

Some  nonlinear partial differential equations (PDEs)  are simple enough and yet they describe deep and nontrivial universal processes in a broad range of physical applications. Well-known examples of such PDEs are the Korteweg-de-Vries and the Gross-Pitaevskii (GP) equations. It would not be an exaggeration to put into such a class
a universal model that allows to understand the concept of spontaneous breaking of global symmetry. Consider for instance the following complex nonlinear Klein-Gordon equation,
\begin{equation}\label{ctkg}
\psi_{tt} - \Delta \psi +(-1+|\psi|^2) \psi =0 \,,
\end{equation}
where $\psi({\bf x},t)\in \mathbb{C}$. The linear part of this equation resembles the standard Klein-Gordon equation, where the square mass coefficient $m^2$ is replaced by $-1$. Such a negative square mass ($m^2 =-1$) was originally believed to lead to superluminal particles. However, it was soon realised that such a field cannot sustain a localised particle-like excitation because the base state $\psi=0$ is unstable. 

Equation (\ref{ctkg}) conserves the total energy
\begin{equation}\label{eq:energy}
E = 
\int \left[  {|\psi_t|^2} +  {|\nabla \psi|^2} -
 {|\psi|^2} + \frac{1}{2} {|\psi|^4} \right] \, d{\bf x} \, ,
\end{equation}
where the last two terms correspond to the famous Mexican hat potential $U(\psi)=  -|\psi|^2 +|\psi|^4/2$; we therefore refer to equation (\ref{ctkg}) as the Klein-Gordon Mexican Hat (KGMH) model.


Equation (\ref{ctkg}) is invariant to a uniform shift in phase of $\psi$, i.e., $\psi \to e^{i \alpha} \psi$, with $\alpha =$~const. The invariant corresponding to this $U(1)$ symmetry is the ``charge'',
\begin{equation}
\label{eq:charge}
Q=\frac{i}{2} \int (\psi \partial_t \psi^* - \psi^* \partial_t \psi ) \, \mathrm{d} {\bf x} \, ,
\end{equation}
where $\psi^*$ denotes the complex conjugate of $\psi$. In this paper we focus mostly on the situation $Q=0$,  $Q \neq 0$ being briefly described in section \ref{finiteq}.

In high energy physics, the KGMH model is a nonlinear $\sigma$-model \cite{gelman,song} with a Mexican-hat potential. It is a simple relativistic model that describes the interaction between two fields: the first one corresponds to a massive particle, while the second one corresponds to a massless Goldstone boson. As such, the KGMH model provides a simplified description of a system of  sigma mesons interacting with pions: the mass of the pions is neglected in this model, as it is much less than the one of the sigma meson.

In this paper we study the KGMH system within the framework of wave turbulence, with direct numerical simulations (DNSs) to check the predictions. Wave turbulence \cite{falkovich1992kolmogorov,nazarenko2011wave} deals with ensembles of weakly-interacting waves. It proceeds through the derivation of a kinetic equation that governs the time evolution of the spectral density of the wave field. Among its solutions are thermal equilibrium and generalized Rayleigh-Jeans distributions. However, the main use of the kinetic equation is to describe out-of-equilibrium situations. Indeed, for scale-invariant systems, one can compute out-of-equilibrium power-law solutions to the kinetic equation: these are the celebrated Kolmogorov-Zakharov spectra. More generally, the kinetic equation governs the evolution of an ensemble of waves (or particles), starting from an arbitrary weakly nonlinear initial condition.

Such out-of-equilibrium initial conditions are encountered in particle physics and cosmology: they correspond for instance to heavy-ion collisions produced experimentally, or to the early universe. There has therefore been a significant recent interest in applying the wave turbulence approach in these domains \cite{Berges:2010ez,Gasenzer:2013era}.

From the point of view of wave turbulence theory, the KGMH model is a new object of study, which is interesting and important for several reasons: spontaneous symmetry breaking leads to two kinds of waves, or particles, with distinct dispersion relations. This contrasts with most wave turbulence systems, where one usually deals with a single kind of waves, like e.g. in the GP equation. In addition to transfers of energy from one scale to another, the KGMH system displays ``reactions": decay of massive particles into massless ones, and fusion of massless particles into massive ones. This brings new features to the wave turbulence dynamics, with strongly nonlocal interactions between different kinds of particles, that prohibit standard Kolmogorov-Zakharov cascades.

Nevertheless, the KGMH model shares some similarities with the simpler GP equation: both have two positive invariants. Indeed, in addition to energy, the KGMH equation has an adiabatic invariant which is similar to the number of particles. Here, ``adiabatic" means that this invariant is conserved exactly in the asymptotic limit of small nonlinearity only. This adiabatic invariance is sufficient to trigger wave condensation \cite{PhysRevLett.95.263901}, a phenomenon that we study both theoretically and numerically. In the particle physics context, this  corresponds to  Bose-Einstein condensation of pions \cite{zimaniy,Bunatian:1983dq,song}.

The decay of the massive mode ($\sigma$-meson) into massless Goldstone bosons (pions) is the leading thread of this paper. In section \ref{analogue} we propose a simple mechanical analogue of the KGMH model (\ref{ctkg}), which provides physical insight into the underlying dynamics. In section \ref{setup} we study the linear and weakly nonlinear dynamics around the minimum of the Mexican-hat potential (the uniform fundamental state). We compute the dispersion relations for the massive and massless modes, together with a reduced Hamiltonian which takes into account the dominant three-wave interaction between these modes. This dominant interaction involves two massless modes and one massive mode. We analyze it in section \ref{decay} by deriving a simple set of ODEs for a single triad of interacting waves, which highlights the decay instability of a monochromatic massive wave into two massless monochromatic waves.

In section \ref{WT} we develop the wave turbulence description of the KGMH model. We compute kinetic equations governing the evolution of the spectra of massive and massless modes. We identify two invariants of these equations: the energy, and an adiabatic invariant somewhat similar to a number of particles. The kinetic equations admit Rayleigh-Jeans equilibrium solutions, but no local Kolmogorov-Zakharov out-of-equilibrium solutions.
We focus on such thermodynamic equilibria in section \ref{bec}. For low enough energy per particle, the thermodynamic state displays wave condensation, the classical counterpart to the Bose-Einstein condensation (BEC) of pions. These findings are confirmed by DNSs, which show unambiguously that BEC is observed over a very long time, even in systems where the number of particles is only an adiabatic invariant.

The out-of-equilibrium dynamics involve strongly nonlocal processes which we consider in sections \ref{statdecay} and  \ref{dyncond}. The first section addresses the decay instability for isotropic ensembles of massive waves: on the one hand, the kinetic equations allow to predict the distribution of the massless waves that appear during the linear stage of the instability. On the other hand, the thermodynamics of section \ref{bec} predict the the reaction yield of the decay instability: starting from massive-waves ($\sigma$-mesons) at a single wavenumber, we predict the fraction of these waves that has decayed into massless waves (pions) in the long-time limit.
 
Section \ref{dyncond} finally addresses the behavior of ensembles of waves coexisting with a strong condensate of massless particles. We derive rigorously a reduced kinetic equation that describes a nonlocal transfer of particles to the condensate, together with a local Kolmogorov-Zakharov energy cascade towards high wave numbers. We see this cascading state as an intermediate asymptotics that describes the dynamical process of condensation: it makes the link between the initial distribution of particles and the final thermodynamic condensed state.

\section{An analogous mechanical system}

\label{analogue}

\begin{figure}
\begin{center}
\includegraphics[width=130 mm]{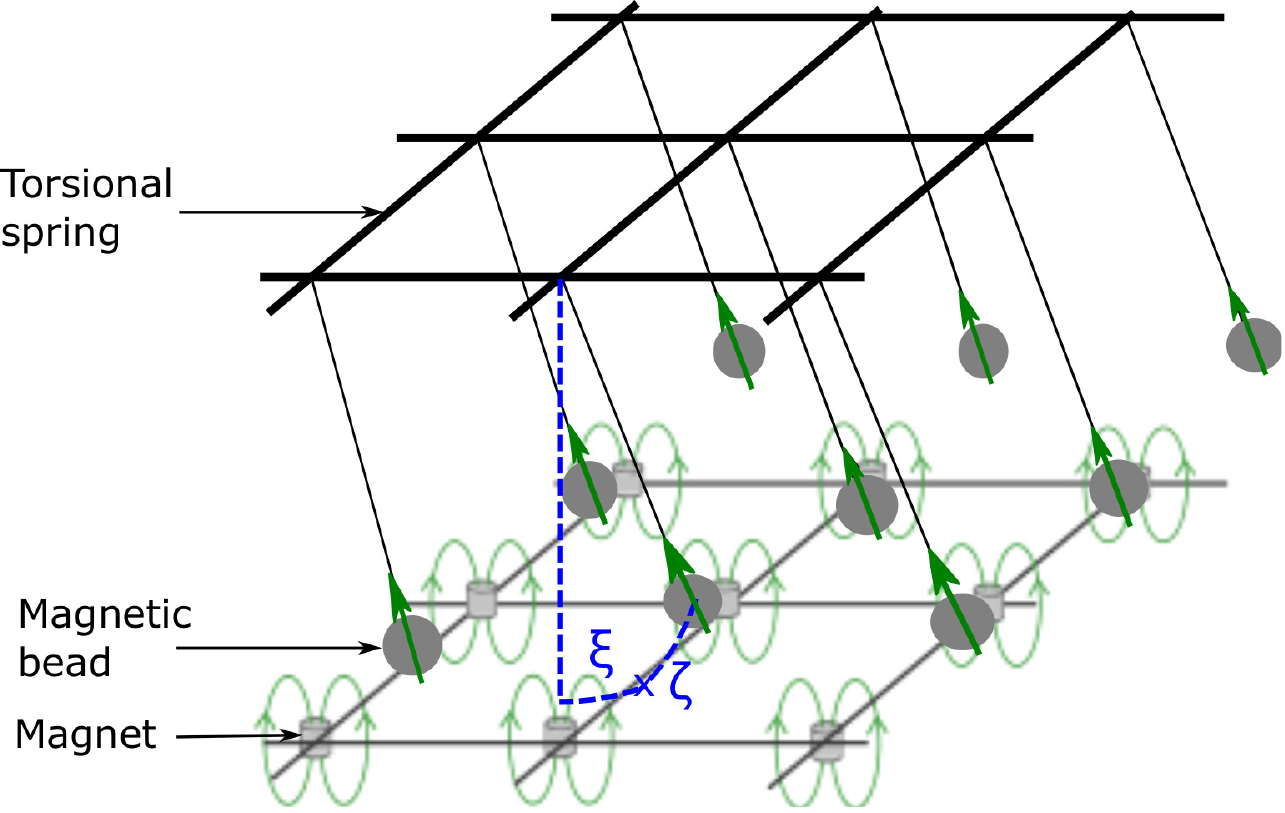}
\end{center}
\caption{\textbf{A lattice of magnetic pendulums:} small magnets placed under the vertical position of each pendulum create the Mexican-hat potential. The interaction between neighboring pendulums results from dipole-dipole interaction and torsional springs between their fixation points. In the continuous limit this system follows the KGMH equation (\ref{ctkg}).\label{fig1}}
\end{figure}

One can gain some insight into the physics described by equation (\ref{ctkg}) by considering a system from classical mechanics that follows the same equation. An example of such system is the square lattice of pendulums sketched in figure \ref{fig1}. The pendulums can move in two angular directions, $\zeta$ and $\xi$. The corresponding kinetic energy of a pendulum is $J (\dot{\zeta}^2+\dot{\xi}^2)/2$, with $J$ the moment of inertia of a pendulum. Each pendulum consists of a magnetized bead attached to a rigid rod, the magnetic dipole of the bead being alined with the rod. Neighboring pendulums interact through two mechanisms: first, two dipoles repulse each other. Second, we assume that the fixation points of the pendulums are connected by torsional springs. Dipole-dipole repulsion efficiently propagates longitudinal perturbations through the lattice, while torsional springs propagate transverse oscillations. The stiffness $C$ of the torsional springs is chosen to match the strength of the dipole-dipole interaction, so that the interaction energy between the pendulums reads
 \begin{equation}
\frac{C}{2} \sum_{(i,j) \, \mathrm{ neighbors}} (\zeta_i-\zeta_j)^2+(\xi_i-\xi_j)^2 \, ,
 \end{equation}
 where $\zeta_n$ and $\xi_n$ are the angular degrees of freedom of the n-th pendulum, and we consider dipole-dipole interactions between nearest neighbors only.
We further assume that the relative displacements of the pendulums are small and consider the lowest order description of the dipole-dipole interaction.

Finally, to create the Mexican-hat potential we place a magnet under the rest position of each bead. Gravity drives a given pendulum towards the vertical, but the repulsion between the bead and the magnet displaces the equilibrium position away from the vertical, at a distance $r=r_0$ from the vertical axis. The combined effects of gravity and of the  magnet result in a Mexican-hat-shaped potential $V(r)$. If the length $\ell$ of the rods is much longer than $r_0$, we can use the approximate relation $r^2 \simeq \ell^2 (\zeta^2+\xi^2)$. In the neighborhood of $r_0$, we approximate the potential by $V(r) \simeq -V_2 r^2 + V_4 r^4= -V_2 \ell^2 (\zeta^2+\xi^2) + V_4 \ell^4 (\zeta^2+\xi^2)^2$. 
The Lagrangian of the system follows from the subtraction of the Mexican-hat potential energy and neighboring interaction term to the kinetic term,
 \begin{equation}
  \mathcal{L}=\sum_n \left[ \frac{J}{2} \left(\dot{\zeta_n}^2+\dot{\xi_n}^2 \right)  - V_4 \ell^4 (\zeta_n^2+\xi_n^2)^2 + V_2 \ell^2 (\zeta_n^2+\xi_n^2) \right] - \frac{C}{2}\sum_{(i,j) \, \mathrm{ neighbors}} (\zeta_i-\zeta_j)^2+(\xi_i-\xi_j)^2 \, .
 \end{equation}
We consider the continuous limit and introduce the complex variable $\psi(x,y,t)=\zeta(x,y,t)+i \xi(x,y,t)$. After a rescaling of time, space, and $\psi$ this Lagrangian takes the following dimensionless form,
 \begin{equation}
 \mathcal{L}=\int \left( |\psi_t|^2 - |\bnabla \psi|^2 + |\psi|^2 - \frac{1}{2} |\psi|^4 \right) \, \mathrm{d}^2 {\bf x}   \, ,
 \end{equation}
which corresponds to equation (\ref{ctkg}) in two spatial dimensions.

In this analogous system, the charge $Q$ corresponds to the sum of the vertical angular momenta of each pendulum around their fixation points. A nonzero value of $Q$ means that the initial condition (and the subsequent evolution) favors one direction of azimuthal rotation of the pendulums.

\begin{figure}
\begin{center}
\includegraphics[width=140 mm]{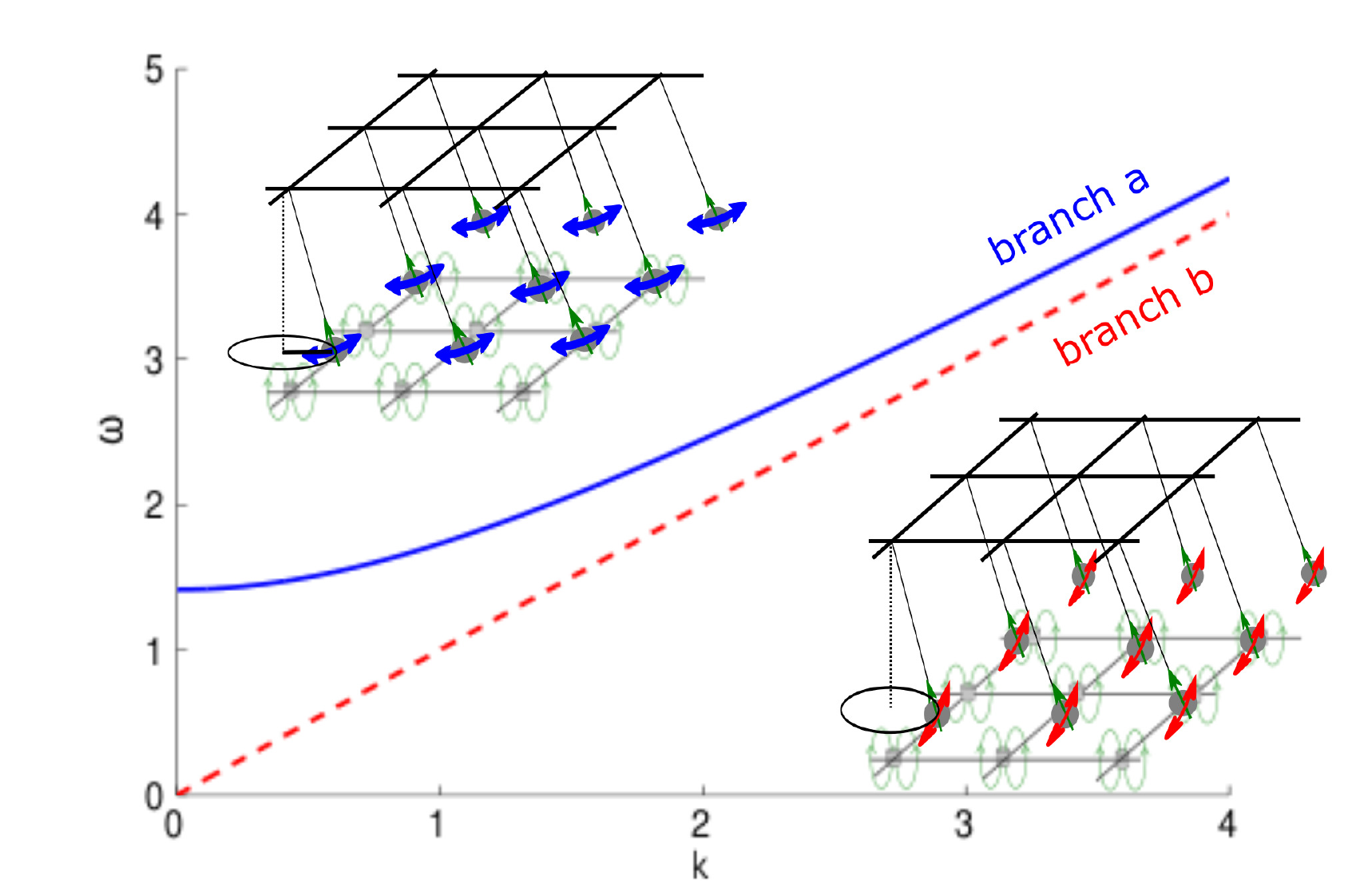}
\end{center}
\caption{Two types of waves propagate in this system: branch $a$ corresponds to radial oscillations of the pendulums, whereas branch $b$ corresponds to azimuthal oscillations of the pendulums, along the minimum of the Mexican-hat potential.\label{dispersion}}
\end{figure}

\section{Weakly nonlinear dynamics around the minimum of the potential.}

\label{setup}

A key feature of the KGMH model is that it displays spontaneous $U(1)$ symmetry-breaking: the fundamental state, defined as the state of lowest energy, is not the usual $\psi=0$. Instead, it is a uniform state $\psi=\psi_0=const.$, with $|\psi_0 |=1$: $\psi$ is uniform, time-independent, and lies in the minimum of the Mexican-hat potential. We wish to study the dynamics of the system around this fundamental state.

Up to a phase-shift in the definition of $\psi$, we can choose $\psi_0=1$ and consider perturbations of this state,
$$
\psi ({\bf x},t)= 1+\phi({\bf x},t) \, .
$$
Substitution into equation (\ref{ctkg}) yields
\begin{equation}\label{ctkg'}
\phi_{tt} - \Delta \phi +\phi + \phi^* + 2 \phi \phi^*+\phi^2+|\phi|^2 \phi =0 \, .
\end{equation}
We focus on the weakly-nonlinear dynamics arising from the quadratic terms of this equation, and therefore neglect the cubic nonlinearity. In variables
$\lambda = \Re(\phi)$ and $\chi = \Im(\phi)$ we have
\begin{eqnarray}\label{ctkg''1}
\lambda_{tt} - \Delta \lambda +2\lambda  + 3\lambda^2+\chi^2 &=&0 ,\label{eqlambda}\\
\chi_{tt} - \Delta \chi +2\lambda  \chi  &=&0. \label{ctkg''2}
\end{eqnarray}
The linear parts of these two equations provide two dispersion relations. For a plane wave proportional to $e^{i(\omega t -{\bf k}\cdot {\bf x})}$, they give respectively
\begin{eqnarray}\label{w1}
\omega^a_{\bf k} = \sqrt{2+k^2}, \\
\omega^b_{\bf k} = k,
\label{w2}
\end{eqnarray}
where $k = |{\bf k}|$.
We refer to these two dispersion relations as branch $a$ and branch $b$. They are plotted in figure \ref{dispersion}. Branch $a$ has a nonzero frequency for vanishing wavenumber, $\omega^a \neq 0$ for $k = 0$. In terms of particle physics, this mode has a nonzero rest energy and thus corresponds to massive particles. By contrast, the frequency of $b$-waves vanishes for $k=0$: branch $b$ corresponds to a massless Goldstone boson.
This is a basic model for coexisting sigma meson (branch $a$) and pion (branch $b$) fields, in which the pion's mass is neglected. 

In the lattice of pendulums sketched in figure \ref{fig1}, the massive mode corresponds to oscillations of the pendulums along their radial direction. In this mode, each pendulum feels two restoring forces,  one from the potential well $V(r)$ and one from the coupling to its neighbors. By contrast, mode $b$ corresponds to oscillations of the pendulums along their azimuthal directions. The restoring force originates only from the coupling with the neighboring pendulums, and if all the pendulums move in phase along their azimuthal direction ($b$ mode with infinite wavelength) then there is no restoring force: mode $b$ is massless. 


\subsection{Hamiltonian formulation}

The system (\ref{ctkg''1}-\ref{ctkg''2}) follows from the following Hamiltonian,
\begin{equation}\label{H}
\tilde{H}=  \int \left[ \frac{1}{2} {p^2} + \frac{1}{2} {q^2} +\frac{1}{2} (\nabla \lambda)^2  +  \lambda^2 +\frac{1}{2}(\nabla \chi)^2 +  \lambda^3 
+\lambda \chi^2 \right] \, d{\bf x} \, ,
\end{equation}
where $p$ and $q$ are the momenta conjugate to the variables $\lambda$ and $\chi$ respectively. Hamilton equations are
\begin{eqnarray}\label{HamEq}
\lambda_{t}  = \frac{\delta \tilde{H}}{\delta p},
&& p_{t}  = - \frac{\delta \tilde{H}}{\delta \lambda},\\
\chi_{t}  = \frac{\delta \tilde{H}}{\delta q}, &&
q_{t}  = - \frac{\delta \tilde{H}}{\delta \chi}.
\end{eqnarray}

We consider these equations in a  $d$-dimensional periodic cube of side $L$ and introduce Fourier series,
\begin{equation}
\left( \begin{array}{c}
\lambda({\bf x},t)  \\
\chi({\bf x},t)  \\
p({\bf x},t)  \\
q({\bf x},t)  \end{array} \right) = \sum_{\bf k} \left( \begin{array}{c}
\hat{\lambda}_{\bf k}(t)  \\
\hat{\chi}_{\bf k}(t)  \\
\hat{p}_{\bf k}(t)  \\
\hat{q}_{\bf k}(t)  \end{array} \right) e^{i {\bf k} \cdot {\bf x}} \, ,
\end{equation}
where the sums are over ${\bf k} \in \frac{2\pi}{L} \mathbb{Z}^d$. Because these four fields are real-valued, $(\hat{\lambda}_{-\bf k},\hat{\chi}_{-\bf k},\hat{p}_{-\bf k},\hat{q}_{-\bf k})=(\hat{\lambda}_{\bf k}^*,\hat{\chi}_{\bf k}^*,\hat{p}_{\bf k}^*,\hat{q}_{\bf k}^*)$. In Fourier space, Hamilton equations therefore become

\begin{eqnarray}\label{HamEq}
\dot  {\hat \lambda}_{\bf k}  = \frac{\delta H}{\delta \hat p^*_{\bf k}},
&& \dot { \hat p}_{\bf k}   = - \frac{\delta H}{\delta \hat \lambda^*_{\bf k}},\\
\dot  {\hat \chi}_{\bf k}  = \frac{\delta H}{\delta \hat q^*_{\bf k}}, &&
\dot  {\hat q}_{\bf k}   = - \frac{\delta H}{\delta \hat \chi^*_{\bf k} },
\end{eqnarray}
with
\begin{equation}\label{Hk}
H=  H_{2} + H_{int},
\end{equation}
where
\begin{equation}\label{Hk2}
H_2 =  \frac{1}{2} \sum_{\bf k} 
 \left[  {|{ \hat p}_{\bf k}|^2} +  (\omega^a_{\bf k})^2 {|{ \hat \lambda}_{\bf k}|^2} +  {|{ \hat q}_{\bf k}|^2} + (\omega^b_{\bf k})^2 {|{ \hat \chi}_{\bf k}|^2}   \right] \, ,
\end{equation}
and
\begin{equation}\label{Hkint}
H_{int} =  \sum_{{\bf k}_1, {\bf k}_2, {\bf k}_3} 
 \left[  \hat {\lambda}_{{\bf k}_1}  \hat {\lambda}_{{\bf k}_2} \hat {\lambda}_{{\bf k}_3} 
+\hat {\lambda}_{{\bf k}_1} \hat {\chi}_{{\bf k}_2} \hat {\chi}_{{\bf k}_3}\right] \delta({\bf k}_1 + {\bf k}_2 + {\bf k}_3),
\end{equation}
where $\delta$  is a Kroenecker symbol, i.e., $\delta({\bf k}=0)=1$ and $\delta({\bf k} \neq 0)=0$.


\subsection{Normal variables}

Let us introduce the normal variables $a_{\bf k}$ and $b_{\bf k}$ that diagonalize $H_2$,
\begin{eqnarray}\label{cantrans}
{\hat \lambda}_{\bf k}  = \frac{a_{\bf k} +a^*_{-{\bf k}}}{
\sqrt{2 \omega^a_{\bf k}}},
&& {\hat p}_{\bf k}  = \frac{\sqrt{ \omega^a_{\bf k} }(a_{\bf k} -a^*_{-{\bf k}})}{
i \sqrt{2 }},\\
\label{cantrans1}
{\hat \chi}_{\bf k}  = \frac{b_{\bf k} +b^*_{-{\bf k}}}{
\sqrt{2 \omega^b_{\bf k}}},
&& {\hat q}_{\bf k}  = \frac{\sqrt{ \omega^b_{\bf k} }(b_{\bf k} -b^*_{-{\bf k}})}{
i\sqrt{2 }}.
\end{eqnarray}
We refer to $a_{\bf k}$ and $b_{\bf k}$ as the respective amplitudes of $a$- and $b$-waves
(or modes, or particles).

In these variables the equations of motion are
\begin{eqnarray}\label{HamEq}
i \dot  {a}_{\bf k}  &=& \frac{\delta H}{\delta a^*_{\bf k}}, \\
i \dot  {b}_{\bf k}  &=& \frac{\delta H}{\delta b^*_{\bf k}},
\end{eqnarray}
where $H=H_2+H_{int} $, with
\begin{equation}\label{Hk2can}
H_2 =   \sum_{\bf k} 
 \left[  \omega^a_{\bf k} |a_{\bf k}|^2 +  \omega^b_{\bf k} |b_{\bf k}|^2 \right].
\end{equation}
and
\begin{eqnarray}
H_{int} &=&  \frac 1{2 \sqrt{2}} \sum_{{\bf k}_1, {\bf k}_2, {\bf k}_3} 
\frac{1}{
\sqrt{ \omega^a_{1} \omega^a_{2} \omega^a_{3}}}
 \left[   ({a}_{1}   {a}_{2}  {a}_{3} \delta({\bf k}_1 + {\bf k}_2 + {\bf k}_3)
+
3 {a}_{1}   {a}_{2}  {a}^*_{3} \delta({\bf k}_1 + {\bf k}_2 - {\bf k}_3)
\right] \nonumber \\
&& +\frac 1{2 \sqrt{2}} \sum_{{\bf k}_1, {\bf k}_2, {\bf k}_3} 
\frac{1}{
\sqrt{ \omega^a_{1} \omega^b_{2} \omega^b_{3}}}
 \left[ 
 {a}_{1}   {b}_{2}  {b}_{3} \delta({\bf k}_1 + {\bf k}_2 + {\bf k}_3)
\right. \nonumber \\
&& \left.
+2{a}_{1}   {b}_{2}  b^*_{3} \delta({\bf k}_1 + {\bf k}_2 - {\bf k}_3)
+{a}^*_{1}   {b}_{2}  b_{3} \delta({\bf k}_1 - {\bf k}_2 - {\bf k}_3)
\right] +c.c.,
\label{HkintCan}
\end{eqnarray}
where $c.c.$ stands for ``complex conjugate", and we introduced the shorthand notations $a_i=a_{{\bf k}_i}$, $b_i=b_{{\bf k}_i}$, $\omega^{a}_i=\omega^{a}_{{\bf k}_i}$, and $\omega^{b}_i=\omega^{b}_{{\bf k}_i}$.

The terms in the Hamiltonian (\ref{HkintCan}) correspond respectively to 3-wave processes of
type $a+a+a \to 0, a+a \to a, a+b +b\to 0, a+b \to b$ and
$b+b \to a$. In a weakly nonlinear description of the system, the only relevant nonlinearities are those for which both a wavenumber and a frequency resonance conditions can be satisfied. The other nonlinearities produce non-resonant terms only, which do not contribute to the weakly nonlinear dynamics. We show in appendix \ref{ap:3wave} that, out of the processes outlined above, the only
one allowed by the frequency resonance condition is
$b+b \to a$. We can therefore discard all the other terms in $H_{int}$, as they do not contribute to the weakly nonlinear dynamics, and simply write
\begin{eqnarray}
H_{int} &=&  \sum_{{\bf k}_1, {\bf k}_2, {\bf k}_3} 
V^1_{23} {a}^*_{1}   {b}_{2}  b_{3}  \delta^{3}_{12} 
 +c.c.,
\label{HkintCan'}
\end{eqnarray}
where we have switched to a shorthand notation of  
 Kroenecker deltas, $ \delta^{1}_{23}  \equiv  \delta({\bf k}_1 - {\bf k}_2 - {\bf k}_3)$,
and introduced the interaction coefficient
\begin{equation}\label{V}
V^1_{23} \equiv  \frac 1{2 \sqrt{2}}\frac{1}{
\sqrt{ \omega^a_{1} \omega^b_{2} \omega^b_{3}}}.
\end{equation}





The equations of motion become
\begin{eqnarray}
\dot a_{\bf k} &=&
- i \omega^a_{\bf k} a_{\bf k}  
-i \sum_{{\bf k}_1, {\bf k}_2}  
V^{\bf k}_{12} \, b_1 b_2
 \delta^{\bf k}_{12} ,\label{ctkg'''1}
 \\
\dot b_{\bf k} &=&
- i \omega^b_{\bf k} b_{\bf k}  
-2i \sum_{{\bf k}_1, {\bf k}_2}   V^1_{2{\bf k}}
\, a_1 b^*_2  \delta^1_{2{\bf k}}.\label{ctkg'''2}
\end{eqnarray}

A similar Hamiltonian system with two types of interacting waves was studied in \cite{Zakharov1985285}. 
However, in that study the dominant process was of the type $a+b \to b$, which leads to very different dynamics than the process $b+b \to a$ considered here. 

\subsection{Finite-charge $Q \neq 0$}

\label{finiteq}

We have considered so far only perturbations around the minimum energy state, which is uniform $\psi$ lying in the minimum of the Mexican-hat potential. This corresponds to a vanishing charge invariant, $Q=0$. We now briefly consider nonzero values of $Q$. Writing the field in polar coordinates, $\psi=r({\bf x},t) e^{i \theta({\bf x},t)}$, the charge reads
\begin{equation}
Q = \int r^2 \dot \theta \, d {\bf x} \, . \label{Qpolar}
\end{equation}
A nonzero charge therefore indicates a preferred direction of rotation in the complex plane. Uniform states with nonzero charge can be sought in the form $\psi_0(t)=R e^{i \omega_0 t}$, where $R$ and $\omega_0$ are real constants. Substitution into the KGMH equation (\ref{ctkg}) leads to
\begin{equation}
R^2=1+\omega_0^2 \, ,
\end{equation}
i.e. there is a continuous family of uniform solutions to the KGMH equation, corresponding to different values of the charge invariant $Q = R^2 \omega_0 L^d$.

Once again, the lattice of magnetic pendulums provides a simple interpretation for these states. From equation (\ref{Qpolar}) the charge is the sum of the vertical angular momenta of the pendulums about their fixation points. The uniform solution with $\omega_0=0$ corresponds to all the pendulums parallel and at rest in the minimum of the Mexican-hat potential, whereas uniform solutions with $\omega_0 \neq 0$ correspond to all the pendulums parallel and rotating at the same angular frequency $\omega_0$ around the vertical. For the latter solutions $R^2=1+\omega_0^2>1$: the field is not exactly in the minimum of the Mexican-hat potential, but instead it is ``bobsleighing'' on its outer edge. In the system of pendulums, this ``bobsleighing'' results from the centrifugal force acting on each pendulum.

To study perturbations around the uniform state with nonzero $Q$, we write $\psi=\psi_0(t)(1+\phi({\bf x},t))=\sqrt{1+\omega_0^2}e^{i \omega_0 t}(1+\phi({\bf x},t))$, and consider infinitesimal perturbations $\phi \ll 1$. Substitution into the KGMH equation gives at linear order in $\phi$
\begin{equation}
 \phi_{tt} +2i\omega_0 \phi_t - \Delta \phi + (1+\omega_0^2) (\phi+{\phi^*}) = 0 \, ,
\end{equation}
and after taking the real and imaginary parts, with $\phi=\lambda+i\chi$,
\begin{eqnarray}
 \lambda_{tt} - \Delta \lambda +2 (1+\omega_0^2) \lambda -2 \omega_0 \chi_t =0 \, , \\
 \chi_{tt} - \Delta \chi +2 \omega_0 \lambda_t =0 \, . 
\end{eqnarray}
As compared to the situation $Q=0$, the fields $\lambda$ and $\chi$ are now coupled at linear order by terms proportional to $\omega_0$. Inserting a plane wave structure for these two fields and asking for non-trivial solutions yields the following dispersion relation
\begin{equation}
\omega^4 - 2 \omega^2 (k^2+1+3\omega_0^2)+k^2(k^2+2+2\omega_0^2)=0 \, ,
\end{equation}
which has two branches of solutions,
\begin{eqnarray}
\omega^2_{\pm}=k^2+1+3\omega_0^2\pm \sqrt{\left(1+3\omega_0^2 \right)^{2}+4k^2\omega_0^2} \, .
\label{branchesQnz}
\end{eqnarray}
In the limit $\omega_0 \to 0$, we see that $\omega_+$ branch corresponds to $a$-modes, whereas $\omega_-$ corresponds to b-modes. For $\omega_0 \neq 0$, the mode $\omega_+$ remains massive, with $\lim_{k \to 0}\omega_+ \neq 0$, and the $\omega_-$ branch remains massless, with $\lim_{k \to 0} \omega_-=0$. Note that the speed of the low-$k$ $b$-particles is now reduced,
$\partial \omega_- /\partial k = \sqrt{(1+ \omega_0^2)/(1+3 \omega_0^2})$, and
the $a$-particle's  mass is $\sqrt{1+3 \omega_0^2}$ times greater than in the $\omega_0=0$ case.
At the linear level, the structure of the problem is nevertheless quite similar to the $Q=0$ situation.

Interesting new phenomena appear at the nonlinear level. As for the $Q=0$ situation, one can compute normal variables $a_{\bf k}$ and $b_{\bf k}$ diagonalizing the quadratic part of the Hamiltonian, $a_{\bf k}$ (resp. $b_{\bf k}$) corresponding to $\omega_+$ (resp. $\omega_-$). These normal variables are now superpositions of the $\lambda$ and $\chi$ fields, together with the corresponding momenta:
\begin{eqnarray} 
{\hat \lambda}_{\bf k} &= &\frac{\omega_+^2-k^2}{ 
\sqrt{2 \omega_+}}\frac{a_{\bf k}+a^*_{-\bf k}}{2}-\frac{\omega_-^2-k^2}{ 
\sqrt{2 \omega_-^3\omega_0^2}}\frac{b_{\bf k} -b^*_{-\bf k}}{2i},\\ \label{cantransgen1} 
{\hat \mu}_{\bf k} &=& \omega_0\sqrt{ \omega_+} 
\frac{a_{\bf k}-a^*_{-\bf k}}{i\sqrt{2}}+\frac{1}{ 
\sqrt{ \omega_-}}\frac{b_{\bf k}+b^*_{-\bf k}}{\sqrt{2}},\\ \label{cantransgen2} 
{\hat p}_{\bf k} &=& \frac{\sqrt{\omega_+}(\omega_+^2-k^2)}{ 
\sqrt{2}}\frac{a_{\bf k}-a^*_{-\bf k}}{2i}+\frac{\omega_-^2-k^2}{ 
\sqrt{2 \omega_-\omega_0^2}}\frac{b_{\bf k} +b^*_{-\bf k}}{2},\\ 
\label{cantransgen3} 
{\hat q}_{\bf k} &=&- \sqrt{\omega_0\omega_-^3}\frac{a_{\bf k}+a^*_{-\bf k}}{\sqrt{2}}+\sqrt{ \omega_-} 
\frac{b_{\bf k}-b^*_{-\bf k}}{i\sqrt{2}}, 
\end{eqnarray}
The weakly nonlinear limit is still governed by three-wave interactions, but the dispersion relations (\ref{branchesQnz}) now allow for new processes: $a+a \to a$  and  $a+b \to b$  are still impossible, but, in addition to $b+b \to a$, the processes $a \to a+b$ and $a+b\to a$ are compatible with the dispersion relation. This possibly triggers interesting new dynamics. However, because the normal variables $a_{\bf k}$ and $b_{\bf k}$ have rather intricate expressions when $Q\neq0$, the computation of the kinetic equation is more involved than for $Q=0$. The wave turbulence analysis of the present paper therefore focuses on $Q=0$, and we leave the case $Q \neq 0$ for future work.

\section{Decay of the massive mode into two Goldstone waves}

\label{decay}

The dominant three-wave process of the weakly-nonlinear KGMH dynamics can be highlighted by studying a single triad of interacting modes. Consider the dynamical equations  in the weak-nonlinearity limit, (\ref{ctkg''1}) and (\ref{ctkg''2}), and expand the fields as
\begin{eqnarray}
\lambda=\epsilon \lambda_0(t,T) + \epsilon^2 \lambda_1(t,T) + \mathcal{O}(\epsilon^3) \, ,\\
\chi=\epsilon \chi_0(t,T) + \epsilon^2 \chi_1(t,T) + \mathcal{O}(\epsilon^3) \, ,\\
\end{eqnarray}
where the slow time scale is $T=\epsilon t$. To order $\epsilon$, equations (\ref{ctkg''1}) and (\ref{ctkg''2}) reduce to
\begin{eqnarray}
\partial_{tt} \lambda_0 - \Delta \lambda_0 +2\lambda_0 &=&0 ,\\
\partial_{tt}\chi_{0} - \Delta \chi_0 &=&0.
\end{eqnarray}
 In Appendix \ref{ap:3wave}, we saw that triads of interacting mode are made of two $b$ modes and one $a$ mode. We consider a single such triad in the solution to the order $\mathcal{O}(\epsilon)$ equations:
\begin{eqnarray}
\lambda_0 & = & A(T) e^{ i(\sqrt{k^2+2}\, t - {\bf k}\cdot{ \bf x})} + c.c. \, , \label{O1lambda}\\
\chi_0 & = & B_1(T) e^{i(k_1 t - {\bf k}_1\cdot{ \bf x})} + B_2(T) e^{i(k_2 t - {\bf k}_2\cdot{ \bf x})} + c.c. \label{O1mu}\, .
\end{eqnarray}
To obtain non-trivial dynamics at next order, we further assume that the resonance conditions $\sqrt{k^2+2}=k_1+k_2$ and ${\bf k}={\bf k}_1+{\bf k}_2$ are met. To order $\epsilon^2$, the equations become
\begin{eqnarray}
\partial_{tt} \lambda_1 - \Delta \lambda_1 +2\lambda_1 &=& -2 \partial_{tT} \lambda_0 -3\lambda_0^2 - \chi_0^2  \, , \label{eq02lambda} \\
\partial_{tt}\chi_{1} - \Delta \chi_1 &=& -2 \partial_{tT} \chi_0 -2 \lambda_0 \chi_0 \label{eq02mu} \, .
\end{eqnarray}
The solvability condition requires the right-hand side of these equations to have no terms that are resonant with the operator on the left-hand side. We insert solutions (\ref{O1lambda}) and (\ref{O1mu}) into equations (\ref{eq02lambda}) and (\ref{eq02mu}), before demanding that the rhs of (\ref{eq02lambda}) has no term proportional to $e^{ i(\sqrt{k^2+2}\, t - {\bf k}\cdot{ \bf x})}$, and that the rhs of (\ref{eq02mu}) has no terms proportional to $e^{i(k_1 t - {\bf k}_1\cdot{ \bf x})}$ or to $e^{i(k_2 t - {\bf k}_2\cdot{ \bf x})}$. These three constraints lead to the following system of ODEs for the slow evolution of the wave amplitudes
\begin{eqnarray}
d_T A & = & \frac{i}{2\sqrt{k^2+2}} B_1 B_2 \, , \label{eqA}\\
d_T B_1 & = & \frac{i}{K_1} A B_2^*\, , \label{eqB1}\\
d_T B_2 & = & \frac{i}{K_2} A B_1^* \, \label{eqB2}.
\end{eqnarray}

Consider a single $b$-wave:  $(A=B_2=0, \,B_1 \neq 0)$ is a time-independent solution to the system of ODEs. We study the stability of this $b$-wave by considering infinitesimal perturbations $A \ll 1$ and $B_2 \ll 1$. The linearized evolution equation for these perturbations is
\begin{equation}
d_{TT} F + \frac{|B_1|^2}{2 k_2 \sqrt{k^2+2}} F = 0 \, , 
\end{equation}
where $F=A$ or $F=B_2$. The $b$-wave is stable, in the sense that the perturbations of $A$ and $B_2$ oscillate but remain small.

By contrast, if one starts off with a single $a$-wave ($A \neq 0$, $B_1=B_2=0$) and perturbs it with infinitesimal $B_1$ and $B_2$, the subsequent evolution of these perturbations is governed by
\begin{equation}
d_{TT} B_{1; 2} - \frac{|A|^2}{k_1 k_2} B_{1; 2} = 0 \, , \label{instA}
\end{equation}
which admits some exponentially growing solutions: the $a$-wave is unstable and decays spontaneously into two $b$-waves. In terms of particles physics, this instability would describe the disintegration of a massive particle (e.g. the $\sigma$-meson) into two massless Goldstone bosons (two pions).

The growth rate $\gamma = {|A|}/\sqrt{k_1 k_2} $ of this instability is a first indication of nonlocal interactions in Fourier space: recalling the resonance conditions, $\gamma$ is maximal when $k_1 k_2 = 1/2$ (see condition (\ref{nl_cond})), i.e., when $k_1 = (\sqrt{2+k^2} + k)/2 $ and $k_2= (\sqrt{2+k^2} -k)/2$ (or vice-versa). Nonlocal interactions occur in at least two cases: if $k \ll 1$, then $k_1 \simeq k_2 \simeq \sqrt 2 \gg k$. If $k \gg 1$, then either  $k_1 \ll k$ or $ k_2 \ll k$.
Such nonlocal interactions are an important feature of wave turbulence in the KGMH model.

\begin{figure}
\begin{center}
\includegraphics[width=180 mm]{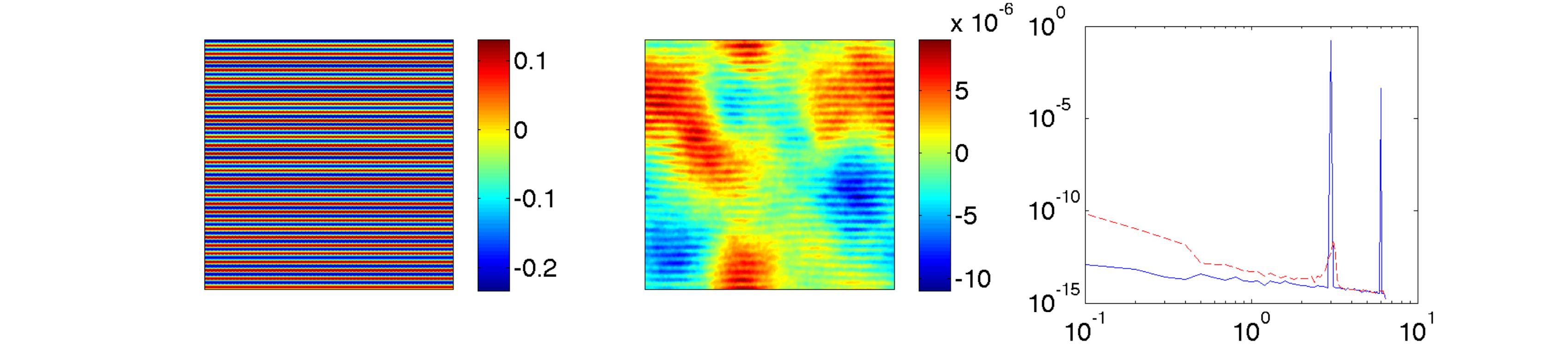}
\includegraphics[width=180 mm]{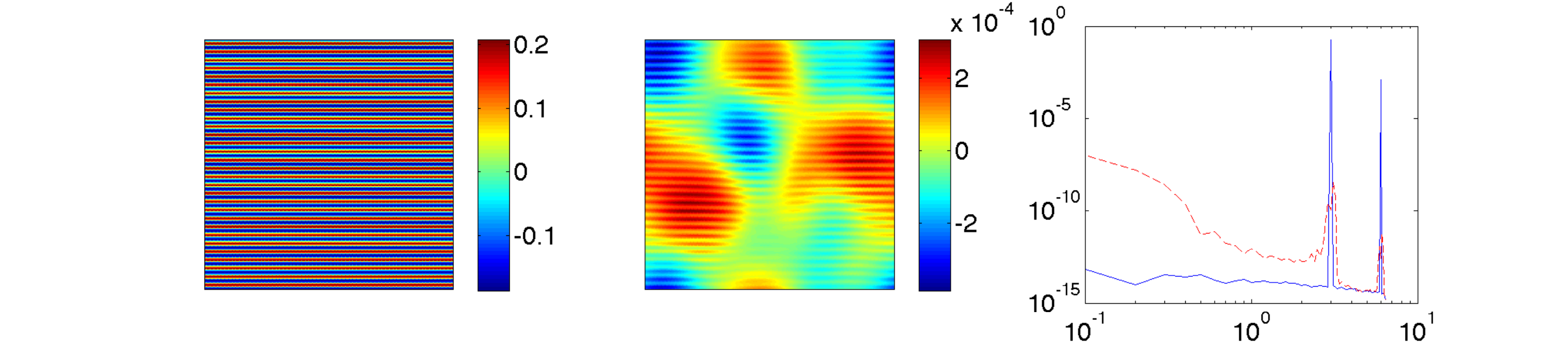}
\includegraphics[width=180 mm]{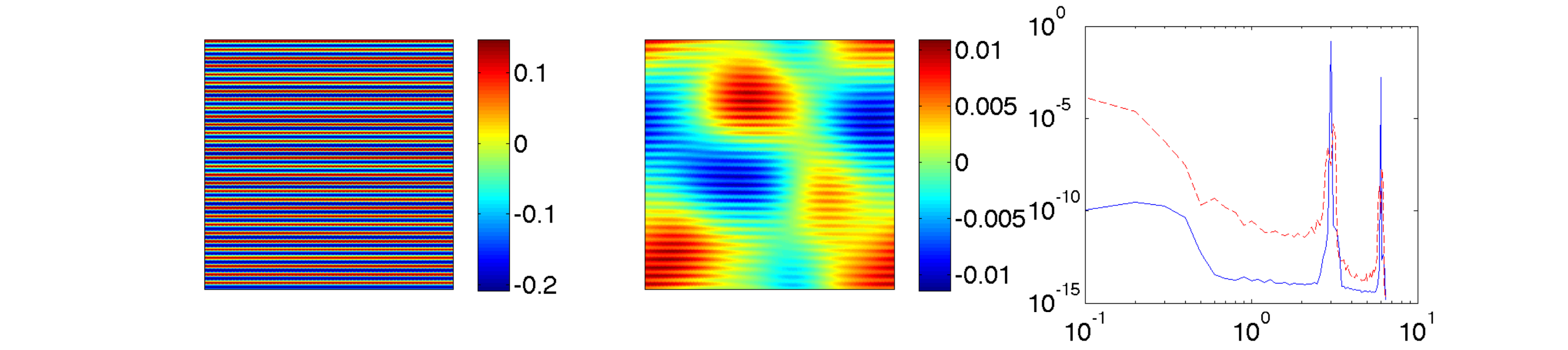}
\includegraphics[width=180 mm]{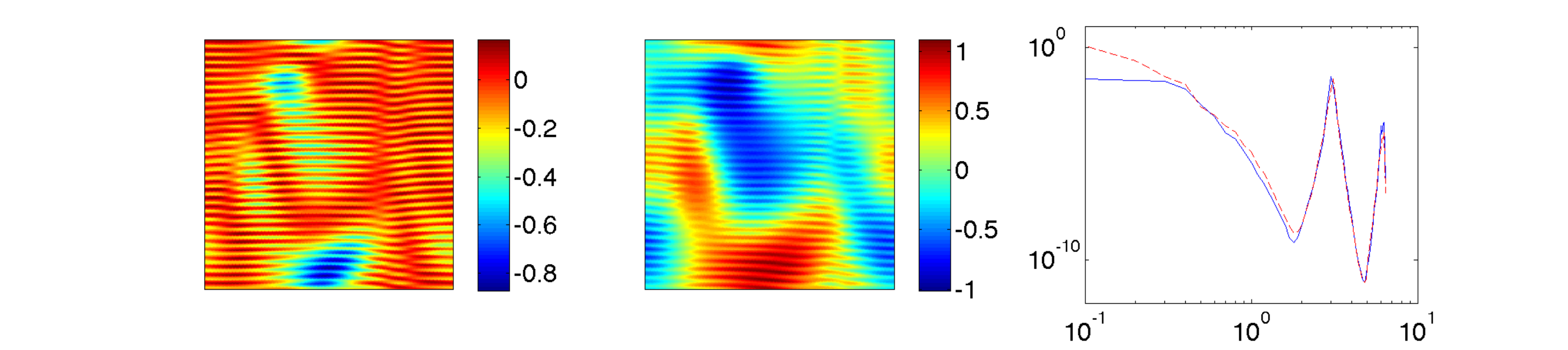}
\caption{Destabilization of a small-scale massive wave into massless ones. The left-hand panels are the $\lambda$ fields, the middle panels are the $\chi$ fields, and the right-hand panels are the 1D spectra of $\lambda$ (blue) and $\chi$ (red-dashed) as a function of $k$. Here $L=20 \pi$, and from top to bottom: $t=0$, $t=30$, $t=60$, and $t=120$. The unstable triads involve a small-scale massless mode and a large-scale one. \label{fig2}}
\end{center}
\end{figure}

\begin{figure}
\begin{center}
\includegraphics[width=180 mm]{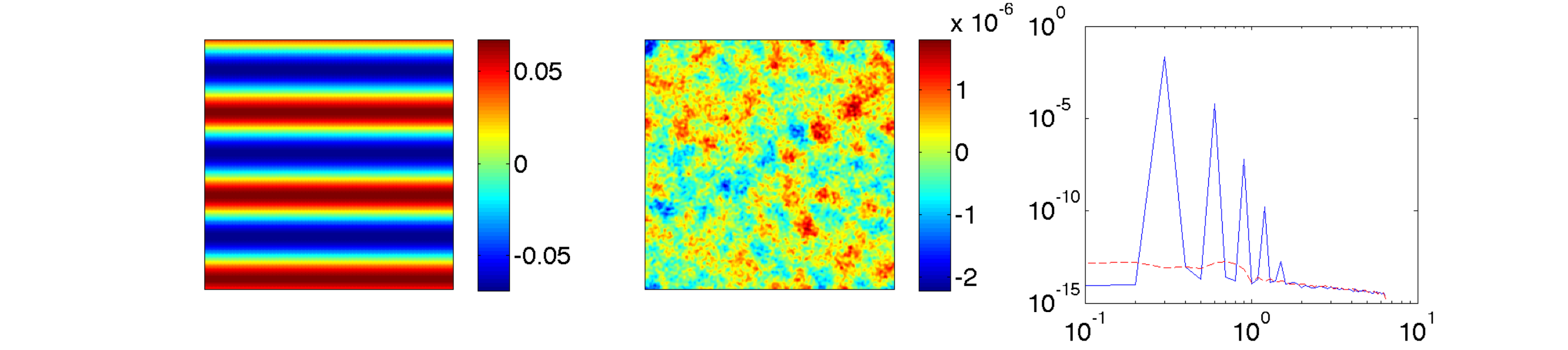}
\includegraphics[width=180 mm]{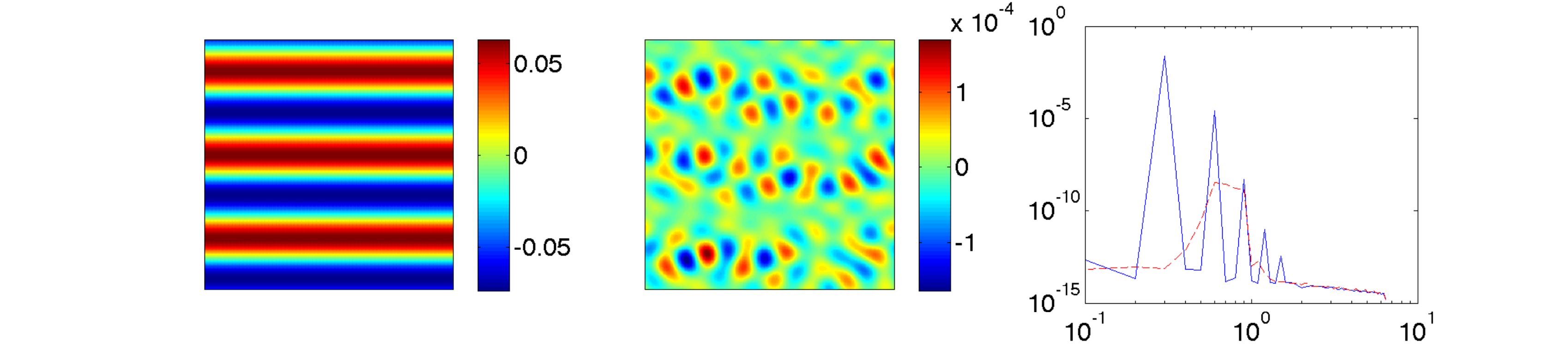}
\includegraphics[width=180 mm]{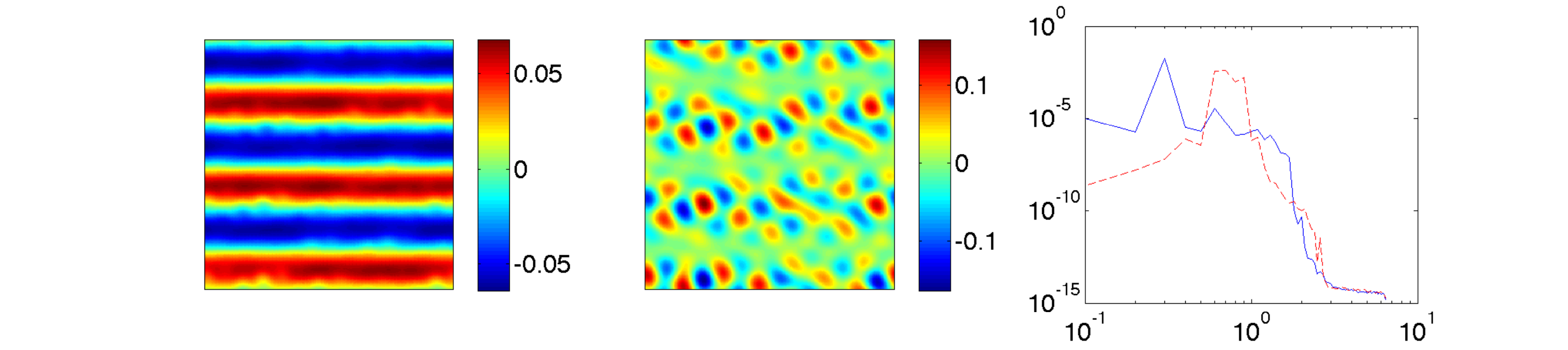}
\includegraphics[width=180 mm]{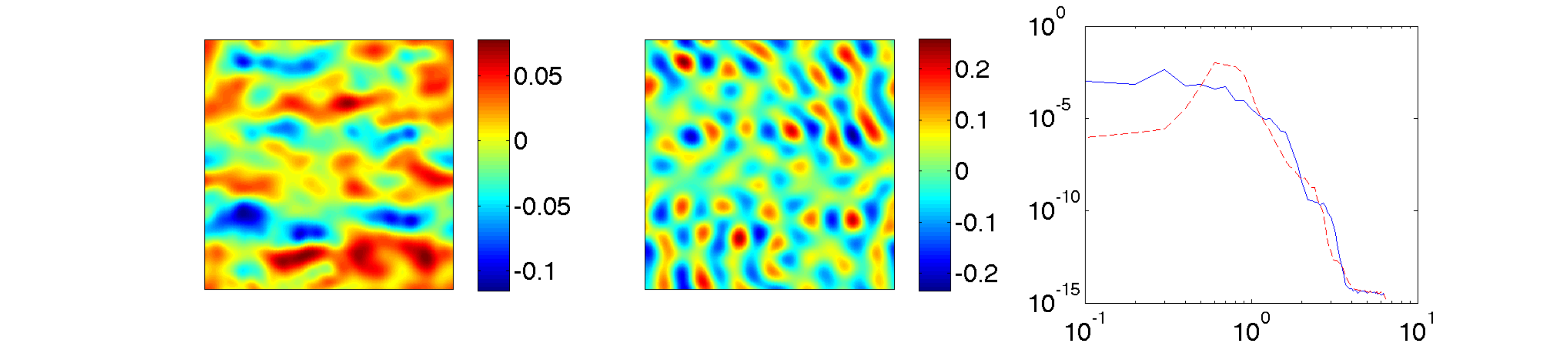}
\caption{Destabilization of a large-scale massive wave into massless ones. The left-hand panels are the $\lambda$ fields, the middle panels are the $\chi$ fields, and the right-hand panels are the 1D spectra of $\lambda$ (blue) and $\chi$ (red-dashed) as a function of $k$. Massless waves appear at the crossover scale $k \simeq 1$. Here $L=20 \pi$, and from top to bottom: $t=0$, $t=120$, $t=270$, and $t=570$. \label{fig2bis}  }
\end{center}
\end{figure}

To study the subsequent evolution of this instability, we solve equation (\ref{ctkg}) numerically. We focus on 2D numerical simulations, which corresponds to the dimension of the equivalent lattice of pendulums. We truncate the system at large wavenumbers. Such a truncation naturally arises from the discreteness of the lattice of pendulums. For more general applications of equation (\ref{ctkg}), the truncation can be seen as a crude modelization of the ultraviolet cut-off arising from Bose statistics.

Note that the domain size is an important parameter of this problem: the dispersion relation $\omega_k^a$ is not self-similar, with non-relativistic waves for $k \ll 1$ and relativistic ones for $k \gg 1$. To accurately describe a mixture of relativistic and non-relativistic waves, we need to choose a domain size that is much larger than $2 \pi$. 

The numerical code is described in appendix \ref{appcode}: it uses a standard pseudo-spectral method, with an exact integration of the stiff linear wave operator and Adam-Bashforth time-stepping to deal with the nonlinear terms. Dealiasing is performed using the one-half rule for this equation with cubic nonlinearity.

In this section the domain size is $[0, 20 \pi]^2$ with a resolution of $256^2$. After dealiasing we keep 128 modes in each direction: ${\bf k} \in \frac{1}{10} \times \{-64;64\}^2$. In figure \ref{fig2} we show snapshots of the $\lambda$ and $\chi$ fields, together with their one-dimensional spectra, i.e., their spectra integrated over the direction of ${\bf k}$. The initial condition is a small-scale progressive $a$-wave with ${\bf k}=(0;3)$, plus some weak background noise. The instability rapidly sets in and $b$-waves appear. As can be seen in equation (\ref{instA}), the maximum growth rate of the instability is achieved when the product $k_1 k_2$ of the two $b$ wavenumbers is small. Minimizing this quantity while satisfying the resonance conditions suggest that $k_1 \ll 1$ and $k_2 \sim k$. Indeed, in figure \ref{fig2} we observe the emergence a combination of large-scale $b$-waves, together with $b$-waves around the wavenumber $k$ of the initial $a$-wave.

By contrast, in figure \ref{fig2bis} we show similar snapshots for a simulation initiated with a large-scale $a$-wave, with ${\bf k}=(0;0.3)$. Once again, $b$-waves rapidly appear as a result of the decay instability. However, for such a large-scale $a$-wave, the  frequency resonance condition imposes that one of the $b$-waves has a frequency (and therefore a wavenumber) of order unity. But then the resonance of the three wave vectors imposes that the second $b$-wave also has a wave number of order unity and almost antiparallel to that of the first $b$-wave. As a consequence, in figure \ref{fig2bis} the instability sets in with $b$-waves having wave numbers of order unity.

In the following we extend the description of these nonlocal interactions to random ensembles of waves, using the framework of wave turbulence.

\section{Wave turbulence}

We now consider ensembles of weakly-nonlinear waves with homogeneous statistics in an infinite domain. For such systems wave turbulence can be used to derive wave kinetic equations that govern the evolution of the wave spectra.

\label{WT}

\subsection{Wave  kinetic equations}

Let us define the wave action spectra of the modes $a$ and $b$:
$$
n^a_{\bf k} = \left( \frac {L}{2\pi} \right)^d \langle |a_{\bf k}|^2 \rangle \, ,
\quad n^b_{\bf k} = \left( \frac {L}{2\pi} \right)^d \langle  |b_{\bf k}|^2 \rangle \, ,
$$
here the angular brackets denote an ensemble average over many realizations. 

In appendix \ref{ap:wt}, we apply standard wave turbulence techniques to equations (\ref{ctkg'''1}) and (\ref{ctkg'''2}), which leads to the following kinetic equations for the evolution of the two spectra,


\begin{eqnarray}\label{kea1}
\dot n^a_{\bf k} & = &4 \pi \int      |V^{\bf k}_{12}|^2 
\, (n^b_1 n^b_2 - 2 n^b_1  n^a_{\bf k})
\delta (\omega_{{\bf k} 12}^{abb} ) 
\delta^{\bf k}_{12} \, d{\bf k}_1 d{\bf k}_2 ,
\\\label{keb1}
\dot n^b_{\bf k} & = & 8\pi \int      |V^1_{2{\bf k}}|^2 
\, (n^a_1 n^b_2 + n^a_1  n^b_{\bf k} -n^b_2  n^b_{\bf k})
\delta (\omega_{12{\bf k} }^{abb} ) 
\delta^1_{2{\bf k}}\, d{\bf k}_1 d{\bf k}_2 .
\end{eqnarray}

\subsection{Conservation laws for the kinetic equations}

Kinetic equations  (\ref{kea1}) and (\ref{keb1}) conserve the
wave energy,
\begin{equation}\label{En}
{\cal E} = L^d \int ( \omega_{\bf k} ^a n_{\bf k}^a + \omega_{\bf k} ^b n_{\bf k}^b) \, d{\bf k},
\end{equation}
as well as the following particle invariant,
\begin{equation}\label{partic}
{\cal N} = L^d \int ( 2 n_{\bf k}^a +  n_{\bf k}^b) \, d{\bf k}.
\end{equation}
This second invariant is nontrivial, and it can be guessed only once the dominant three-wave process is identified. Indeed, the process $a \rightleftharpoons b+b$ is similar to a chemical reaction where $a$ is a ``molecule" made of two $b$ ``atoms", and ${\cal N}$ would be the total number of such $b$ ``atoms": 1 per $b$-wave, and 2 per $a$-wave. ${\cal N}$ is also the number of $b$ particles that would be in the system if the reaction $a \rightarrow b+b$ were complete. For brevity, in the following we simply refer to ${\cal N}$ as the number of particles.

An important comment to make here is that ${\cal N}$ is only an adiabatic invariant: it is conserved by the weakly nonlinear dynamics, but there is no similar invariant for the strongly nonlinear KGMH equation. By contrast, ${\cal E}$ is the quadratic part of the global invariant $E$, the latter being conserved even in the strongly nonlinear regime.

\subsection{Thermodynamic equilibrium\label{thermoeq}}

One can easily check that the kinetic equations admit solutions corresponding to energy equipartition in ${\bf k}$-space,
\begin{equation}\label{RJe}
n_{\bf k}^a = \frac T{\omega_{\bf k} ^a}, \quad  n_{\bf k}^b = \frac T{\omega_{\bf k} ^b} \, ,
\end{equation}
for any temperature $T=$~const  \footnote{This definition of $T$ is common in wave turbulence. In the energy equipartition state, it corresponds to an energy $k_B T$ per oscillatory degree of freedom, where $k_B$ is a dimensionless Boltzmann constant, $k_B=(2 \pi)^d$ in $d$ dimensions.}, and solutions corresponding to an equipartition of the particle invariant,
\begin{equation}\label{RJn}
n^b_{k} =2n^a_{k} = \hbox{const}.
\end{equation}

More generally, the equilibrium Rayleigh-Jeans (or thermodynamic) solution is given by
\begin{equation}\label{RJg}
n_{\bf k}^a = \frac T{\omega_{\bf k} ^a + 2 \mu}, \quad  n_{\bf k}^b = \frac T{\omega_{\bf k} ^b +\mu},
\end{equation}
where $\mu=$~const is a chemical potential.

%
%

Finally, we remark that any initial condition with a vanishing spectrum $n_{\bf k}^b = 0$ and an arbitrary spectrum $n_{\bf k}^a $ is a stationary solution to the kinetic equations  (\ref{kea1}) and (\ref{keb1}). One can also show that spectra with $b$-waves at low wavenumber only are also steady solutions to the kinetic equations. Such distributions with only a single type of particles are described in details in sections \ref{statdecay} and \ref{dyncond}.

\subsection{Isotropic systems}

\begin{figure}
\begin{center}
\includegraphics[width=60 mm]{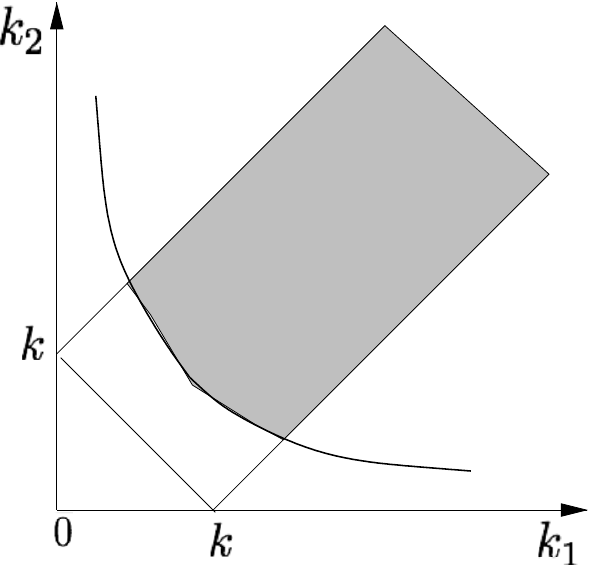} \qquad
\includegraphics[width=60 mm]{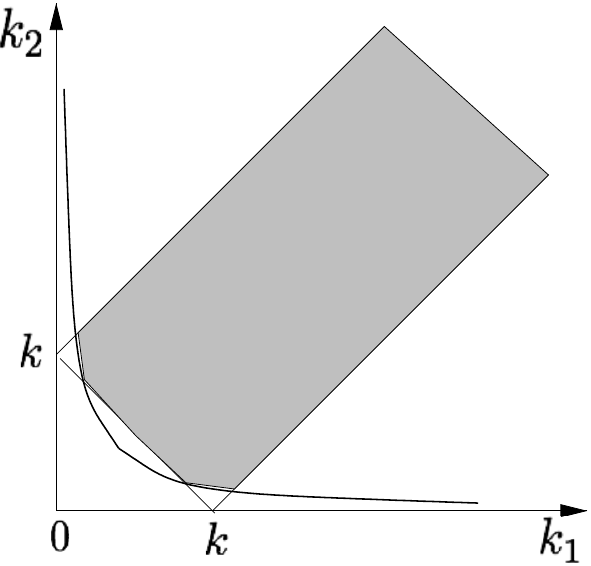}
\caption{Integration domain ${\cal D}_a(k)$ for $k < 1/\sqrt{2}$ (left-hand panel) and $k >1/\sqrt{2}$ (right-hand panel).}\label{fig:int1a}
\end{center}
\end{figure}

\begin{figure}
\begin{center}
\includegraphics[width=60 mm]{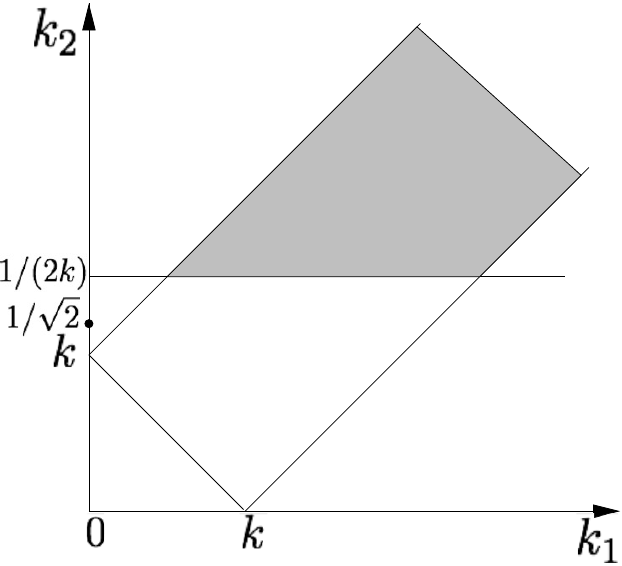} \qquad
\includegraphics[width=60 mm]{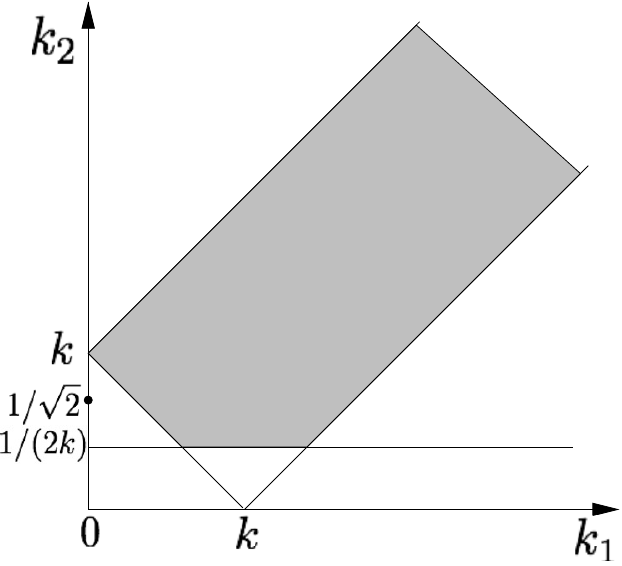}
   \caption{Integration domain ${\cal D}_b(k)$ for $k < 1/\sqrt{2}$ (left-hand panel) and $k > 1/\sqrt{2}$ (right-hand panel).}\label{fig:int1b}
\end{center}
\end{figure}

Consider the class of isotropic spectra, $n_{\bf k}^a \equiv n_{ k}^a $ and $n_{\bf k}^b \equiv n_{ k}^b $, where 
$k =|{\bf k}| $, and integrate out the angular variables describing the directions of ${\bf k}_1$ and ${\bf k}_2$ in the kinetic equations  (\ref{kea1}) and (\ref{keb1}).
Because $V^{\bf k}_{12} \equiv \frac{1}{
\sqrt{ \omega^a_{k} \omega^b_{2} \omega^b_{3}}}$ and $\delta (\omega_{{\bf k} 12}^{abb} ) \equiv \delta (\omega_{{k} 12}^{abb} ) $ are independent of the directions of ${\bf k}_1$ and ${\bf k}_2$, all that needs to be integrated
over the angular variables is $\delta^{\bf k}_{12}$. This leads to
\begin{eqnarray}\label{keaI}
\dot n^a_{ k} & = & 4 \pi \int_{{\cal D}_a(k)}      \frac{S_{12k} \delta (\omega_{k12{ } }^{abb} ) }{
{ \omega^a_{k} \omega^b_{1} \omega^b_{2}}}
\, (n^b_1 n^b_2 - 2 n^b_1  n^a_{ k})
\, ({ k}_1 { k}_2)^{d-1} \, d{ k}_1 d{ k}_2 \, ,
\\ \label{kebI}
\dot n^b_{k} & = & 8\pi \int_{{\cal D}_b(k)}       \frac{S_{12k} \delta (\omega_{12{ k} }^{abb} ) }{
{ \omega^a_{1} \omega^b_{k} \omega^b_{2}}}
\, (n^a_1 n^b_2 + n^a_1  n^b_{ k} -n^b_2  n^b_{k})
\, ({ k}_1 { k}_2)^{d-1}  \, d{ k}_1 d{ k}_2 \, ,
\end{eqnarray}
where
\begin{equation}\label{S2}
S_{12k} = \frac 1{2
\sqrt{ 2(k^2 k_1^2 +  k^2 k_2^2 + k_1^2 k_2^2) -k^4 - k_1^4 -k_2^4}}
\quad \hbox{for} \; d=2 \, ,
\end{equation}
and
\begin{equation}\label{S3}
 S_{12k} = \frac 1{2 k k_1 k_2} \quad \hbox{for} \; d=3 \, .
\end{equation}
The integration domains ${\cal D}_a(k)$ and ${\cal D}_b(k)$ in equations (\ref{keaI}) and (\ref{kebI}) are determined by the triangle inequalities for 
$k, k_1$ and $k_2$ and the additional restriction of type (\ref{nl_cond}):
\begin{eqnarray}
{\cal D}_a(k) & = & \left\{(k_1,k_2) \in (\mathbb{R}^+)^2 \, | k_1 k_2 \geq \frac{1}{2}; \, k_1 + k_2 \leq k ; \, k +k_1 \leq k_2;  k +k_2 \leq k_1 \right\} \, ,\\
{\cal D}_b(k) & = & \left\{(k_1,k_2) \in (\mathbb{R}^+)^2 \,  | \, k_2 \geq \frac{1}{2k}; \, k_1 + k_2 \leq k ; \, k +k_1 \leq k_2;  k +k_2 \leq k_1 \right\}  \, .
\end{eqnarray}
They are sketched respectively in figures \ref{fig:int1a} and \ref{fig:int1b}.

%
%
%
%

Substituting $\omega^b_{1} \omega^b_{2} = { k}_1 { k}_2$ and $\omega^b_{k} \omega^b_{2} = { k} { k}_2$
we have
\begin{eqnarray}\label{keaIb}
\dot n^a_{ k} & = &\frac{4 \pi} { \omega^a_{k}} \int_{{\cal D}_a(k)}    S_{12k}    \delta (\omega_{k12{ } }^{abb} ) 
\, (n^b_1 n^b_2 - 2 n^b_1  n^a_{ k})
\, ({ k}_1 { k}_2)^{d-2} \, d{ k}_1 d{ k}_2 \, ,
\\ \label{kebIb}
\dot n^b_{k} & = & \frac{8 \pi} { {k}}  \int_{{\cal D}_b(k)}     \frac{k_1}{\omega_1^a} S_{12k}    \delta (\omega_{12{ k} }^{abb} ) 
\, (n^a_1 n^b_2 + n^a_1  n^b_{ k} -n^b_2  n^b_{k})
\, ({ k}_1 { k}_2)^{d-2}  \, d{ k}_1 d{ k}_2 \, .
\end{eqnarray}
Substitution of $S_{12k}$ for $d=3$ leads to an even simpler system,
\begin{eqnarray}\label{keaIb3}
\dot n^a_{ k} & = & \frac{2 \pi} { k \omega^a_{k}} \int_{{\cal D}_a(k)}        \delta (\omega_{k12{ } }^{abb} ) 
\, (n^b_1 n^b_2 - 2 n^b_1  n^a_{ k})
\, d{ k}_1 d{ k}_2 \, ,
\\ \label{kebIb3}
\dot n^b_{k} & = & \frac{4 \pi} { k^2}  \int_{{\cal D}_b(k)}      \frac{k_1}{\omega_1^a}   \delta (\omega_{12{ k} }^{abb} ) 
\, (n^a_1 n^b_2 + n^a_1  n^b_{ k} -n^b_2  n^b_{k})
 \, d{ k}_1 d{ k}_2 \, .
\end{eqnarray}

\subsection{Absence of Kolmororov-Zakharov spectra}
\label{sec:kz}


The central feature of wave turbulence theory is that one can compute power-law spectra satisfying the kinetic equations. Such Kolmororov-Zakharov (KZ) spectra are obtained for scale-invariant systems, i.e., when the dispersion relations and interaction coefficients are homogeneous functions of their arguments. In the KGMH model, this is the case only in the relativistic limit, $k \gg 1$. Let us focus on this limit and exclude 2D systems: wave turbulence kinetic equations are known to be ill-posed in 2D for non-dispersive waves, $\omega \sim k$, see e.g. \cite{2DacousticWT}. In 3D, the relativistic limit of the isotropic kinetic equations (\ref{keaIb3}) and (\ref{kebIb3}) takes the simple form

\begin{eqnarray}\label{keaIr}
\dot n^a_{ k} =\frac{2 \pi}{k^2} \int      
\, (n^b_1 n^b_2 - 2 n^b_1  n^a_{ k})\, \delta (k-k_1-k_2)
  \, d{ k}_1 d{ k}_2 ,
\\ \label{kebIr}
\dot n^b_{k} =\frac{4\pi}{k^2} \int       
\, (n^a_1 n^b_2 + n^a_1  n^b_{ k} -n^b_2  n^b_{k}) \, \delta (k_1-k-k_2)
\, d{ k}_1 d{ k}_2 ,
\end{eqnarray}


In appendix \ref{ap:kz}, we consider power law spectra and show that there exists no KZ solutions with local interactions for these relativistic kinetic equations. The only
power-law solutions to these equations are the relativistic limits of the thermodynamic spectra corresponding to equipartition of energy (\ref{RJe}) or of the particle invariant (\ref{RJn}).

We are thus left with two routes to investigate the spectral evolution of the  KGMH system: in the next section we characterize the thermodynamic states reached in the long-time limit and study the wave condensation phenomenon. In sections \ref{statdecay} and \ref{dyncond} we come back to the nonlocal interactions that govern the out-of-equilibrium dynamics and we derive rigorously a nonlocal KZ state characterized by a nonlocal transfer of particles to a strong condensate, together with a local energy cascade in $k$-space. This ``dynamical condensation" describes the approach to the thermodynamic condensed state, and therefore makes the link between the two routes.


\section{Bose-Einstein condensation}

\label{bec}

\begin{figure}
\begin{center}
\includegraphics[width=120 mm]{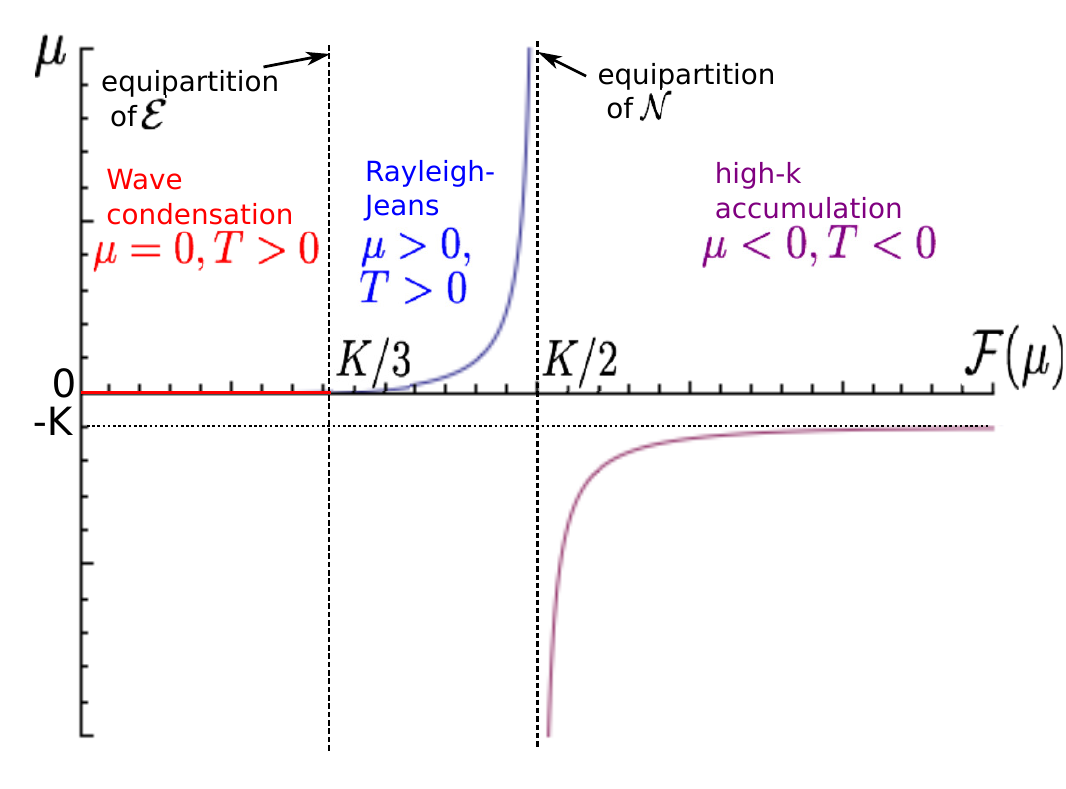} 
\caption{The chemical potential $\mu$ as a function of the ``energy per particle'' $\mathcal{F}=\mathcal{E}/\mathcal{N}$. The usual Rayleigh-Jeans spectrum is a solution for intermediate values of $\mathcal{F}$ only. For low $\mathcal{F}$ condensation occurs at low $k$, whereas for large $\mathcal{F}$ energy and particles accumulate at the cutoff wavenumber $K$.\label{fig7}}
\end{center}
\end{figure}

We now study equilibration in truncated conservative systems.
Non-dissipative wave systems are known to exhibit a condensation phenomenon when (i) there are two conserved quantities, (ii) there is a truncation at some finite wavenumber $K$, and (iii) the spatial dimension $d$ is large enough. In discrete systems such as the array of pendulums described in section \ref{analogue}, a truncation in $k$-space naturally arises. In quantum systems, the Bose statistics ensures that modes with large enough energy remain unoccupied, hence an effective truncation.

Condensation of classical waves was originally reported for the Gross-Pitaevskii equation \cite{PhysRevLett.95.263901,During}. Further studies on condensation also considered equations of the Klein-Gordon type \cite{Berges}. For all these systems one can compute KZ solutions corresponding to an inverse cascade of the particle invariant. Condensation then naturally occurs, as the inverse cascade induces an accumulation of particles at low $k$. In contrast with these systems, we did not find any such KZ inverse cascade for the MHKG equation. Furthermore, the particle invariant is only an adiabatic one. In the following,  we follow an approach similar to Connaughton et al. \cite{PhysRevLett.95.263901} to show theoretically and numerically that condensation nevertheless occurs.

We consider decaying solutions with fixed initial number of particles $\mathcal{N}$ and several values of the initial energy $\mathcal{E}$. After some transient, we expect the system to reach the thermodynamic spectra (\ref{RJg}), with a pair $(T,\mu)$ corresponding to a pair $(\mathcal{E},\mathcal{N})$. However, we will see that there are some values of $(\mathcal{E},\mathcal{N})$ for which no pair $(T,\mu)$ can be found with positive $\mu$.

From the thermodynamic spectra, we compute the numbers $N_a$ and $N_b$ of $a$ and $b$ particles, as well as the energy $E_a$ and $E_b$ contained in these two fields. Assuming for simplicity that $k \in [0,K]$ is a continuous variable and considering the 2D problem, we obtain
\begin{eqnarray}
N_a & = & \! 2 \pi L^2 T \int_0^K \frac{k}{2 \mu + \sqrt{2+k^2}} \mathrm{d}k \label{eq:bec1} \\
\nonumber & = &  2 \pi L^2 T \left[ \sqrt{2+K^2}-\sqrt{2}+ 2\mu \log \left( \frac{\sqrt{2}+2\mu}{\sqrt{2+K^2}+2\mu} \right)   \right]  \!\! \, ,  \\
N_b & = & 2 \pi L^2 T \int_0^K \frac{k}{ \mu +k} \mathrm{d}k = 2 \pi L^2 T \left[ K + \mu \log \left( \frac{\mu}{K+\mu} \right)   \right]  , \label{eq:bec2} \\
E_a & = & 2 \pi L^2 T \int_0^K \sqrt{2+k^2} \frac{k}{2 \mu + \sqrt{2+k^2}} \mathrm{d}k \\
\nonumber & = & 2\pi L^2 T \left[ \frac{K^2}{2}+2\mu(\sqrt{2}-\sqrt{K^2+2})+4\mu^2\log \left( \frac{\sqrt{2+K^2}+2\mu}{\sqrt{2}+2\mu} \right)   \right] \, , \label{eq:bec3} \\
E_b & = & 2 \pi L^2 T \int_0^K \frac{k^2}{ \mu +k} \mathrm{d}k = 2 \pi L^2 T \left[\frac{K}{2}(K-2\mu)+\mu^2 \log \left( 1+\frac{K}{\mu} \right) \right] \, .\label{eq:bec4}
\end{eqnarray}
We define the average energy per particle as
\begin{equation}
\mathcal{F}(\mu)=\frac{\mathcal{E}}{\mathcal{N}}=\frac{E_a+E_b}{2N_a+N_b} \, . \label{EsurN}
\end{equation}
$\mathcal{F}$ is independent of temperature, and can be specified at the beginning of a numerical simulation by an appropriate choice of the initial condition. However, not all values of $\mathcal{F}$ are compatible with the thermodynamic spectra. Indeed, we show in Figure \ref{fig7} a plot of $\mu$ as a function of $\mathcal{F}$. There is a minimum value of $\mathcal{F}$ under which $\mu$ becomes negative: the thermodynamic spectrum is not a valid solution anymore, and wave condensation occurs. In the limit of large $K$, the threshold is $\mathcal{F}_{cr} = K/3$.

We stress the fact that in the present 2D system, condensation occurs even in the infinite domain limit. This does not contradict the Mermin-Wagner theorem, because the dispersion relation of the $b$-waves goes to zero slow enough ($\omega^b_k \sim k$ as $k\to 0$) for the integrals (\ref{eq:bec1}) and (\ref{eq:bec2}) to converge at $k=0$, even when $\mu =0$: for a given value of the energy, only a finite number of particles can be in the Rayleigh-Jeans spectrum, and the rest of the particles have to condense. By contrast, in the 2D Gross-Pitaevskii equation the dispersion relation is $\omega_k = k^2$, and the number of particles contained in the Rayleigh-Jeans distribution involves a diverging integral for $\mu=0$: for a given value of the energy, an arbitrary large number of particles can be in the Rayleigh-Jeans spectrum, and condensation does not occur in 2D. For a system with dispersion relation $\omega_k \sim k^\sigma$ in dimension $d$, the criterion for Bose-Einstein condensation to occur is $\sigma < d$ \cite{Lacour}.

There is also a limiting value of $\mathcal{F}(\mu)$ that is attained for $\mu \to \infty$. In the limit of large $K$, this limit is $\mathcal{F}(\mu \to \infty)=K/2$. As one approaches this limiting value of $\mathcal{F}$, the spectrum tends to the equipartition of $\mathcal{N}$.
Obviously, for a finite number of particles, taking the limit $\mu \to \infty$ also requires $T \to \infty$.  If the initial values of $\mathcal{N}$ and $\mathcal{E}$ are such that $\mathcal{F}>\mathcal{F}(\mu \to \infty)$, then there is an accumulation of energy in the high $k$ modes. This accumulation corresponds to Rayleigh-Jeans thermodynamic spectra (\ref{RJg}) with negative values of $T$ and $\mu$.

To investigate numerically these different thermodynamic states, we perform numerical simulations with an initial condition $\phi_{\bf k}\sim k^{\alpha}$ and the same initial value ${\cal N} \simeq 44$. Large negative values of $\alpha$ correspond to low ${\cal F}$ and steep decreasing spectra, whereas positive $\alpha$ correspond to large values of $\mathcal{F}$. In the following we describe the stationary state attained after a long integration time. The size of the domain is $[0, 12 \pi]^2$ and the resolution is $128^2$. After dealiasing, we keep 64 modes in each direction, i.e., we keep wave numbers ${\bf k}$ with $k \leq K=32/6$ where the step is $\delta k = 1/6$ for the two components of ${\bf k}$. We plot the spectra of $a$ and $b$ modes, which we evaluate using $|a_{\bf k}|^2 \simeq  \omega_k^{(a)} |\hat{\lambda}_{\bf k}|^2$ and $|b_{\bf k}|^2 \simeq  \omega_k^{(b)} |\hat{\chi}_{\bf k}|^2$. More specifically, we show the 1D spectra $S_a(k)$ and $S_b(k)$ defined in such a way that $\int S_{a;b}(k) \mathrm{d}k$ is the variance of the corresponding field.

\begin{figure}
\begin{center}
\includegraphics[width=60 mm]{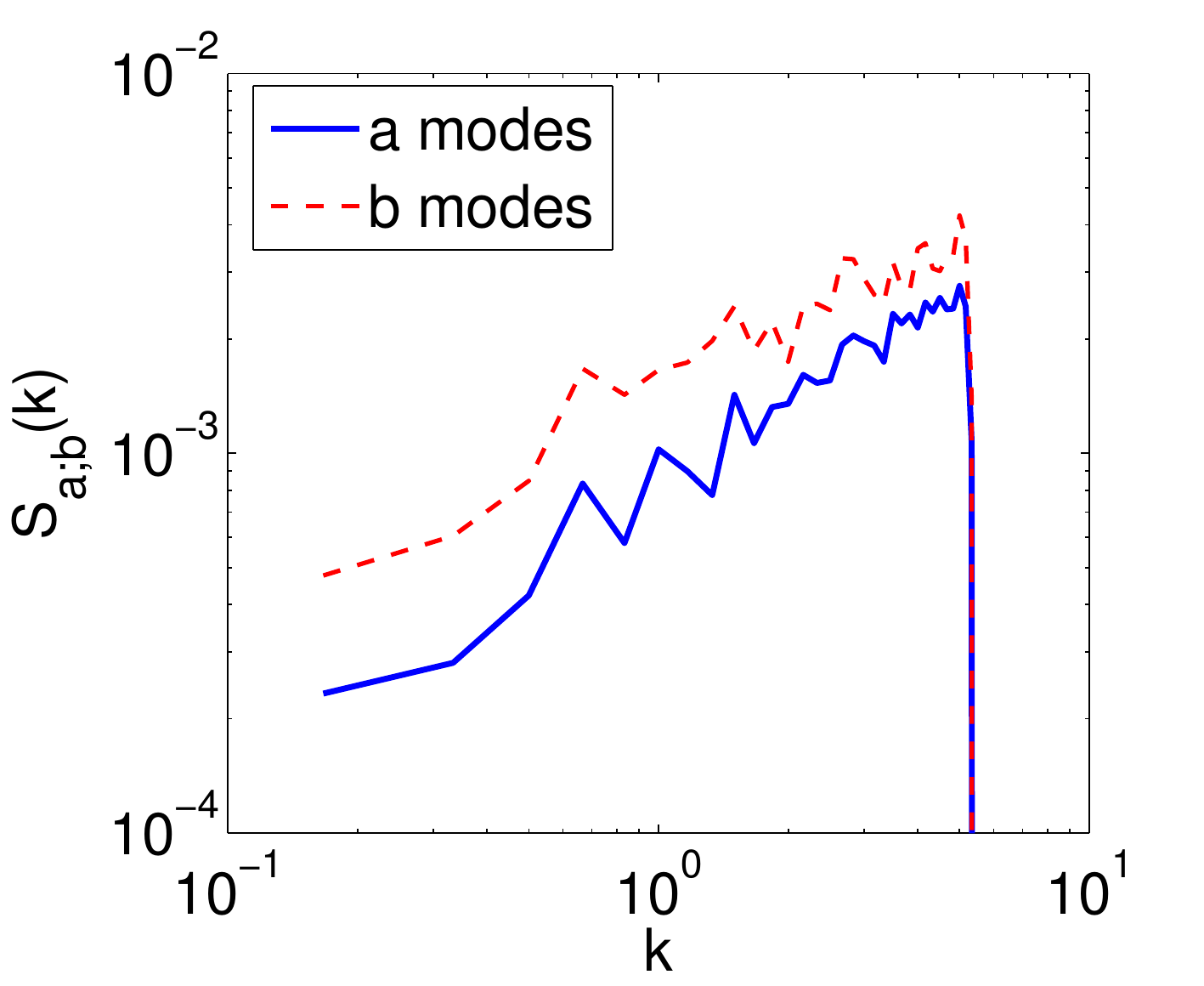} 
\includegraphics[width=60 mm]{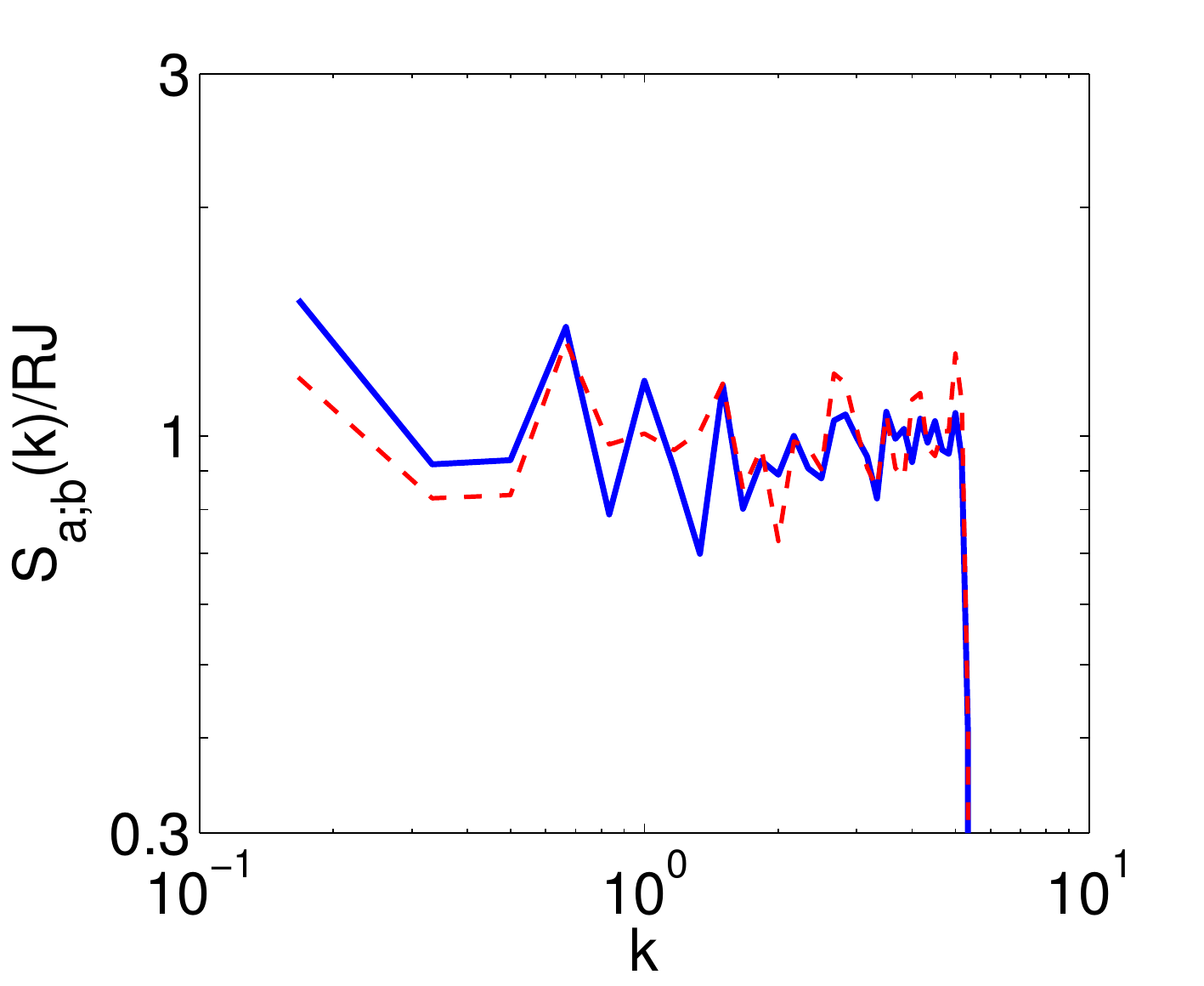} \\
\includegraphics[width=60 mm]{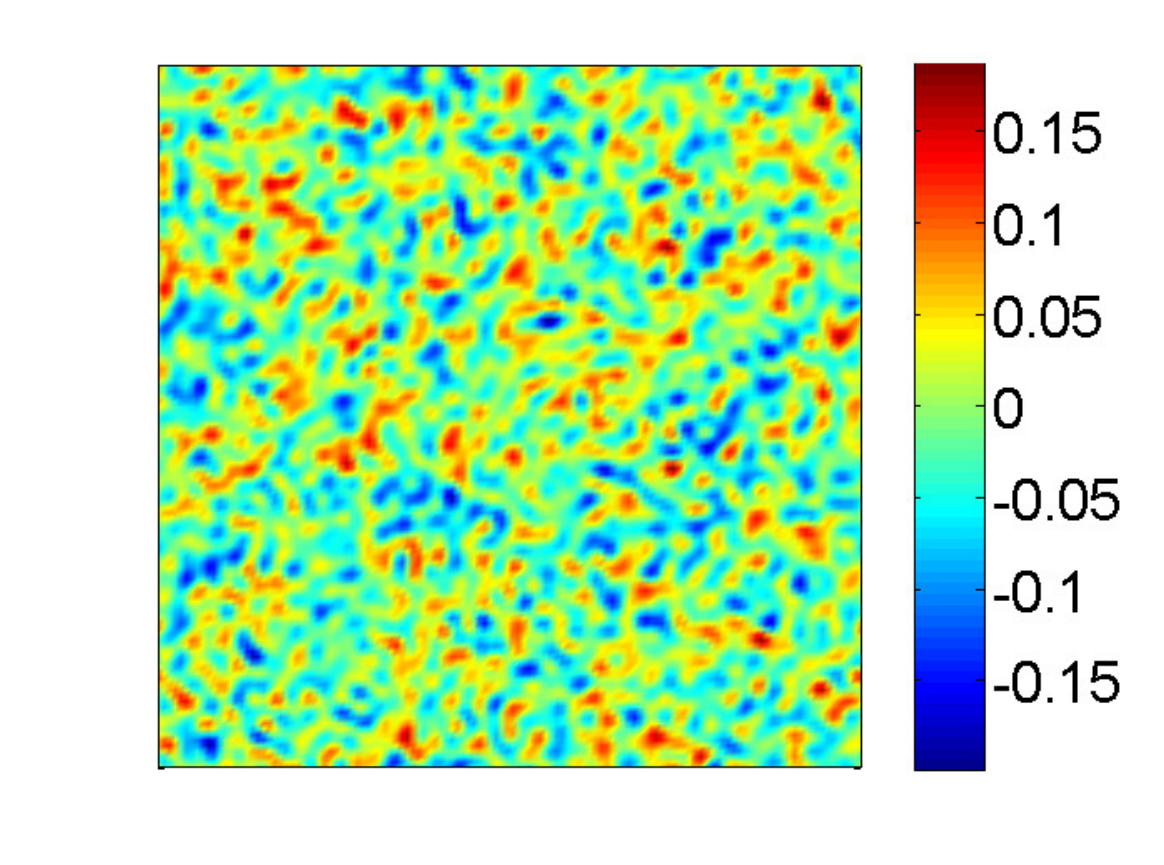} 
\includegraphics[width=60 mm]{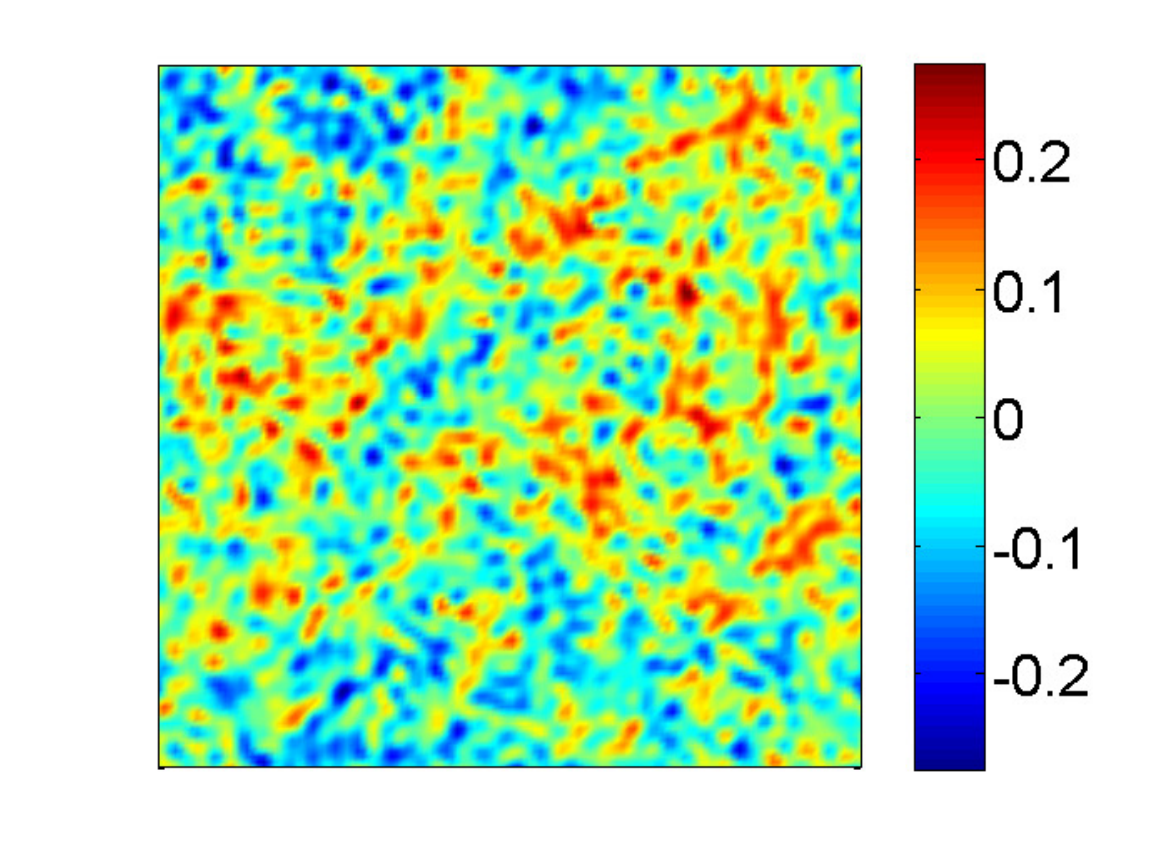} 
\caption{$\alpha=-0.5$: in the stationary state, $\mathcal{F}= 2.42 \in(K/3,K/2)$, so we expect a Rayleigh-Jeans distribution. Top: spectra of a and b modes (left), and spectra normalized by a Rayleigh-Jeans spectra with $\mu=1.67$ and $T=7 \times 10^{-4}$ (right). Bottom: snapshots at final time of the $\lambda$ and $\chi$ fields. \label{run3}}
\end{center}
\end{figure}

The four runs described in figures \ref{run3} to \ref{run10} illustrate respectively (i) a Rayleigh-Jeans state, (ii) the  threshold for wave condensation, (iii)  wave condensation, and (iv) accumulation at large $k$. In the latter simulation the system has not settled to a steady state: the ratio $\mathcal{F}=\mathcal{E}/\mathcal{N}$ very slowly decreases in time until it reaches its critical value and a Rayleigh-Jeans state. The drift  of $\mathcal{F}$ probably originates from the fact that $\mathcal{N}$ is only an adiabatic invariant: it is not conserved by the higher order nonlinearities that we discarded in the present weakly nonlinear study.

\begin{figure}
\begin{center}
\includegraphics[width=60 mm]{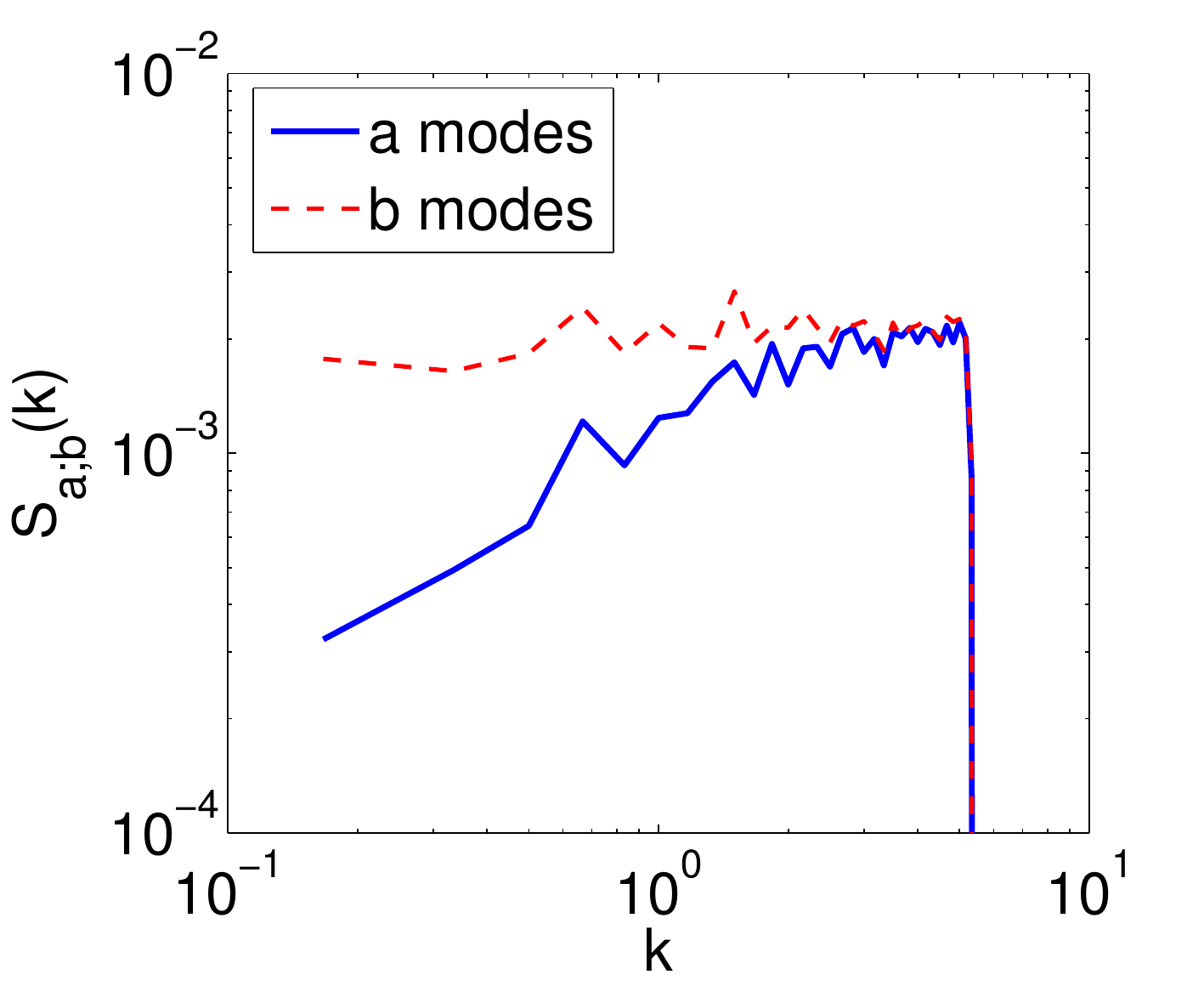} 
\includegraphics[width=60 mm]{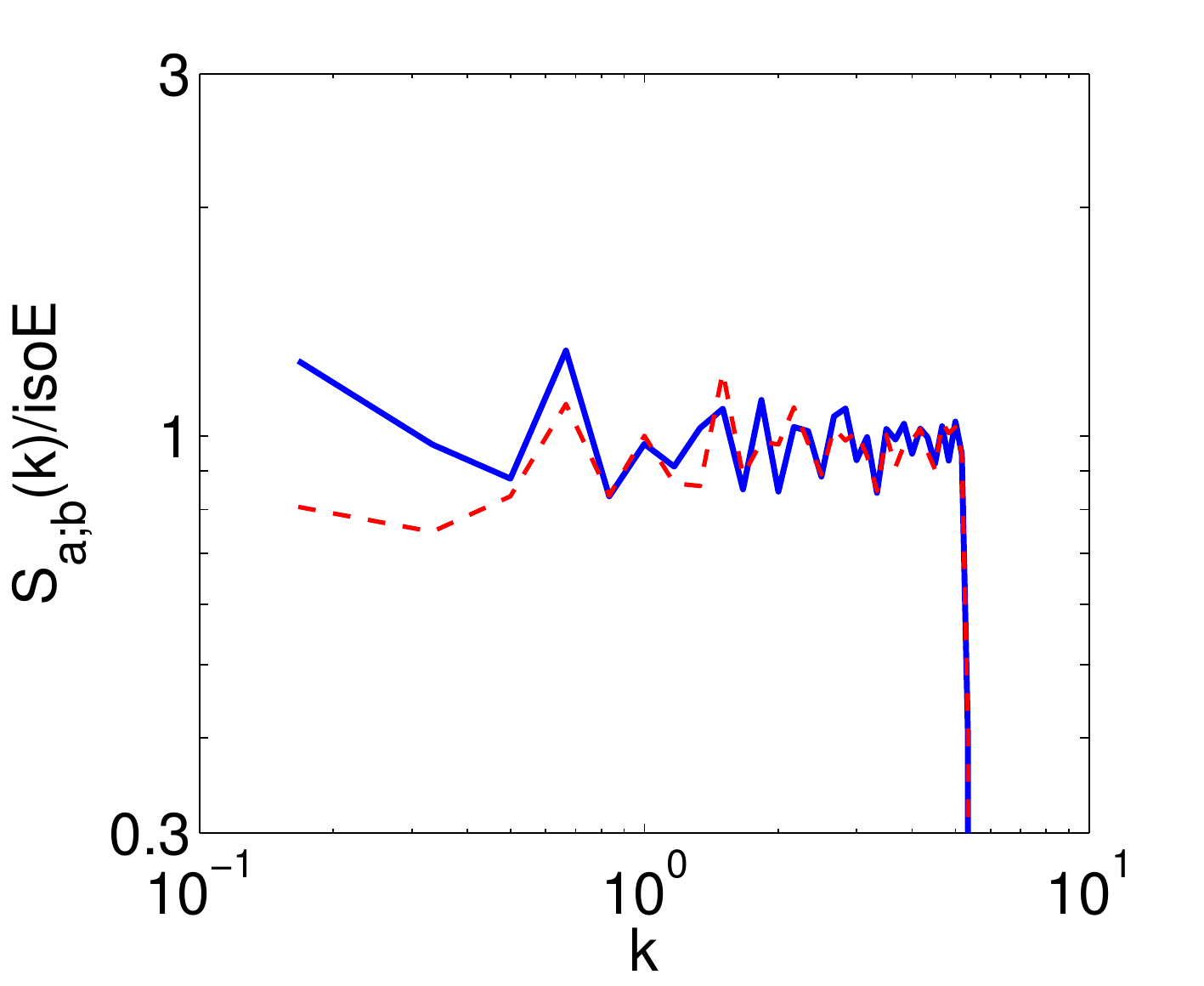} \\
\includegraphics[width=60 mm]{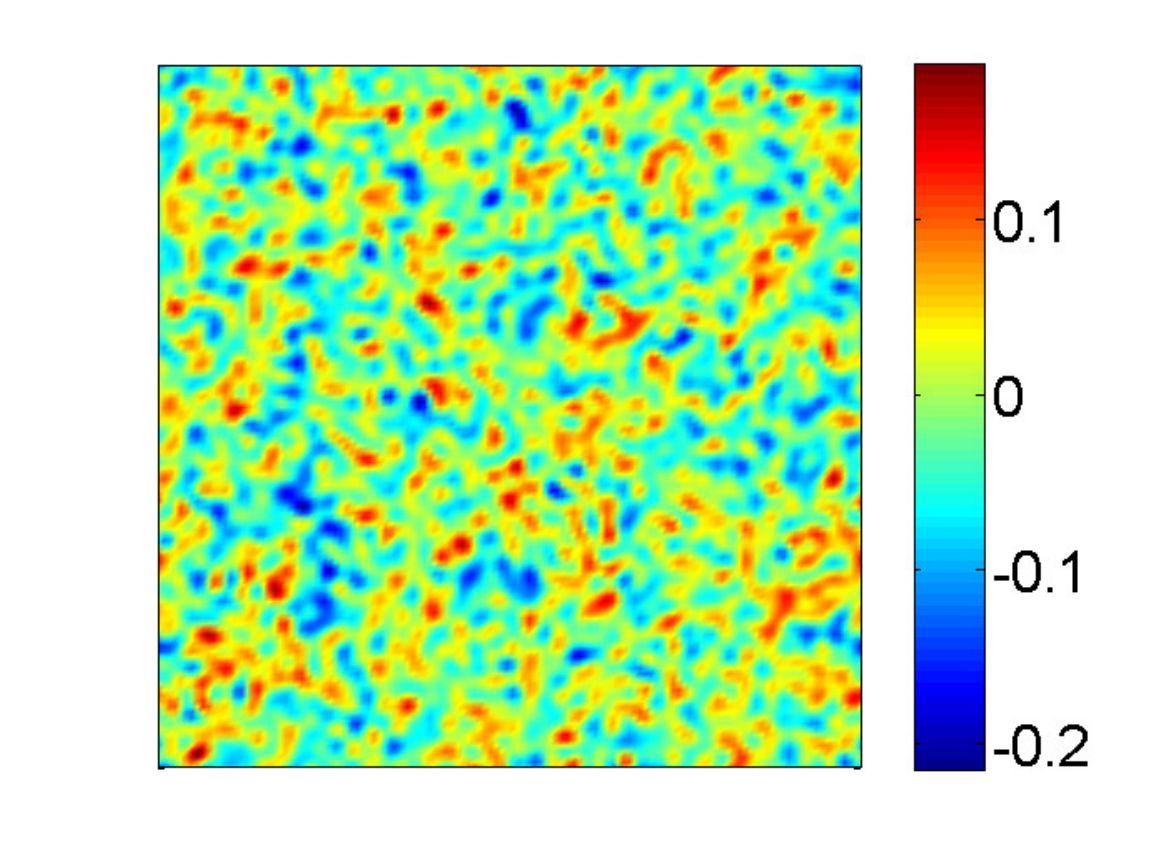} 
\includegraphics[width=60 mm]{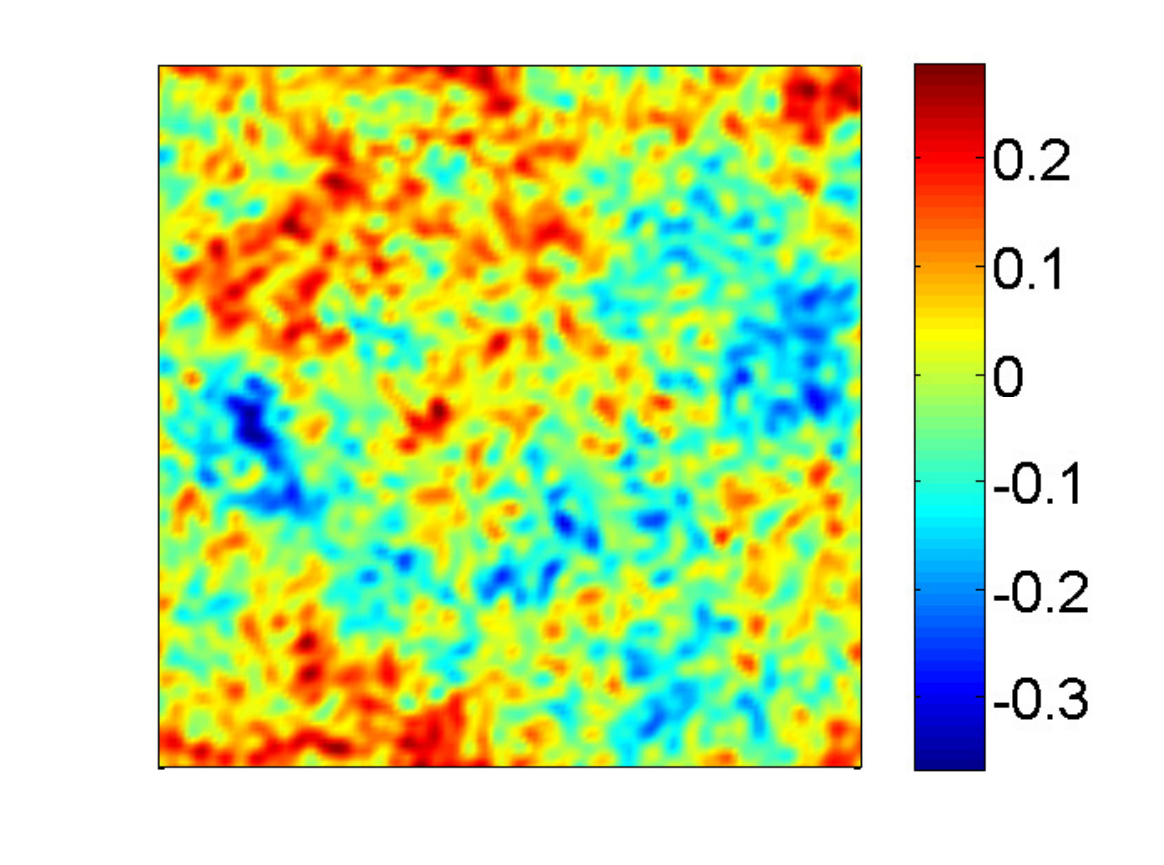} 
\caption{$\alpha=-0.8$: in the stationary state, $\mathcal{F}= 2.12$, which is close to the critical value (approximately $K/3$) for wave condensation on this $K=32/6$ grid. Top: spectra of a and b waves (left), and spectra normalized by pure energy equipartition spectra with $T=3.5 \times 10^{-4}$ (right). Bottom: snapshots at final time of the $\lambda$ and $\chi$ fields. }
\end{center}
\end{figure}

\begin{figure}
\begin{center}
\includegraphics[width=60 mm]{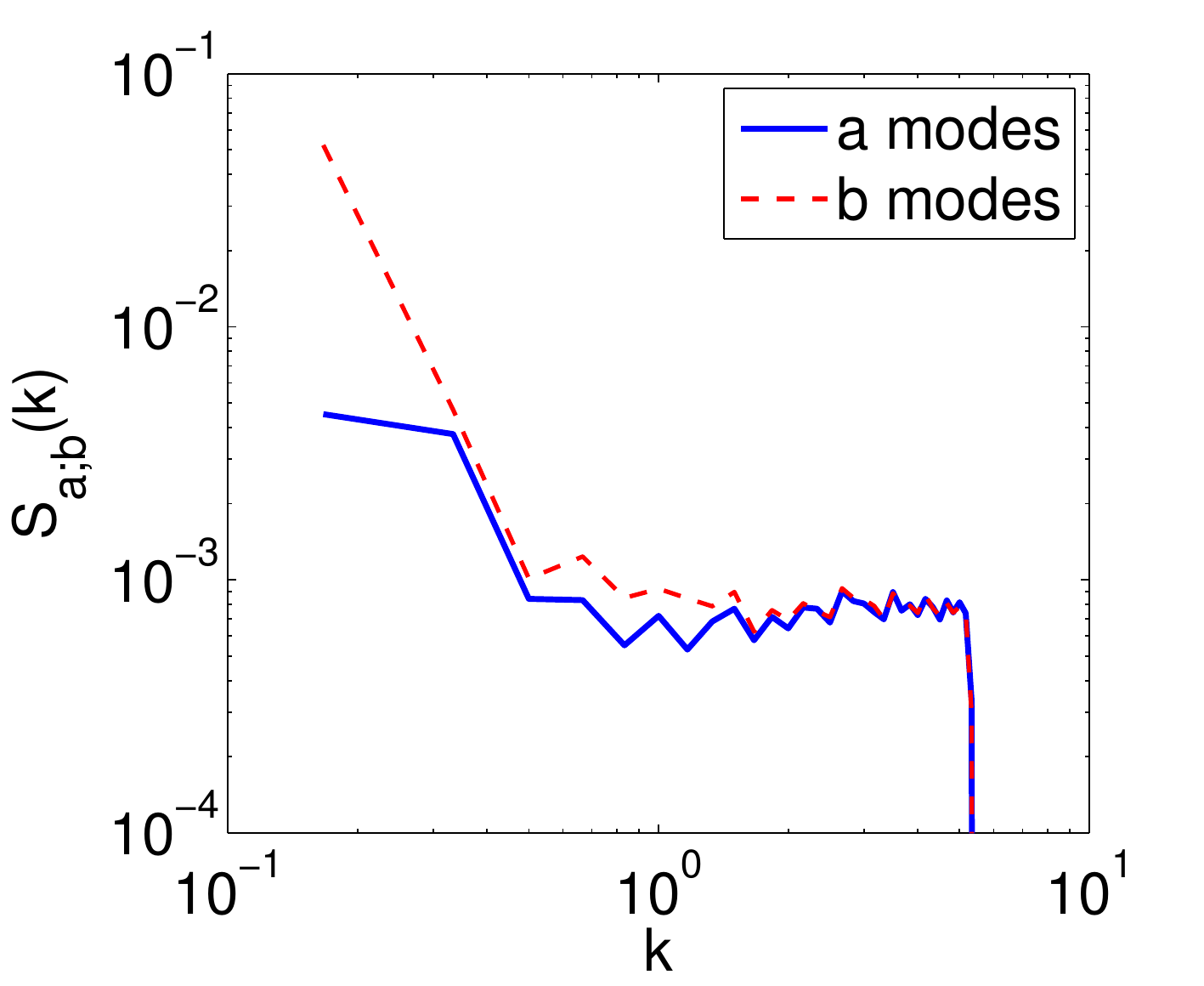} 
\includegraphics[width=60 mm]{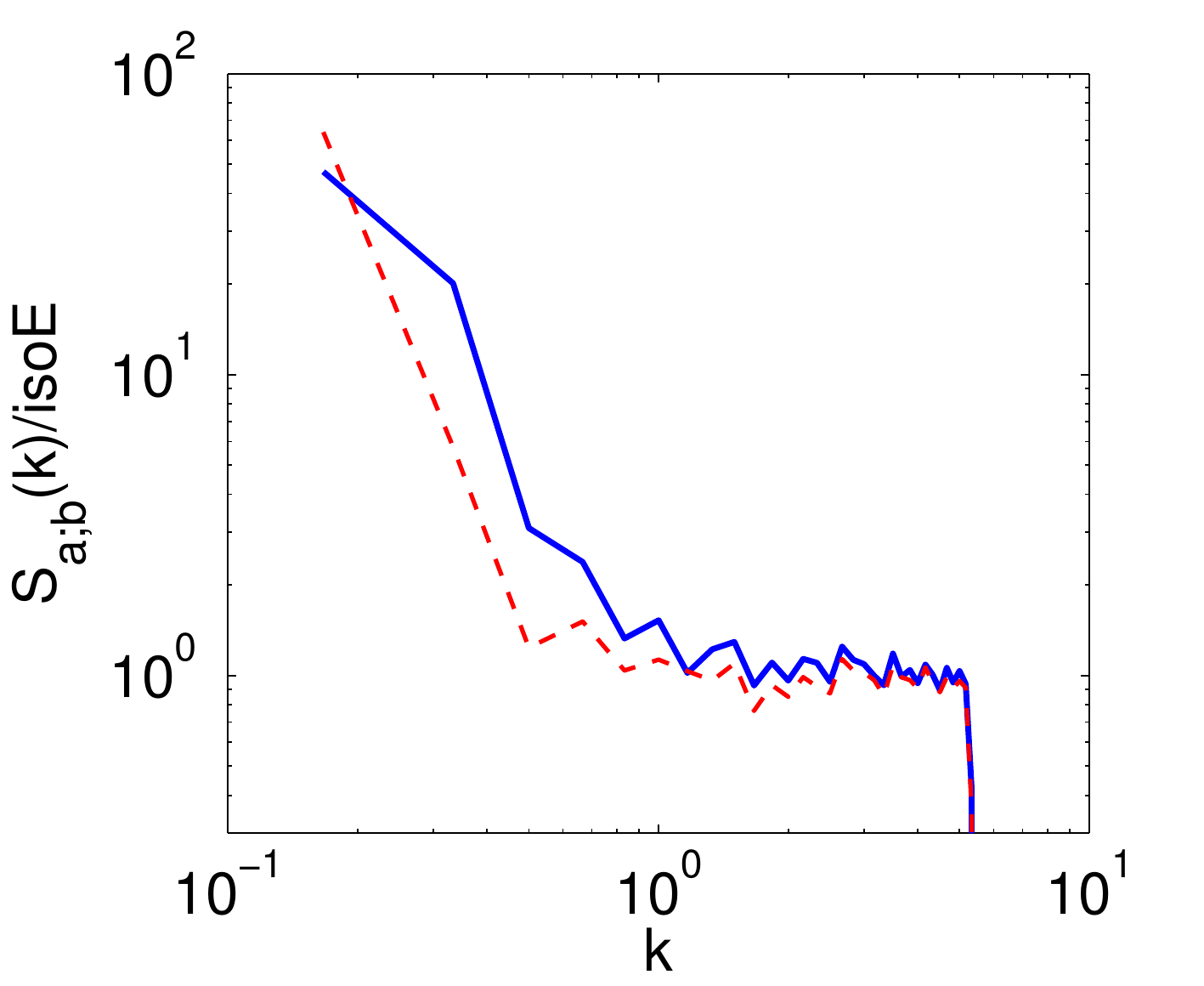} \\
\includegraphics[width=60 mm]{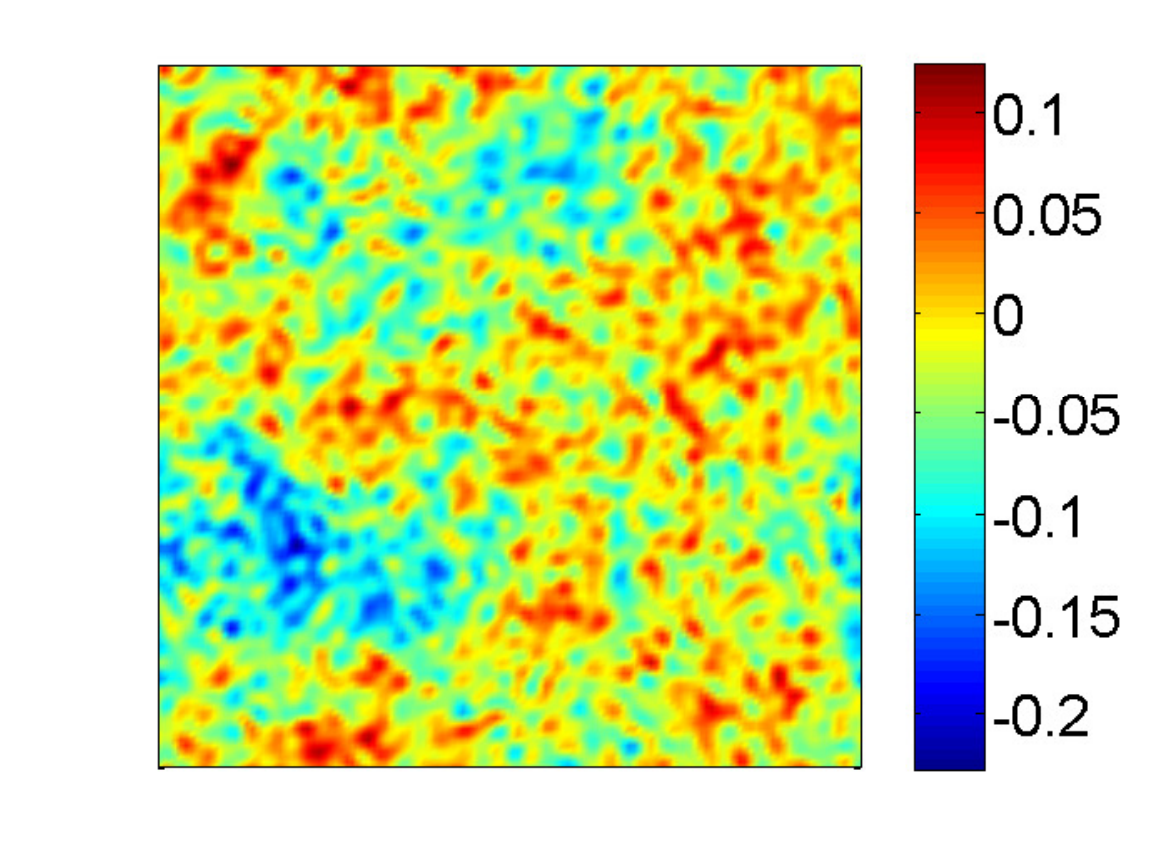} 
\includegraphics[width=60 mm]{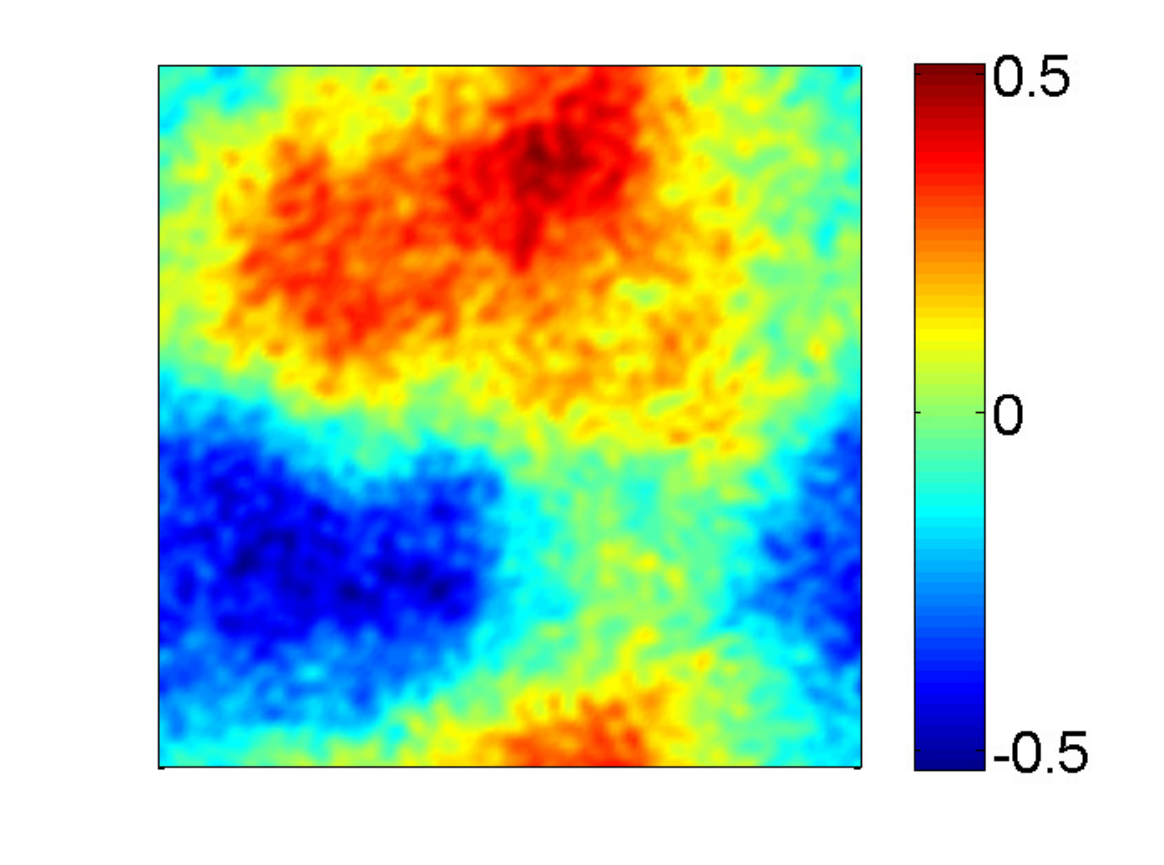} 
\caption{$\alpha=-1.3$: in the stationary state, $\mathcal{F}= 1.12$, which is below the value for wave condensation. Top: spectra of a and b waves (left), and spectra normalized by energy equipartition spectra with $T=3.3 \times 10^{-5}$. The flat part at high $k$ is in thermal equilibrium, and the bump at low $k$ indicates condensation. Bottom: snapshots at final time of the $\lambda$ and $\chi$ fields.\label{BECfig}}
\end{center}
\end{figure}

\begin{figure}
\begin{center}
\includegraphics[width=58 mm]{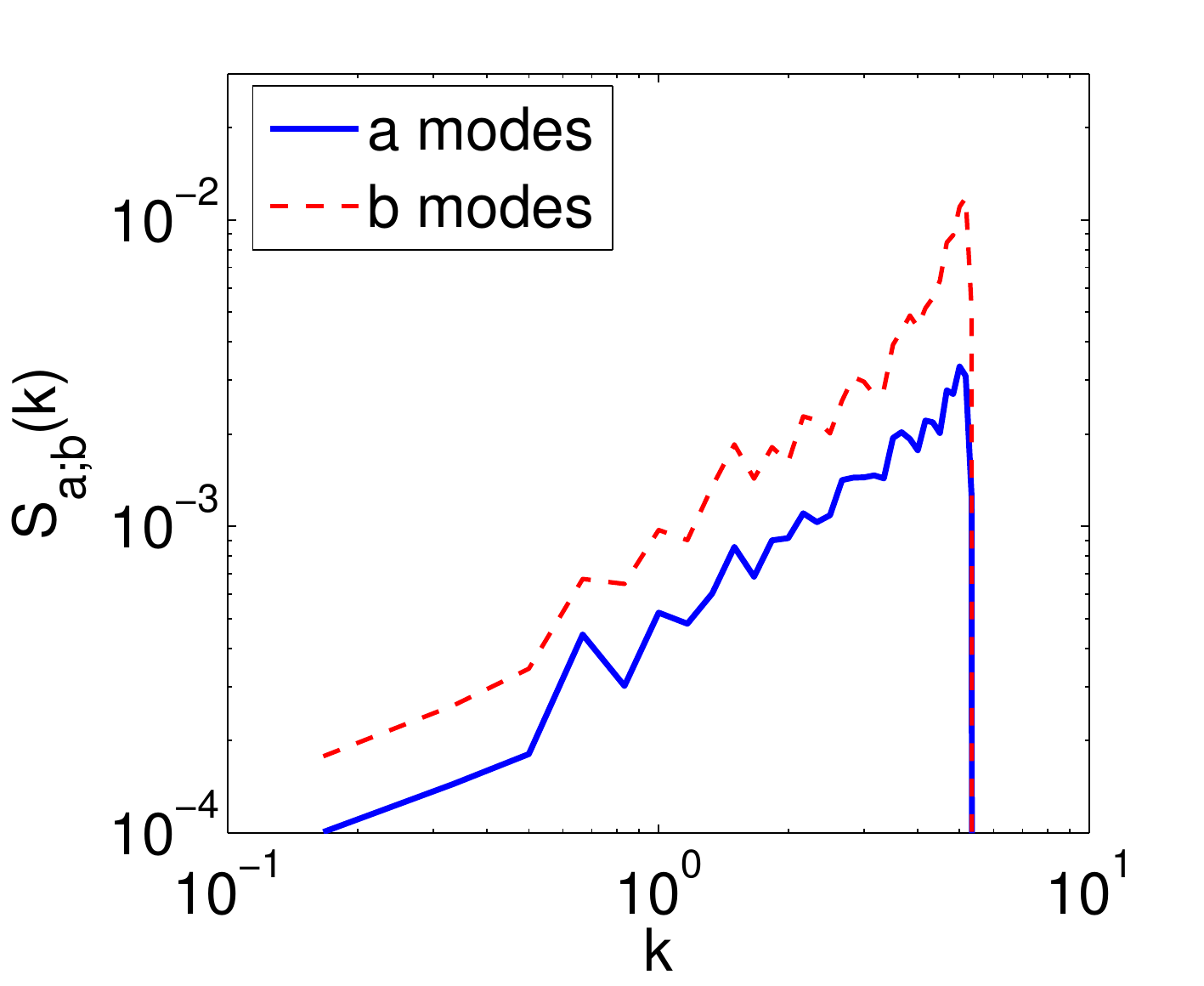} 
\includegraphics[width=58 mm]{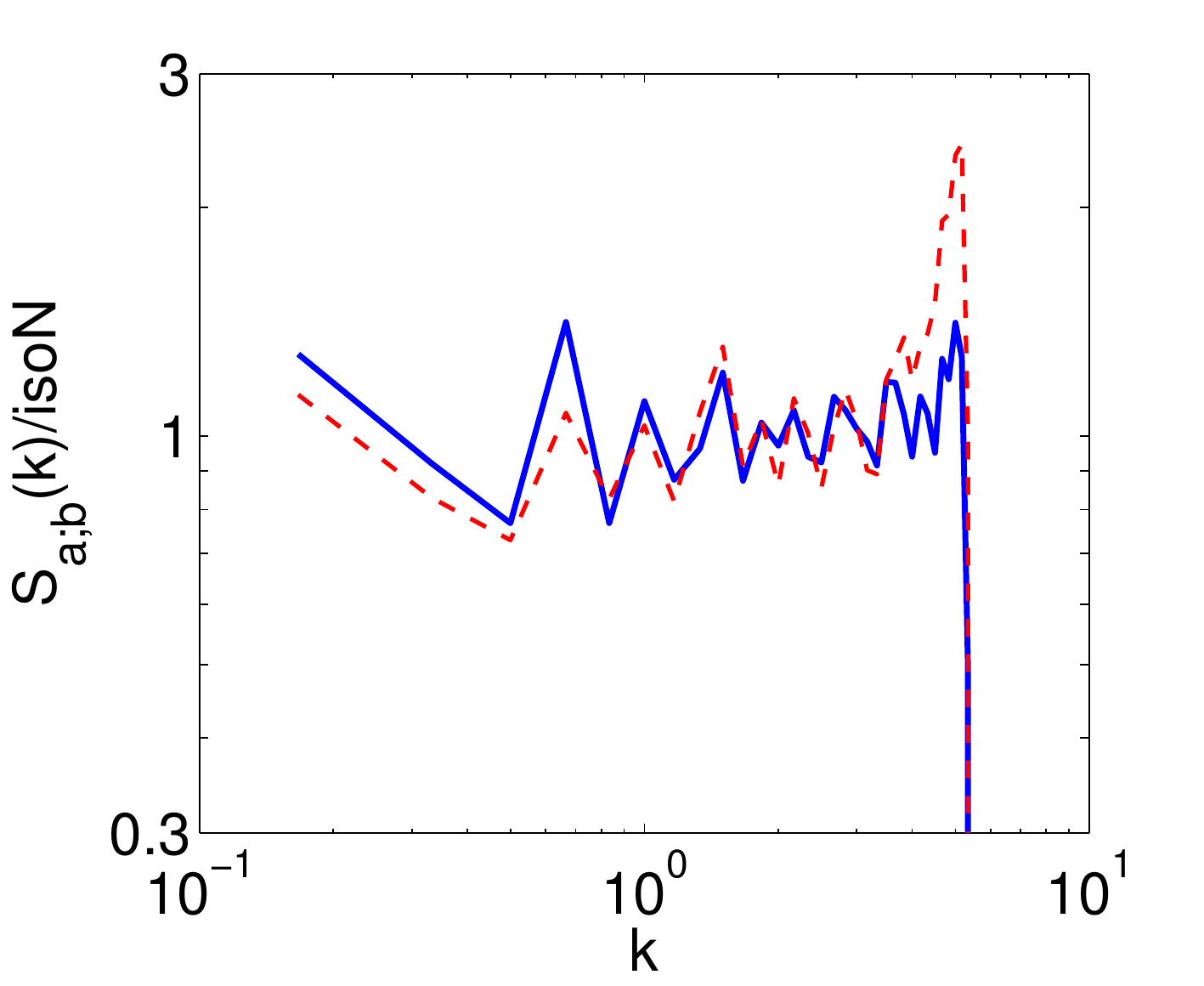} 
\includegraphics[width=58 mm]{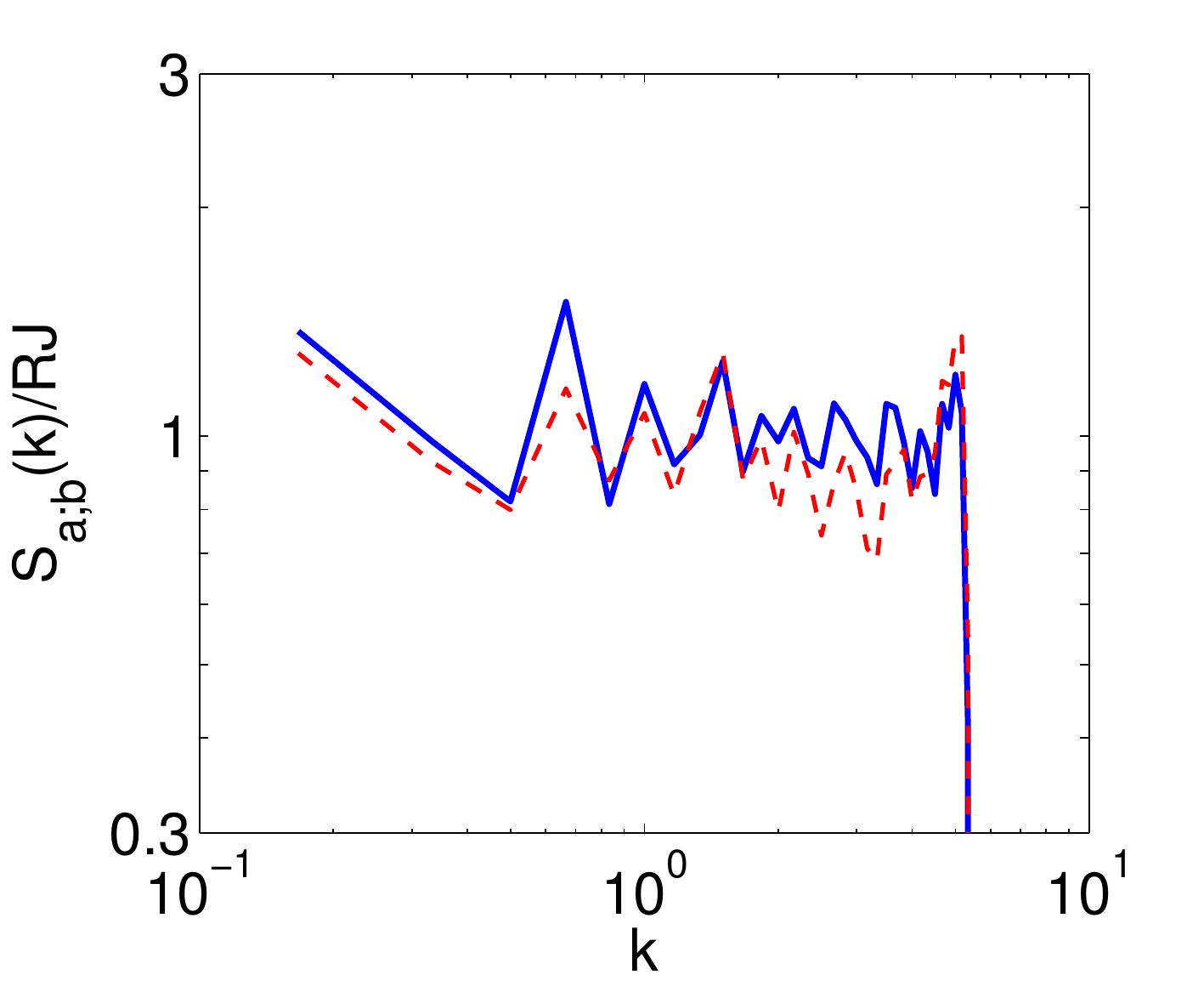} \\
\includegraphics[width=60 mm]{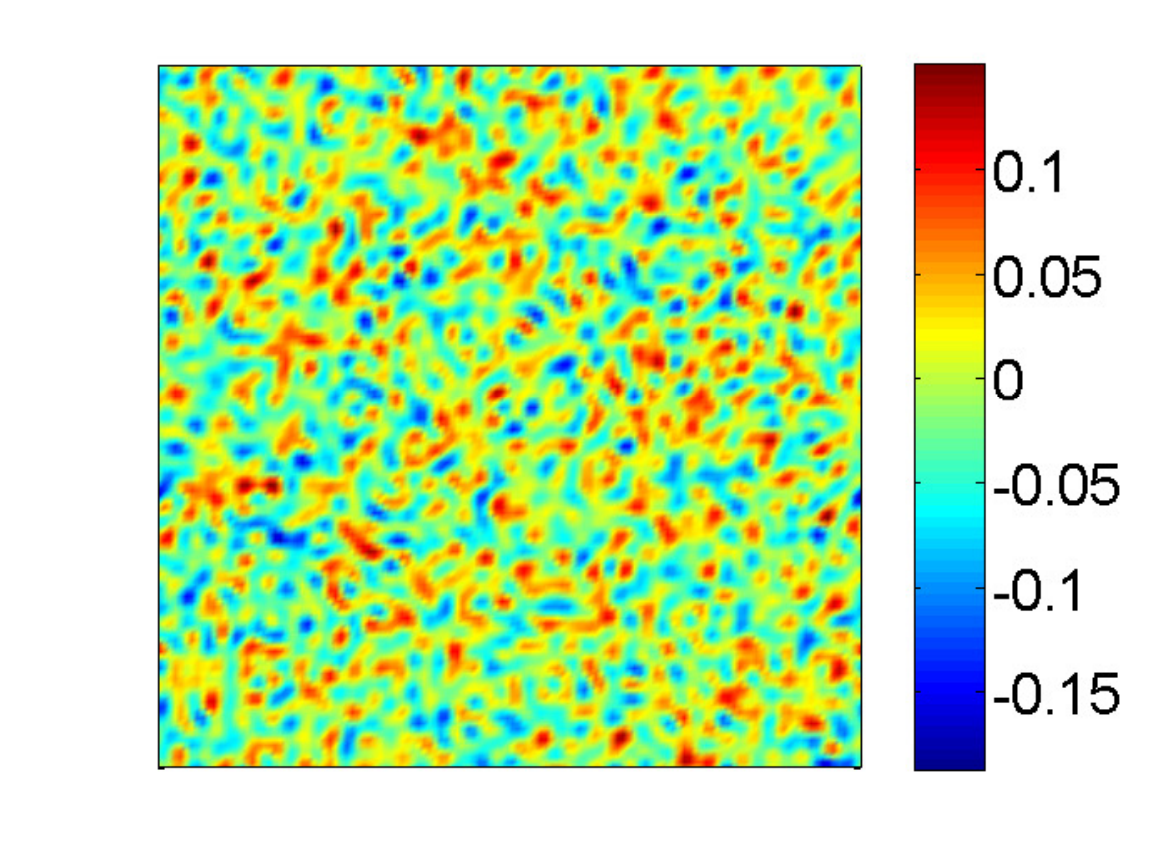} 
\includegraphics[width=60 mm]{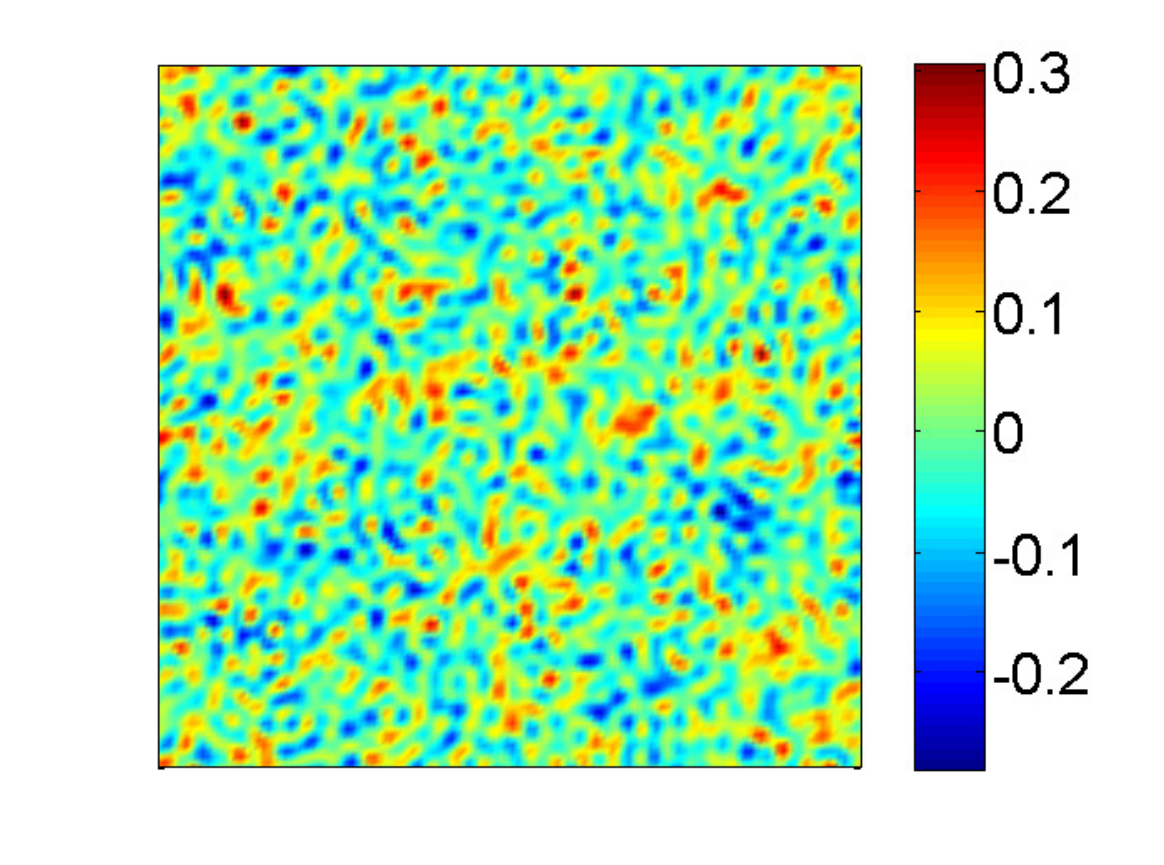} 
\caption{$\alpha=-2$: after $6000$ time units, $\mathcal{F}=3.08$, which is above the critical value for high-$k$ accumulation. Top-left: spectra of a and b waves. Top-center:  spectra normalized by equipartition of $\mathcal{N}$, i.e., Rayleigh-Jeans spectra in the limit $(T,\mu) \to (\infty,\infty)$, with $T/\mu=1.5 \times 10^{-4}$. The flat part at low $k$ is in equipartition of $\mathcal{N}$, and the bump at high $k$ indicates accumulation. Top-right: spectra normalized by thermodynamic spectra with negative temperature $T=-1.3 \times 10^{-3}$ and chemical potential $\mu=-10$. Bottom: snapshots after integration over $7200$ time units of the $\lambda$ and $\chi$ fields.\label{run10}}
\end{center}
\end{figure}

\subsection{Non-uniform condensates}

To study the condensation transition, we performed a set of numerical simulations with decreasing initial values of ${\cal F}$: as ${\cal F}$ decreases, so does $\mu$, until the denominator of the Rayleigh-Jeans spectrum (\ref{RJg}) vanishes for some value of $k$. This happens first for a $b$-mode, since $\omega_k^b < \omega_k^a$ for a given $k$. As a consequence, close to threshold we expect condensation of $b$-waves, but no condensation of $a$-waves: particles accumulate in the lowest-$k$ $b$-modes, which in the equivalent particle physics system corresponds to Bose-Einstein condensation of pions.

These $b$-modes cannot have vanishing wavenumber, because $b_{{\bf k}=0}$ is fixed in the leading order. Indeed, inserting the weakly nonlinear expansion $\psi({\bf x},t)=1+\epsilon(\lambda({\bf x},t)+i \chi({\bf x},t))$ in the expression for the vanishing charge $Q$, we obtain to lowest order that $\frac{d}{dt}\left(\int \chi \, d{\bf x}\right)=0$, i.e., the uniform phase mode remains constant in time, and it can be set to zero.
As a consequence, in a finite periodic domain of size $L$ in $d$ dimensions, there are $d$ $b$-modes corresponding to the lowest non-vanishing frequency $\omega^b=k=2 \pi / L$. For such a finite-size system, condensation occurs as soon as $\mu=-2 \pi / L$, and it manifests as an accumulation of particles in these large-scale (but non-uniform) $b$-modes.

As the condensate fraction increases, the weakly-nonlinear assumption rapidly breaks down in the low-$k$ modes. Indeed, the level of nonlinearity is determined by the ratio of the typical intensity to the typical frequency of the waves. Hence nonlinearity increases during condensation, because $b$-particles accumulate in the low-frequency modes. Note that conservation of the particle invariant ${\cal N}$ may even fail because of this increased nonlinearity. Nevertheless, we observe in the numerics that condensation is a robust phenomenon, that occurs even when the particle invariant is only an adiabatic one. As condensation proceeds, the enhanced nonlinearity arrests further accumulation and sharpening of the spectrum in the lowest (non-zero) frequency modes, resulting in a condensate that spreads over a few low-frequency modes.
 Because dynamical separation of the normal modes of the linear system is lost for strong nonlinearity, condensation of the $b$-modes ``spills" into the low-$k$  $a$-modes, even though the weakly nonlinear theory predicts no $a$-mode condensation. The resulting condensation behavior is displayed in figure \ref{BECfig}.

\subsection{Condensed fraction}

\begin{figure}
\begin{center}
\includegraphics[width=110 mm]{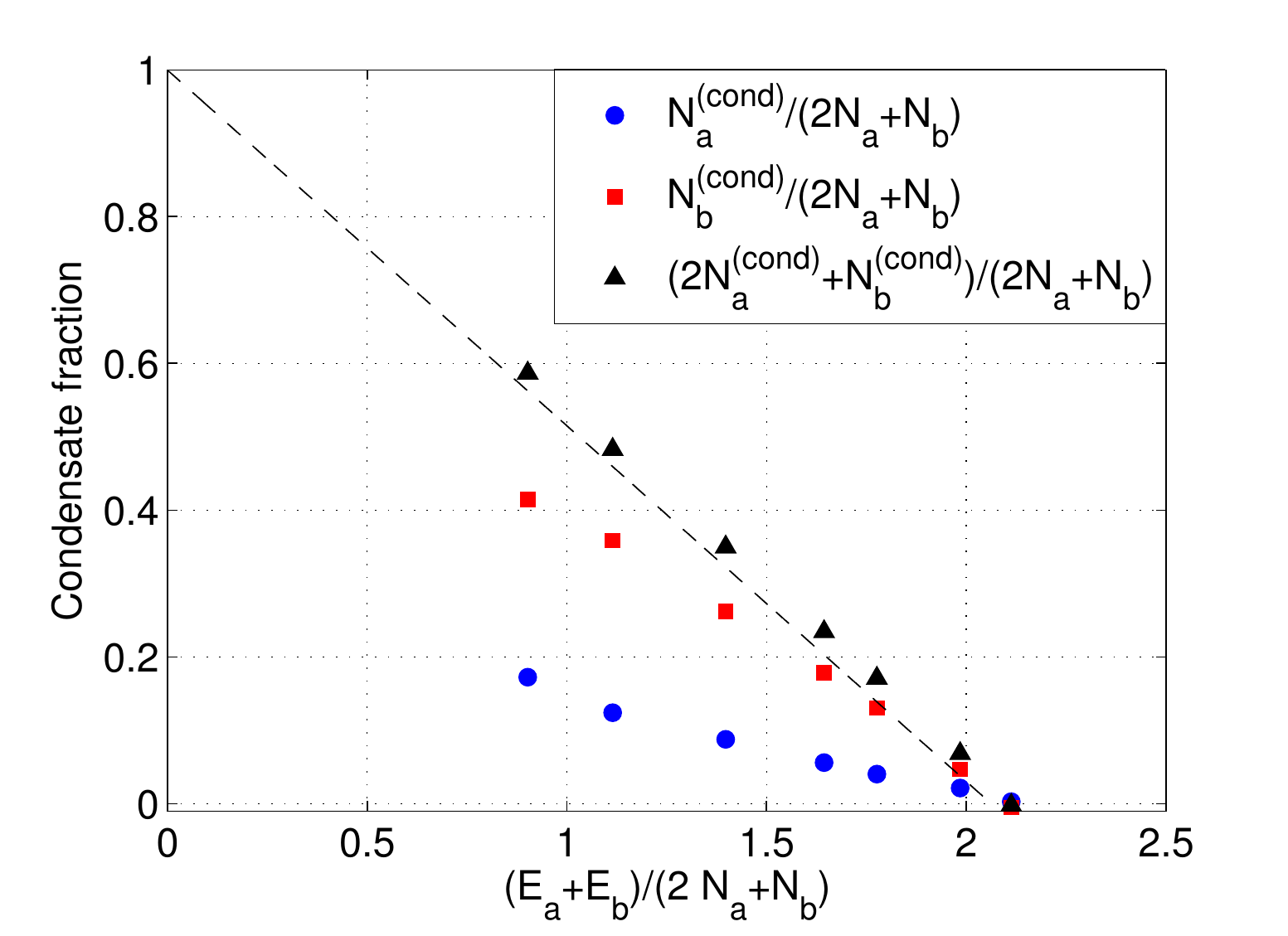} 
\caption{Condensate fraction as a function of the energy per particle ${\cal E}/{\cal N}$. We show the condensate fraction of $a$ particles ($\bullet$), $b$ particles ($\square$), and of $2 N_a+ N_b$ ($\triangle$). The dashed-line is the theoretical prediction (\ref{thpred}). \label{bifurcation}}
\end{center}
\end{figure}

To extract the condensate fraction from the numerical simulations, we first determine the temperature of the system using a fit to the high-$k$ part of the spectra. These spectra consist of a flat isothermal level, plus a bump at low-$k$ corresponding to the condensate. By subtracting the isothermal part of the spectra before integrating over $k$ we access the number of condensed particles. The condensed fraction is the ratio of the number of condensed particles to the total number of particles. We plot it in figure \ref{bifurcation}. From the numerics, the threshold for condensation is estimated to be ${\cal F} \simeq 2.1$. We can predict this value accurately if we take into account the discreteness of the lattice of wave numbers. Replacing the integrals in equations (\ref{eq:bec1}) to (\ref{eq:bec4}) by a sum over the discrete lattice of wave numbers, we obtain a critical value for condensation ${\cal F}_{cr}=2.06$, hence a very good agreement between the theory and the numerics.

The condensed fraction can be predicted theoretically as follows: under the threshold for condensation, the chemical potential vanishes and the thermodynamic spectrum of $b$ particles diverges for small $k$: slightly below the onset ${\cal F}_{cr}$, a macroscopic number ${\cal N}_b^{\mathrm{(cond)}}$ of $b$ particles condenses, while the $a$ particles and the remaining $b$ particles are (approximately) in energy equipartition, i.e., they have $\mu \simeq 0$ thermodynamic spectra. The non-condensed particles carry (almost) all the energy ${\cal E}$ of the system, and their number is $2 {\cal N}_a + {\cal N}_b-{\cal N}_b^{\mathrm{(cond)}}$. The ratio of these two quantities is ${\cal F}_{cr}$, from which we deduce the condensed fraction of $b$ particles,
\begin{equation}
\frac{{\cal N}_b^{\mathrm{(cond)}}}{{\cal N}}=\frac{{\cal F}_{cr}-{\cal F}}{{\cal F}_{cr}} \, .\label{thpred}
\end{equation}
This prediction is in good agreement with the numerical simulations (see figure \ref{bifurcation}).

\section{Statistical description of the decay instability\label{statdecay}}

\label{sec:nonlocal}

Kinetic equations allow to compute the out-of-equilibrium evolution of some initial distribution of waves. In particle physics, such an initial distribution can be given in some cosmological model, or it can correspond to particles appearing at $t=0$ in an experiment. 
As we mentioned earlier, for the KGMH system the transfers governed by the kinetic equations can be very nonlocal in $k$-space. 
As an example of such nonlocal transfers, we come back to the decay instability described in section \ref{decay}, which we now study for isotropic ensembles of waves.

\subsection{Decay instability for an isotropic ensemble of $a$-waves\label{isotropicdecay}}

\begin{figure}
\begin{center}
\includegraphics[width=160 mm]{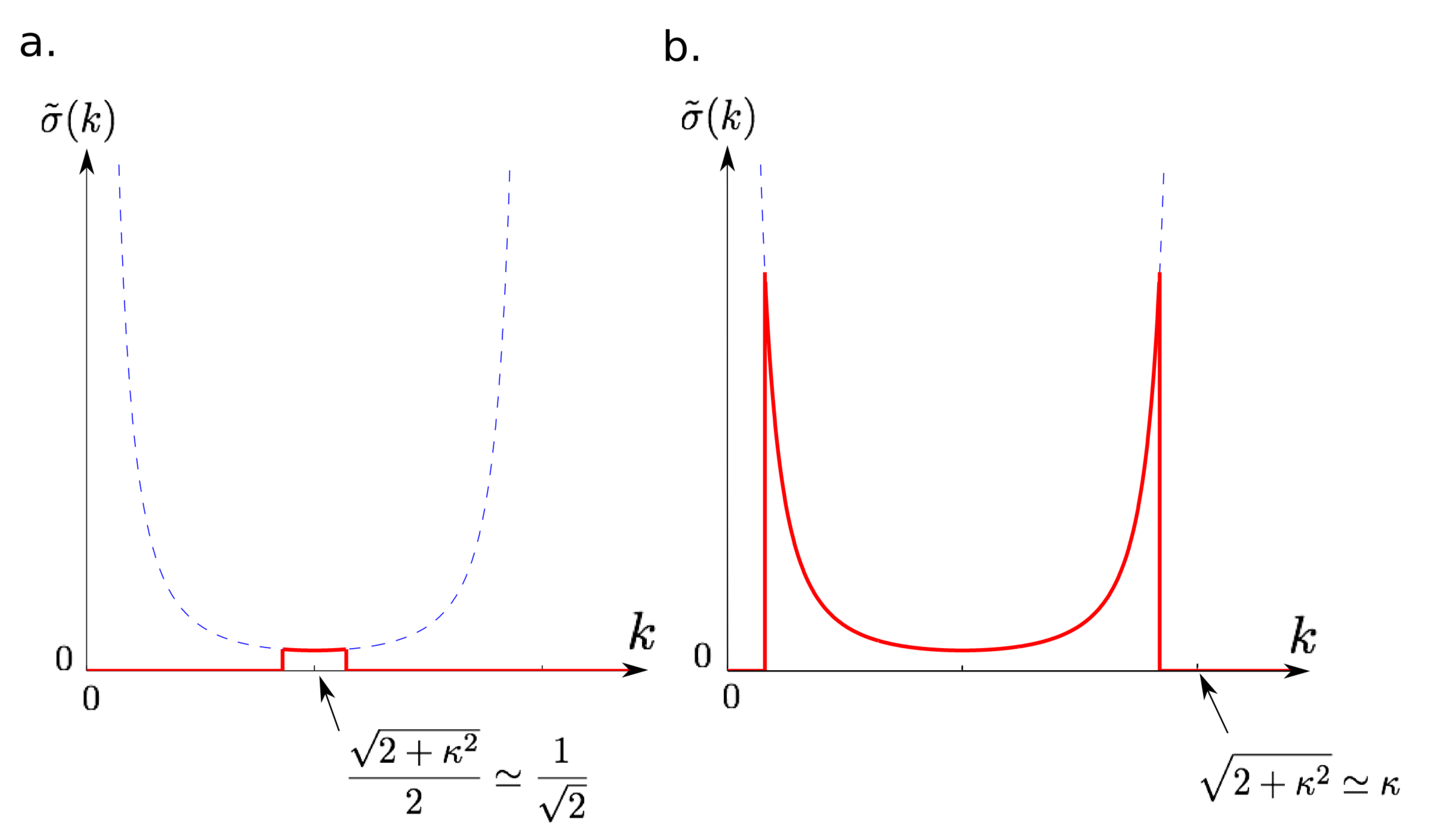}
\caption{Reduced growthrate of $b$-waves as a function of wavenumber $k$, for $a$-waves at wavenumber $\kappa$. The dashed curve is $\left( \sqrt{2+\kappa^2}-k \right)^{-2} + k^{-2}$. Left-hand panel: for $\kappa\ll1$, $b$-waves grow at wavenumber $k\simeq 1/\sqrt{2}$. Right-hand-panel: for $\kappa \gg 1$, the growthrate has two strong maxima, one at $k \simeq 1/\kappa$ and one at $k \simeq \kappa$.  \label{figsigma}}
\end{center}
\end{figure}

Consider a 3D isotropic ensemble of $a$-waves with wavenumber $\kappa$. These waves spontaneously induce some $b$-waves through the decay instability. A key question is to know the distribution of $b$-waves that is produced by this instability: what is the wavenumber (or energy) of the $b$-waves? In the particle physics analogous system, we use the kinetic equation to predict the distribution of pions that arises from the decay of $\sigma$-mesons with wavenumber $\kappa$.

The initial distribution is 
\begin{eqnarray}
n^a_{k} & = & \frac{N_a}{4 \pi \kappa^2 L^3} \delta (k-\kappa) \, , \label{distributiona}\\
n^b_{k} & = & 0 \, .
\end{eqnarray}

As already mentioned in section \ref{thermoeq}, this distribution is a steady solution to the kinetic equations because $n^b_{k}=0$. However, it is an unstable solution, as we can see by considering an infinitesimal perturbation $n^b_{k}(t) \ll 1$. Inserting (\ref{distributiona}) into kinetic equation (\ref{kebIb3}) and retaining only the linear terms in the infinitesimal perturbation $n^b_{k}$, we obtain
\begin{eqnarray}
\dot{n}^b_{k} & = & \frac{N_a}{k^2 L^3 \kappa \sqrt{2+\kappa^2}} \left(n^b_{k} +n^b_{\sqrt{2+\kappa^2}-k} \right) \mbox{ if } k \in {\cal I} \equiv \left[ \frac{\sqrt{2+\kappa^2}-\kappa}{2}  ; \frac{\sqrt{2+\kappa^2}+\kappa}{2} \right] \, , \label{evolutionnb}\\
\dot{n}^b_{k} & = & 0 \qquad \mbox{ otherwise } \, ,
\end{eqnarray}
where the condition in (\ref{evolutionnb}) arises from the integration domain ${\cal D}_b(k)$.

We write the same equation at wavenumber $\sqrt{2+\kappa^2}-k$ to obtain the following $2\times 2$ system of linear ODEs,
\begin{eqnarray}
\dot{n}^b_{k} & = & \frac{N_a}{k^2 L^3 \kappa \sqrt{2+\kappa^2}} \left(n^b_{k} +n^b_{\sqrt{2+\kappa^2}-k} \right) \, , \\
\dot{n}^b_{\sqrt{2+\kappa^2}-k} & = & \frac{N_a}{L^3 (\sqrt{2+\kappa^2}-k)^2 \kappa \sqrt{2+\kappa^2}} \left(n^b_{k} +n^b_{\sqrt{2+\kappa^2}-k} \right) \, ,
\end{eqnarray}
if $k \in {\cal I}$, and $\dot{n}^b_{k}=\dot{n}^b_{\sqrt{2+\kappa^2}-k}=0$ otherwise. 

Consider an eigenmode $({n}^b_{k},{n}^b_{\sqrt{2+\kappa^2}-k}) = (\hat{n}^b_{k},\hat{n}^b_{\sqrt{2+\kappa^2}-k}) e^{\sigma t}$, where $\sigma$ is the growth rate. Substitution into the system of equations yields a $2 \times 2$ system in $\hat{n}^b_{k}$ and $\hat{n}^b_{\sqrt{2+\kappa^2}-k}$, that has nontrivial solutions when the determinant vanishes. This gives the following expression for the reduced growth rate $\tilde{\sigma}=\sigma \kappa \sqrt{2+\kappa^2} \, L^3/N_a$,
\begin{eqnarray}
\tilde{\sigma}(k) & = & \frac{1}{\left( \sqrt{2+\kappa^2}-k \right)^2} + \frac{1}{k^2} \qquad \mbox{ if } k\in {\cal I}\, , \\
\tilde{\sigma}(k) & = & 0 \qquad \mbox{otherwise. }
\end{eqnarray}
Plots of $\tilde{\sigma}(k)$ are provided in figure \ref{figsigma}, in the limiting cases $\kappa \ll 1$ and $\kappa \gg 1$. For $\kappa \ll 1$, ${\cal I}$ is a narrow interval centered around $\frac{\sqrt{2+\kappa^2}}{2} \simeq \frac{1}{\sqrt{2}}$, and $b$-waves appear at this wavenumber: there is a strong nonlocal transfer of massive particles with $\kappa\ll 1$ into massless particles with $k \simeq \frac{1}{\sqrt{2}}$. For $\kappa \gg 1$, the growth rate has a maximum at $k \simeq \kappa$ and a maximum at $k = \sqrt{2+\kappa^2}-\kappa \simeq \frac{1}{\kappa} \ll 1$: massless particles will appear preferentially at these two wave numbers, hence a strong nonlocal transfer from $a$-particles with $k=\kappa$ into $b$-particles with $k=\frac{1}{\kappa} \ll 1$.

These predictions for isotropic initial distributions of $a$-waves are qualitatively compatible with the numerical results of section \ref{decay} on the decay of a single $a$-mode. If the initial $a$-mode has wavenumber $\kappa\gg1$, $b$-waves appear at both wave numbers $k \ll 1$ and $k \simeq \kappa$ (see figure \ref{fig2}), whereas if the initial $a$-mode has wavenumber $\kappa \ll 1$, then $b$-waves appear at wavenumber $k\simeq \frac{1}{\sqrt{2}}$ (see figure \ref{fig2bis}).

\subsection{Reaction yield of the decay instability}

\begin{figure}
\begin{center}
\includegraphics[width=100 mm]{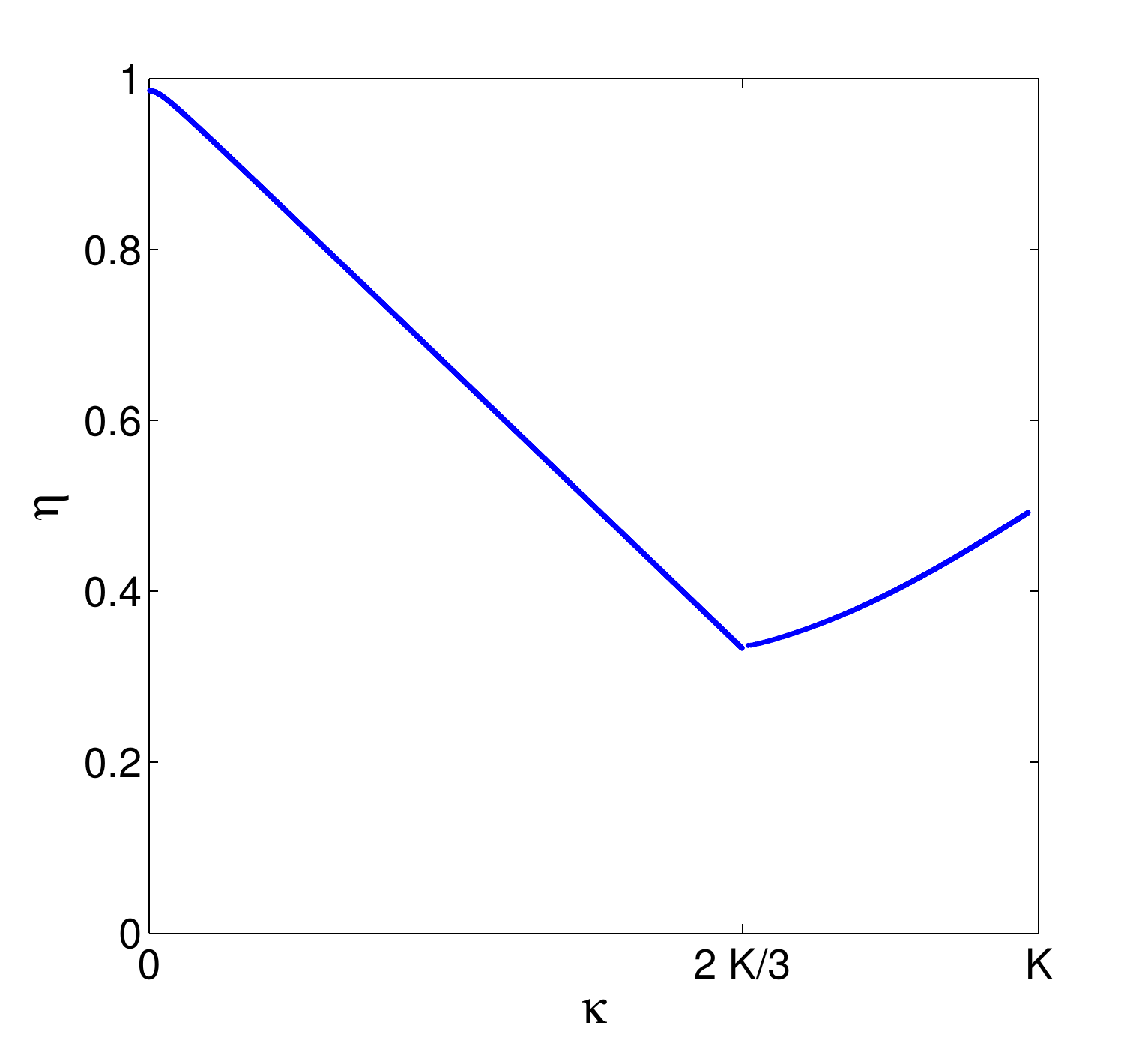}
\end{center}
\caption{Reaction yield $\eta$ for $d=2$ as a function of the initial wavenumber $\kappa$ of the $a$-waves, for a maximum allowed wavenumber $K \gg 1$ (here $K=100$). In the decreasing part, some fraction of the produced $b$ particles condenses. In the increasing part such condensation does not occur.
\label{etavskappa}}
\label{fig5}
\end{figure}

Starting from an isotropic distribution of $a$-particles only with wavenumber $\kappa$, together with weak background noise on the $a$ and $b$ fields, the decay instability sets in and $b$-waves appear through nonlocal transfers. The intensity of the $a$-waves decreases accordingly, to ensure conservation of the energy and particle invariants. When $b$-waves become strong enough, the quadratic term in $\tilde{n}_k^b$ of kinetic equation (\ref{keaIb3}) cannot be neglected anymore. This term modifies the spectral distribution of $a$-waves, the system enters the nonlinear regime and eventually it reaches thermodynamic equilibrium, with both spectra of $a$- and $b$-waves spreading over the entire range of wave numbers.

We can thus determine the saturated state of the decay instability using the thermodynamics described in section \ref{bec}. Indeed, the initial distribution (\ref{distributiona}) corresponds to an energy per particle
\begin{equation}
{\cal F}=\frac{\sqrt{2+\kappa^2}}{2} \, . \label{Fvskappa}
\end{equation}
 An $a$-wave can be seen as a molecule made of two $b$ atoms, and the decay instability corresponds to the reaction $a \rightarrow b + b$. In the saturated state, one can compute the yield $\eta$ of this reaction using thermodynamics: out of the initial ${\cal N}$ number of $b$ atoms (two per $a$-wave, and one per $b$-wave), what fraction $\eta$ decays into free $b$ particles? The reaction yield is given by the equilibrium value of $N_b/{\cal N}$, and it depends on the initial wavenumber $\kappa$ of the $a$-waves. We now compute it in 2D, as a function of $\kappa$.

For ${\cal F}>{\cal F}_{cr}$, no condensation occurs, and the yield is simply given by
\begin{eqnarray}
\nonumber  \eta & = & \frac{N_b(\mu)}{2 N_a(\mu)+N_b(\mu)} \\
 & = &  \frac{K+\mu \ln \left( \frac{\mu}{K+\mu} \right)}{K+\mu \ln \left( \frac{\mu}{K+\mu} \right) + 2 \left[ \sqrt{2+K^2}-\sqrt{2}+2\mu  \ln \left( \frac{\sqrt{2}+2\mu}{\sqrt{2+K^2}+2\mu} \right)  \right]} \, , \label{yieldmuneq0}
\end{eqnarray}
where the chemical potential $\mu$ is related to $\kappa$ through the relation (\ref{Fvskappa}), i.e., $\kappa =\sqrt{4 {\cal F}(\mu)^2-2}$ where ${\cal F}(\mu)$ is given by (\ref{EsurN}). 

For ${\cal F}<{\cal F}_{cr}$, condensation of $b$-waves occur and the spectra consist of thermodynamic spectra with $\mu=0$, together with $b$-particles condensed at low $k$. The reaction yield becomes
\begin{eqnarray}
\nonumber  \eta & = & \frac{N_b(\mu=0)}{{\cal N}}+\frac{N_b^{\mathrm{(cond)}}}{{\cal N}} \\
\nonumber & = & {\cal F} \, \frac{N_b(\mu=0)}{E_a(\mu=0)+E_b(\mu=0)} + \frac{{\cal F}_{cr}-{\cal F}}{{\cal F}_{cr}} \\
\nonumber & = & \frac{{\cal F}}{K}+\frac{{\cal F}_{cr}-{\cal F}}{{\cal F}_{cr}} = 1 + {\cal F} \left(\frac{1}{K} - \frac{1}{{{\cal F}_{cr}}} \right)\\
& = & 1 + \sqrt{2+\kappa^2} \left(\frac{1}{K} - \frac{1}{{{\cal F}_{cr}}} \right) \, .\label{yieldmu0}
\end{eqnarray}
In figure \ref{etavskappa}, we plot the yield given by expressions (\ref{yieldmuneq0}) and (\ref{yieldmu0}) as a function of the initial wavenumber $\kappa$ of the $a$-particles, for a maximum wavenumber $K \gg 1$. In this limit, ${\cal F}_{cr}\simeq K/3$, and we observe the following behavior: the reaction yield is a non monotonous function of $\kappa$. It is close to unity for $\kappa \ll 1$, i.e., the reaction is almost complete, with most of the $a$-particles eventually decaying into $b$ particles. In this regime a large fraction of the produced $b$-particles form a condensate. As $\kappa$ increases the yield decreases, with a minimum attained at the threshold of wave condensation $\kappa_{cr}=\sqrt{4 {\cal F}_{cr}^2-2}\simeq 2 {\cal F}_{cr} \simeq 2 K/3$. For $\kappa>\kappa_{cr}$ the yield increases with $\kappa$. For $\kappa=K$ the yield reaches $0.5$, i.e., half of the $b$ ``atoms" only are free $b$-waves, the other half remaining in the form of $a$-waves.

\section{Dynamical condensation as a nonlocal KZ cascade\label{dyncond}}

In this section we describe the dynamical process by which condensation is achieved. We first show that ensembles of $b$-waves at low-enough wave numbers are steady solutions to the kinetic equations. We refer to such solutions as ``condensates". We consider an infinitesimal perturbation of the $a$- and $b$-wave spectra and study the linear stability of a condensate at $k=0$. This analysis indicates that there is a rapid transfer of particles from the perturbation to the condensate. However, this transfer is only partial, and some marginally stable perturbations remain. We study the evolution of these remaining perturbations through a weakly nonlinear expansion and derive a kinetic equation that governs the nonlinear evolution of the perturbation to the condensate. This kinetic equation admits a KZ solution, with a net energy flux towards high wave numbers. In a infinite system, this solution corresponds to a condensate that increases in amplitude, while the energy of the remaining waves cascades to larger and larger wave numbers. In a finite system, the cut-off in wavenumber arrests this cascade and the system eventually reaches thermodynamic equilibrium.

\subsection{Steady solutions with $b$-waves only}

Consider the kinetic equations for $n^a_{k}=0$, and $n^b_{k} \neq 0$, and further restrict attention to distributions of $b$-waves at low-wavenumbers: we assume that there is a maximum wavenumber ${\cal K}$ that supports the $b$-waves, i.e., $n^b_{\cal K} \neq 0$, and $n^b_{k} = 0$ for $k > {\cal K}$. The evolution of this initial distribution of waves is governed by the kinetic equations (\ref{keaIb3}) and (\ref{kebIb3}). At $t=0^+$, only the quadratic terms in $n_k^b$ may be nonzero. In the right-hand-side of equation (\ref{keaIb3}), the quadratic term involves triads of wave numbers $(k_1,k_2,k=\sqrt{(k_1+k_2)^2-2}\,)$ such that $n_{k_1}^b \neq 0$, $n_{k_2}^b \neq 0$, $k_1 k_2>1/2$, $k_1+k_2 \geq \sqrt{2}$. Such $k_1$ and $k_2$ cannot be found if ${\cal K}\leq \frac{1}{\sqrt{2}}$: provided the distribution of $b$-waves is restricted to wave numbers smaller than $\frac{1}{\sqrt{2}}$, the right-hand-side of (\ref{keaIb3}) vanishes. We assume that this condition is satisfied by the initial distribution, and we consider the right-hand-side of kinetic equation (\ref{kebIb3}). Because $n_k^a=0$, only the quadratic term in $n_k^b$ may be nonzero. However, it is nonzero only if we can find two wave numbers $k$ and $k_2$ such that $n_k^b \neq 0$, $n_{k_2}^b \neq 0$, and $k k_2 \geq 1/2$. But the latter condition cannot be satisfied if ${\cal K} < \frac{1}{\sqrt{2}}$, hence the right-hand-side of (\ref{kebIb3}) vanishes.

To summarize, we have shown that an initial distribution of $b$-waves only ($n_k^a=0$) is a steady solution to the kinetic equations (\ref{keaIb3}) and (\ref{kebIb3}) provided these waves have only wave numbers smaller than $\frac{1}{\sqrt{2}}$.

\subsection{Dynamics of waves on a condensate}

For simplicity, we consider a $b$-wave condensate at very low-wavenumber, which we simply write as $A \, \delta (k) $, and we denote as $\tilde{n}^a(k)$ and $\tilde{n}^b(k)$ the remainders of the wave spectra. Hence,
\begin{eqnarray}
n^a_{k} & = & \tilde{n}^a_k \, , \label{distributiona}\\
n^b_{k} & = &A \delta (k) + \tilde{n}^b_k \, .
\end{eqnarray}
Inserting this decomposition into the 3D kinetic equations for $k \neq 0$ leads to
\begin{eqnarray}
\dot{\tilde{n}}^a_{k} & = & \frac{4 \pi A}{  k \sqrt{2+k^2}} \left( \tilde{n}^b_{\sqrt{2+k^2}} - \tilde{n}^a_k \right) + \frac{2 \pi} { k \omega^a_{k}} \int_{{\cal D}_a(k)}        \delta (\omega_{k12{ } }^{abb} ) 
\, ( \tilde{n}^b_1  \tilde{n}^b_2 - 2  \tilde{n}^b_1   \tilde{n}^a_{ k}) 
\, d{ k}_1 d{ k}_2 \, ,\label{decompa} \\
\dot{\tilde{n}}^b_{k} & = & \frac{4 \pi A}{   k^2} \left( \tilde{n}^a_{\sqrt{k^2-2}} - \tilde{n}^b_k  \right) + \frac{4 \pi} { k^2}  \int_{{\cal D}_b(k)}      \frac{k_1}{\omega_1^a}   \delta (\omega_{12{ k} }^{abb} ) 
\, ( \tilde{n}^a_1  \tilde{n}^b_2 +  \tilde{n}^a_1   \tilde{n}^b_{ k} - \tilde{n}^b_2   \tilde{n}^b_{k})
 \, d{ k}_1 d{ k}_2 \, .\label{decompb}
\end{eqnarray}
The terms proportional to $A$ arise from the interaction between the condensate and the remainders, while the integral terms correspond to interactions of the remainders. The latter terms are exactly the collision integrals of the original kinetic equations, with $n$ replaced by $\tilde{n}$. We stress the fact that if $\tilde{n}^a_k$ and $\tilde{n}^b_k$ are the isothermal distributions (\ref{RJe}), we obtain a steady solution to (\ref{decompa}) and (\ref{decompb}), i.e., both the terms proportional to $A$ and the integral terms vanish: a condensate at $k=0$ together with remainders in isothermal equilibrium constitute a steady solution to the kinetic equations, in perfect agreement with the thermodynamic treatment of Bose-Einstein condensation.

To study the dynamical process of condensation, we perform a multiple-scale expansion of equations (\ref{decompa}) and (\ref{decompb}), to describe the evolution of remainders much weaker than the condensate, $\tilde{n} \ll A$. Consider a small parameter $\epsilon \ll 1$ and $A={\cal O}(1)$, and expand the remainders as
\begin{eqnarray}
 \tilde{n}^a_k & = & \epsilon \, \tilde{n}^{a;0}(k,t,T)+\epsilon^2 \, \tilde{n}^{a;1}(k,t,T)+\dots \, ,\\
 \tilde{n}^b_k & = & \epsilon \, \tilde{n}^{b;0}(k,t,T)+\epsilon^2 \, \tilde{n}^{b;1}(k,t,T)+\dots \, ,
\end{eqnarray}
where the slow time is $T=\epsilon t$.

Inserting this decomposition into (\ref{decompa}) and (\ref{decompb}) gives at order $\mathcal{O}(\epsilon)$ the linear stability analysis for the condensate,
\begin{eqnarray}
\partial_t{\tilde{n}}^{a;0}_{k} & = & \frac{4 \pi A}{  k \sqrt{2+k^2}} \left( \tilde{n}^{b;0}_{\sqrt{2+k^2}} - \tilde{n}^{a;0}_k \right)  \\
\partial_t{\tilde{n}}^{b;0}_{k} & = & \frac{4 \pi A}{   k^2} \left( \tilde{n}^{a;0}_{\sqrt{k^2-2}} - \tilde{n}^{b;0}_k  \right) 
\end{eqnarray}
Considering the first of these two equations at wavenumber $\sqrt{k^2-2}$ instead of $k$ leads to the following $2\times2$ system of first-order linear ODEs,
\begin{eqnarray}
\partial_t{\tilde{n}}^{a;0}_{\sqrt{k^2-2}} & = & \frac{4 \pi A}{ {\sqrt{k^2-2}} \,  k} \left( \tilde{n}^{b;0}_{k} - \tilde{n}^{a;0}_{\sqrt{k^2-2}} \right) \, , \\
\partial_t{\tilde{n}}^{b;0}_{k} & = & \frac{4 \pi A}{ k^2} \left( \tilde{n}^{a;0}_{\sqrt{k^2-2}} - \tilde{n}^{b;0}_k  \right) \, .
\end{eqnarray}
Solutions to this system of ODEs can be sought in the form $\left( {\tilde{n}}^{a;0}_{\sqrt{k^2-2}}, {\tilde{n}}^{b;0}_{k} \right) \sim e^{\sigma t}$, which leads to the following growth rates,
\begin{eqnarray}
\sigma_1=0 \qquad \mbox{ and } \qquad \sigma_2(k)=-{4 \pi A} \left( \frac{1}{k^2} + \frac{1}{k\sqrt{k^2-2}}\right) < 0 \, .
\end{eqnarray}
Decomposing on the corresponding eigenmodes we obtain
\begin{eqnarray}
{\tilde{n}}^{a;0}_{\sqrt{k^2-2}} & = & C_1(k,T)+C_2(k,T) e^{\sigma_2(k) t} \\
{\tilde{n}}^{b;0}_{k} & = & C_1(k,T) - \frac{\sqrt{k^2-2}}{k} C_2(k,T) e^{\sigma_2(k) t} \, .
\end{eqnarray}
Because $\sigma_2 < 0$, the contribution in $C_2$ rapidly decays on the fast time $t$,
and after a short transient,
\begin{equation}
{\tilde{n}}^{a;0}_{\sqrt{k^2-2}} = {\tilde{n}}^{b;0}_{k} \equiv n_k(T) \, . \label{defn}
\end{equation}
One can check that the $C_2$ part of the spectra has vanishing total energy at any time. Its rapid decay occurs at constant energy, as it should. By contrast, this $C_2$ contribution carries a nonzero number of particles ${\cal N}_2=\int C_2(k) \sqrt{k^2-2} k  e^{\sigma_2(k) t} \, dk$. The decay of the $C_2$ contribution on the fast time $t$ therefore corresponds to a rapid transfer of particles between the remainders and the condensate. These rapid transfers of particles ensure that (\ref{defn}) is satisfied adiabatically. The dynamics of the system is then characterized by a single slowly-evolving spectrum $n_k(T)$ from which both spectra of $a$ and $b$ particles can be deduced. We pursue the expansion to determine the kinetic equation that governs the evolution of $n_k(T)$.

To order $\mathcal{O}(\epsilon^2)$, we consider a linear combination of equation (\ref{decompa}), considered at wavenumber $\sqrt{k^2-2}$, and of equation (\ref{decompb}), to obtain
\begin{eqnarray}
\partial_t \left( \sqrt{k^2-2} \, {\tilde{n}}^{a;1}_{\sqrt{k^2-2}} + k \, {\tilde{n}}^{b;1}_{k} \right) & = & -\left(k+{\sqrt{k^2-2}}\right) \, \partial_T n_k  \\
\nonumber & + & \frac{2 \pi}{k} \left[ \int_{\frac{k-\sqrt{k^2-2}}{2}\leq k_1 \leq \frac{k+\sqrt{k^2-2}}{2}} n_{k_1} n_{k-k_1} - 2 n_{k_1}n_k \, dk_1   \right. \\
\nonumber & + & \left. 2 \int_{k_1 \geq k+\frac{1}{2k}} n_{k_1}n_{k_1-k} + n_{k_1} n_k - n_{k_1-k} n_k \, dk_1  \right] \, .
\end{eqnarray}
To obtain this form for the second integral, we have made the change of variables $\sqrt{k_1^2+2} \to k_1$ in the integrand.
The solvability condition demands that the right-hand side vanish, which leads to the following kinetic equation,
\begin{eqnarray}
\partial_T n_k  & = & \frac{2\pi}{k\left(k+{\sqrt{k^2-2}}\right)}  \left[ \int_{\frac{k-\sqrt{k^2-2}}{2}\leq k_1 \leq \frac{k+\sqrt{k^2-2}}{2}} n_{k_1} n_{k-k_1} - 2 n_{k_1}n_k \, dk_1   \right. \\
\nonumber & + & \left. 2 \int_{k_1 \geq k+\frac{1}{2k}} n_{k_1}n_{k_1-k} + n_{k_1} n_k - n_{k_1-k} n_k \, dk_1  \right] \, .
\end{eqnarray}
In the relativistic limit $k \gg 1$ the equation becomes scale-invariant,
\begin{eqnarray}
\partial_T n_k  & = & \frac{\pi}{k^2}  \left[ \int_{k_1 \leq k} n_{k_1} n_{k-k_1} - 2 n_{k_1}n_k \, dk_1  +  2 \int_{k_1 \geq k} n_{k_1}n_{k_1-k} + n_{k_1} n_k - n_{k_1-k} n_k \, dk_1  \right] \, .\label{relKEcond}
\end{eqnarray}
This equation conserves the total energy, but it does not conserve the number of particles in the remainders. Indeed, these remainders rapidly exchange particles with the condensate, and only the total number of particles contained in both the remainders and the condensate is conserved.

In appendix \ref{secKZnonlocal} we determine the steady power-law solutions to this equation $n_k \sim k^x$. As already mentioned, one such solution to the kinetic equation is the isothermal spectrum $n_k \sim k^{-1}$ that coexists with the condensate when wave-condensation occurs. But this kinetic equation also admits a solution of the KZ type, with exponent $x=-\frac{3}{2}$. This turbulent spectrum carries a flux ${\cal P}$ of energy through the scales of the system. It reads (see appendix \ref{secKZnonlocal} for details)
\begin{eqnarray}
n_k=\frac{{\cal P}^{\frac{1}{2}} \, k^{-\frac{3}{2}}}{4 \pi \sqrt{2 \pi - 8 \ln 2}}  \, .\label{nlKZ}
\end{eqnarray}
In this cascading state, the flux of energy is due to local interactions in $k$-space, while nonlocal interactions continuously transfer particles from the remainders to the condensate to satisfy the adiabatic constraint $n_k=n_k^a=n_k^b$ (one could highlight the corresponding slow growth of the condensate by considering the $\mathcal{O}(\epsilon^2)$ equations at $k=0$ and introducing an evolution of the amplitude $A$ of the condensate on an even slower timescale $\epsilon^2 t$). 
\section{Discussion}

\label{summary}

\begin{figure}
\begin{center}
a.\\
\includegraphics[width=70 mm]{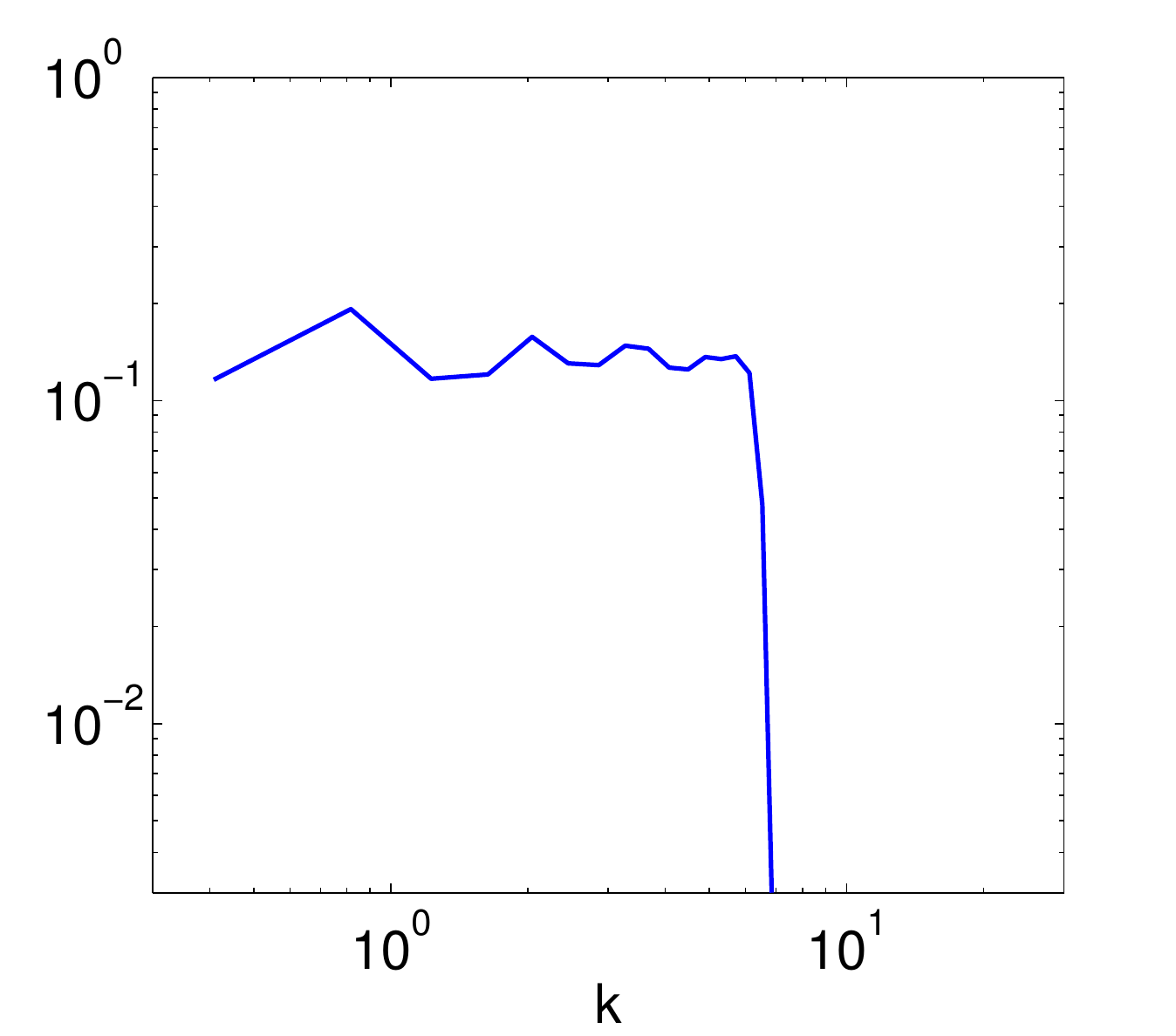}\\
b.\\
\includegraphics[width=70 mm]{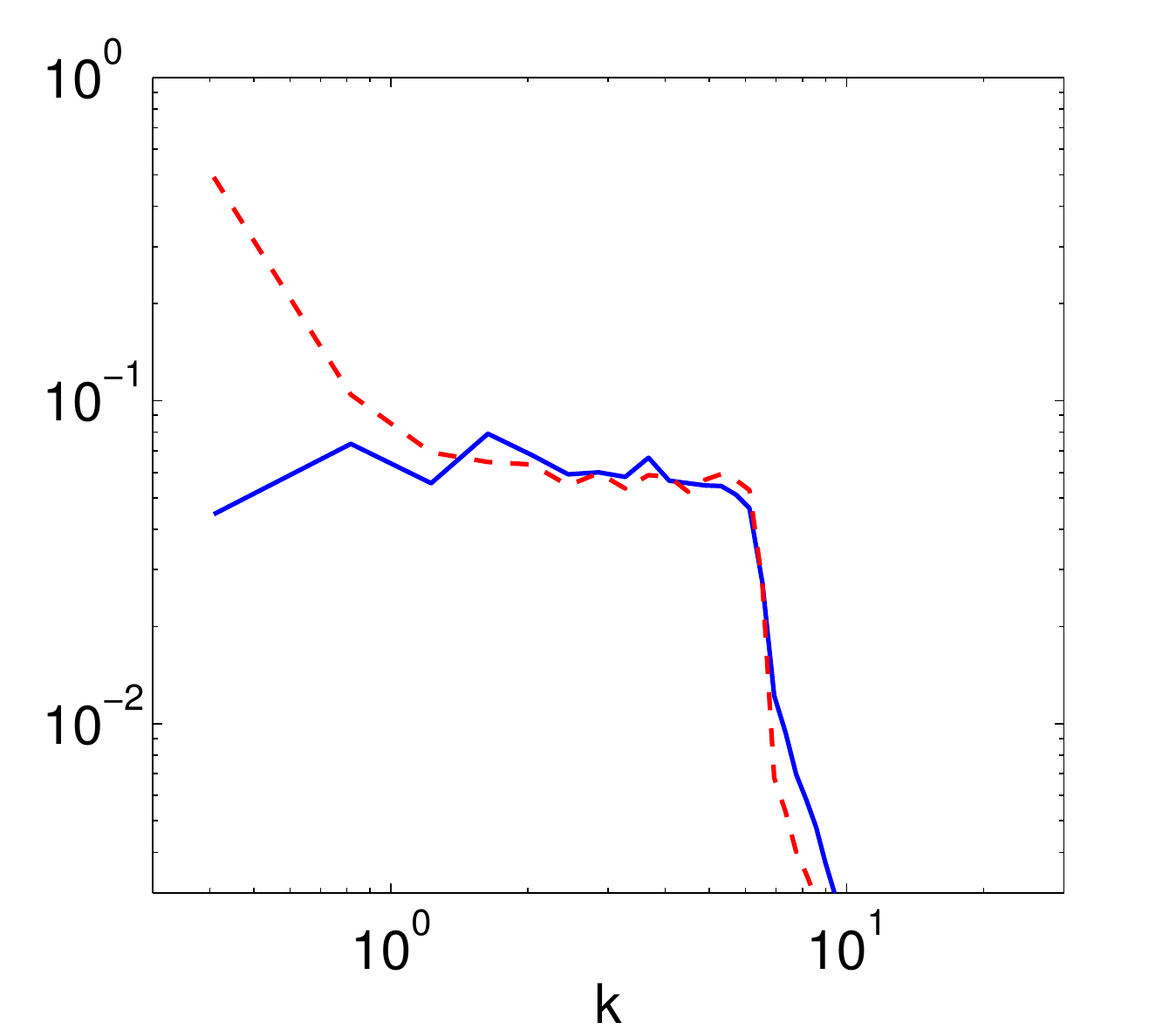}\\
c.\\
\includegraphics[width=70 mm]{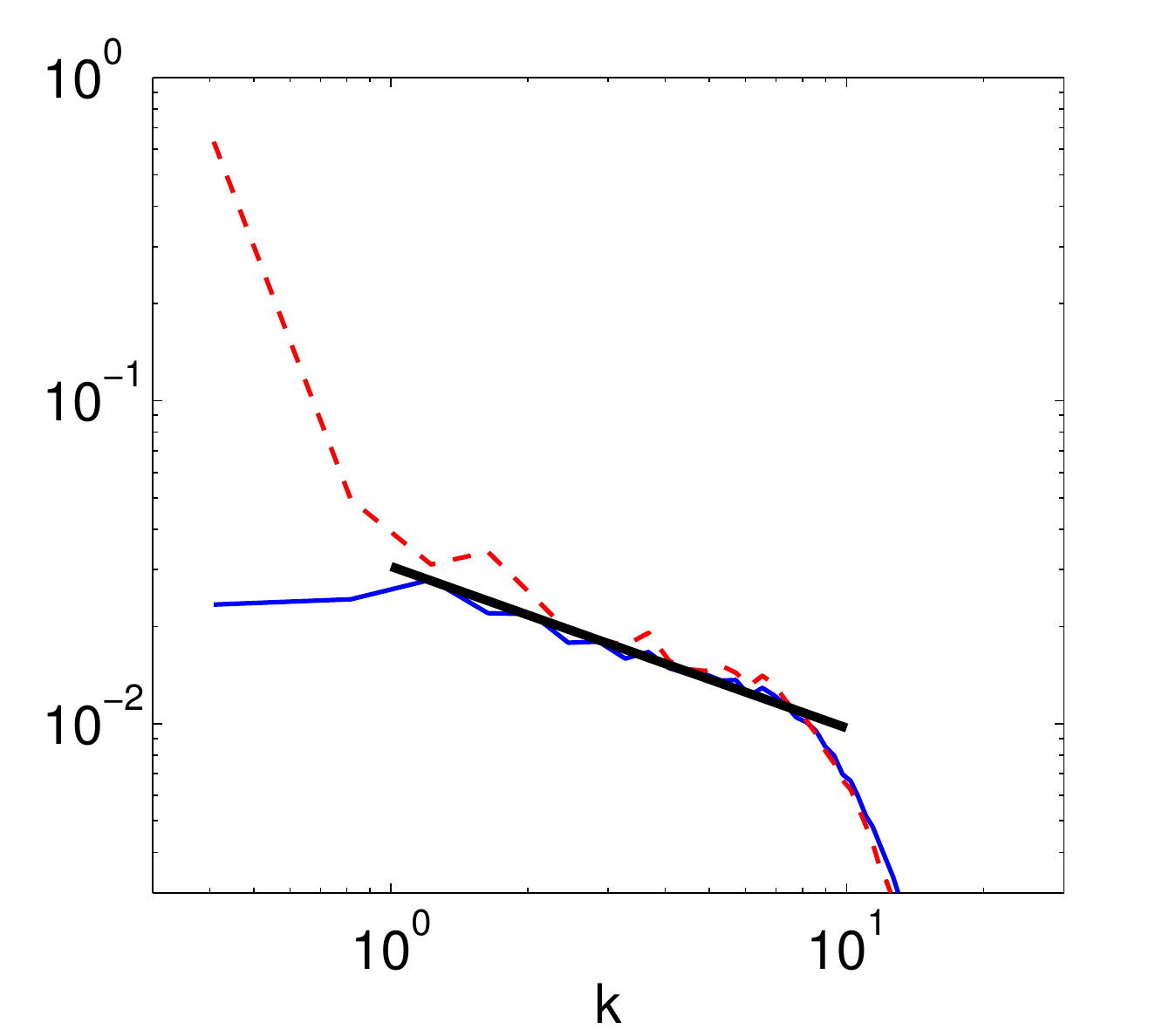}
\caption{Out-of-equilibrium spectra $S_\lambda$ (blue solid line) and $S_\chi$ (red dashed line) in a 3D numerical simulation of the KGMH equation. Panel $a$ shows the initial condition ($S_\chi(k) \simeq 2\, 10^{-3}$ is out of the figure). Panel $b$ corresponds to $t=40$: after a fast transient, part of the massive waves decayed into massless waves. A strong condensate of $b$-waves appear, and the small-scale remaining spectra follow the adiabatic constraint $S_\lambda \simeq S_\chi$. Panel $c$ corresponds to $t=475$: energy cascades from low to high wave numbers, in quantitative agreement with the KZ prediction (\ref{predictionS}), shown as a thick solid line.   \label{3steps}}
\end{center}
\end{figure}

We have developed a weak turbulence theory for the Klein-Gordon equation with a Mexican-hat potential  (\ref{ctkg}), a universal nonlinear equation which displays spontaneous symmetry-breaking. The fundamental state of this equation is a uniform field at the minimum of the Mexican-hat potential. The weak-turbulence theory describes the interaction of random weakly nonlinear waves propagating over this uniform background state. Such wave disturbances are of two-kinds, with two branches in the dispersion relation: in the particle physics context, these branches correspond to a massive and a massless particle fields, and the KGMH equation provides the simplest particle physics model for a system of interacting sigma mesons and pions.

In the weak-nonlinearity limit, the dominant interaction between these fields is a three-wave process by which two massless waves can merge into a massive one, or inversely one massive wave can decay into two massless ones. This process conserves energy and an equivalent of the number of particles.

We have derived the wave turbulence kinetic equations describing the coupled evolution of the spectra of massive and massless waves under this three-wave interaction. They admit a two-parameter family of thermodynamic equilibrium spectra of the Rayleigh-Jeans type. When the initial condition corresponds to little energy per particle, the thermodynamic state reached in the long-time limit displays massless-wave condensation, the classical analogue of Bose-Einstein condensation.

The kinetic equations admit no Kolmogorov-Zakharov (KZ) cascading states in the usual sense. Instead, the out-of-equilibrium dynamics is dominated by nonlocal interactions in scale-space. We have highlighted two cases of such nonlocal evolution. The first one is when a massive mode spontaneously decays into massless waves, and the second one is when small-scale $a$ and $b$ fields evolve in presence of an intense large-scale $b$ condensate. In the latter regime some particles are rapidly transferred to the condensate, and the small-scale remaining waves follow a single reduced kinetic equation which admits a KZ solution corresponding to an energy cascade.

These different regimes are illustrated in the 3D numerical simulation reported in figure \ref{3steps}: the domain is $[0,  L]^3$ with$L=2 \pi\, \sqrt{6} \simeq 15.4$, with a resolution of $192^3$. The wave vector ${\bf k}$ is on the 3D grid $\frac{1}{\sqrt{6}} \times [-96, 96]^3$, but after dealiasing we keep only the modes with $k \leq \frac{48}{\sqrt{6}}\simeq 20$. We plot the 1D wave spectra $S_\lambda(k)=\left( \frac{L}{2\pi} \right)^3 4\pi k^2 |\lambda_k|^2$ and $S_\chi(k)=\left( \frac{L}{2\pi} \right)^3 4\pi k^2 |\chi_k|^2$.

The simulation starts off with a distribution of massive waves only, together with a weak level of massless waves. The initial spectrum $S_\lambda$ is flat up to wavenumber $k=6$, i.e., the massive waves are mostly relativistic (see panel \ref{3steps}a). The initial evolution of this spectrum corresponds to a fast decay instability. As discussed in section \ref{statdecay}, the decay of relativistic massive particles with wavenumber $k_a$ produces massless waves with $k_b \simeq k_a$ and with $k_b \ll 1$. The latter waves accumulate in a strong condensate of massless particles. The subsequent evolution corresponds to waves superposed to a strong condensate, the dynamics of which we presented in section \ref{dyncond}: the spectra of the small-scale remaining $a$ and $b$-waves rapidly become identical (see panel \ref{3steps}b), and follow the reduced kinetic equation (\ref{relKEcond}). After some transient, we obtain the quasi-static spectral distribution plotted in panel \ref{3steps}c, which corresponds to the KZ energy cascade (\ref{nlKZ}). Indeed, we computed the energy leaving the domain $k\in [1;10]$ per unit time to determine the value of the energy flux. We obtained ${\cal P} \simeq 7. \,10^{-4}$. In the relativistic regime, the KZ spectra corresponding to (\ref{nlKZ}) are given by $k S_\lambda(k)=k S_\chi (k)=4\pi k^2 n_k$, i.e.,
\begin{eqnarray}
S_\lambda(k)=S_\chi(k)=\frac{{\cal P}^{\frac{1}{2}} \, k^{-\frac{1}{2}}}{\sqrt{2 \pi - 8 \ln 2}} \label{predictionS}
\end{eqnarray}
This theoretical prediction is plotted in figure \ref{3steps}c. It has no fitting parameters and is in good quantitative agreement with the spectra obtained numerically in the inertial range. This confirms that the system displays a nonlocal Kolmogorov-Zakharov cascading state, where nonlocal transfers of particles with the condensate coexist with a local energy cascade. This state is therefore the intermediate asymptotics that describes the evolution between the initial distribution of waves and the thermodynamic equilibrium attained in the long-time limit.

\section{Acknowledgements}

We thank Jean-Paul Blaizot for insightful discussions. During the time this work was conducted, Sergey Nazarenko was supported by Labex PALM ANR-10-LABX-0039 at SPEC laboratory, CEA, France.

\appendix

\section{Three-wave resonances}
\label{ap:3wave}

The interaction Hamiltonian (\ref{HkintCan}) corresponds to three-wave processes of
type $a+a+a \to 0, a+a \to a, a+b +b\to 0, a+b \to b$ and
$b+b \to a$. Out of these processes, the only ones allowed are those that can satisfy both the wavenumber and the frequency resonance conditions.

\subsection{$a+a+a \to 0$ and $a+b+b \to 0$ processes}

Clearly, these processes cannot be resonant because the sum of three positive frequencies cannot be zero.
Thus we neglect the corresponding terms in the Hamiltonian. 

\subsection{$a+a \to a$ process}

The frequency resonance condition for this process is $\omega_{{\bf k}_1} ^a + \omega_{{\bf k}_2} ^a =\omega_{{\bf k}_3} ^a$. After the substitution ${\bf k}_3 = {\bf k}_1 + {\bf k}_2 $, it becomes
$$
\sqrt{2 +k_1^2} + \sqrt{2 +k_2^2} = \sqrt{2 +({\bf k}_1 + {\bf k}_2)^2}.
$$
Taking the square of both sides, we obtain
$$
4 + k_1^2 +k_2^2 + 2 \sqrt{(2 +k_1^2)(2 +k_2^2)} = 2 + k_1^2 +k_2^2 + 2 ({\bf k}_1 \cdot {\bf k}_2),
$$
or
$$
1 +   \sqrt{(2 +k_1^2)(2 +k_2^2)} =   ({\bf k}_1 \cdot {\bf k}_2).
$$
This equation has no solutions because the left hand side is strictly greater than $k_1 k_2$ whereas
the right hand side is less than $k_1 k_2$.
Thus we neglect the corresponding terms in the Hamiltonian.

\subsection{$a+b \to b$ process}

For this process the resonance conditions give
$$
\sqrt{2 +k_1^2} + k_2 = \sqrt{({\bf k}_1 + {\bf k}_2)^2}.
$$
Squaring both sides yields
$$
2 + k_1^2 +k_2^2 + 2 k_2 \sqrt{2 +k_1^2} =  k_1^2 +k_2^2 + 2 ({\bf k}_1 \cdot {\bf k}_2),
$$
or
$$
 1+ k_2 \sqrt{2 +k_1^2} =  ({\bf k}_1 \cdot {\bf k}_2) .
$$
Again, this equation has no solution because the left hand side is strictly greater than $k_1 k_2$ whereas
the right hand side is less than $k_1 k_2$, so we neglect the corresponding terms in the Hamiltonian.

\subsection{$b+b \to a$ process}

For this process the resonance conditions give
$$
k_1 + k_2 = \sqrt{2 +({\bf k}_1 + {\bf k}_2)^2}.
$$
Squaring both sides, we obtain
$$
k_1^2 +k_2^2 + 2 k_2 k_1 = 2 + k_1^2 +k_2^2 + 2 ({\bf k}_1 \cdot {\bf k}_2),
$$
or
$$
 2 k_2 k_1 =  2 + 2 ({\bf k}_1 \cdot {\bf k}_2) = 2 + 2 k_2 k_1 \cos \theta ,
$$
where $\theta$ is the angle between ${\bf k}_1$ and ${\bf k}_2$.
We have
\begin{equation}
\label{cos}
-1 \le \cos \theta = \frac{  k_2 k_1 -1 }{k_2 k_1} \le 1   ,
\end{equation}
i.e., such three-wave resonances exist if
\begin{equation}
\label{nl_cond}
  k_1 k_2 \ge \frac 1{2}   .
\end{equation}
This condition predicts nonlocal interactions 
if the initial spectrum of $a$-waves is at $k \ll 1$: then the initial evolution
 generates b-waves at $k \approx 1/\sqrt{2}$.
Indeed, $k \ll 1$ implies $k_1 \approx k_2$ and $\cos \theta \approx -1$, which
according to (\ref{cos}) gives $k_1 \approx k_2 \approx 1/\sqrt{2}$.
Conversely, two short b-waves can produce a long a-wave through a wave-beating process.

On the other hand, if the a-particles are relativistic then the condition (\ref{nl_cond})
is not restrictive: for $k, k_1, k_2 \gg 1$ we have $\cos \theta \approx 1$,
which is typical of acoustic-like systems: the wave vectors are almost colinear in all
resonant triads.

\section{Wave turbulence derivation}
\label{ap:wt}

Wave turbulence deals with statistical ensembles of waves. For such wave fields, a small amount of dispersion rapidly leads to the complex amplitudes of the different waves being uncorrelated. We can therefore consider the amplitudes and the phases of the waves to be random independent variables to start with. Such a random phase and amplitude approximation is a useful shortcut to obtain the wave kinetic equations. The brackets $\left< \dots \right>$ denote an ensemble average over these random phases and amplitudes.

\subsection{Separating the time scales}

Consider the interaction representation variables
$$
a_{\bf k} = \epsilon e^{i \omega^a_{\bf k} t} \,  c_{\bf k},
\quad
b_{\bf k} = \epsilon  e^{i \omega^b_{\bf k} t} \,  d_{\bf k},
$$
where the small parameter $\epsilon \ll 1$ has been introduced to study the weakly-nonlinear behavior.
These variables are steady in the absence of nonlinearity, and they evolve slowly for weak nonlinearity, with

\begin{eqnarray}\label{intrepeq1}
\dot c_{\bf k} &=& 
-i \epsilon \sum_{{\bf k}_1, {\bf k}_2}  
V^{\bf k}_{12} \, d_1 d_2
e^{-i \omega_{{\bf k} 12}^{abb}  t} \delta^{\bf k}_{12} \, ,
 \\
\dot d_{\bf k} &=& \label{intrepeq2}
-2i \epsilon \sum_{{\bf k}_1, {\bf k}_2}   V^1_{2{\bf k}}
\, c_1 d^*_2 e^{i \omega_{12 {\bf k} }^{abb} t} \delta^1_{2{\bf k}}  \, ,
\end{eqnarray}
where
$$
 \omega_{{\bf k} 12}^{abb}  =
\omega^a_{\bf k} -\omega^b_1 - \omega^b_2 \, , \quad
\omega_{12 {\bf k} }^{abb}  =
\omega^a_1 -\omega^b_2 - \omega^b_{\bf k} \, .
$$

Consider a time $T$ which is intermediate
between the (short) period of the linear oscillations
and the (long) nonlinear time-scale,
$$
\frac{2 \pi} {\omega} \ll T \ll
\frac{2 \pi} {\epsilon^2 \omega} \, ,
$$
where the inequality must be satisfied for both
$\omega = \omega^a$ and $\omega = \omega^b$.

Seek the solutions at time $T$ via a regular expansion in $\epsilon$
\begin{eqnarray}\label{wn}
 c_{\bf k}(T) &=& c^{(0)}_{\bf k} + \epsilon c^{(1)}_{\bf k} + \epsilon^2 c^{(2)}_{\bf k} + \dots , \\
 d_{\bf k} (T)  &=& d^{(0)}_{\bf k} + \epsilon d^{(1)}_{\bf k} + \epsilon^2 d^{(2)}_{\bf k} + \dots .
\end{eqnarray}
Below we neglect the orders of $\epsilon$ higher than two.
Substitution into the evolution equations (\ref{intrepeq1}) and
(\ref{intrepeq2}) gives at leading order $\dot c_{\bf k}^{(0)}=\dot d_{\bf k}^{(0)}=0$, i.e.,
\begin{equation}\label{c0d0}
c^{(0)}_{\bf k} = c_{\bf k} (0), \quad
d^{(0)}_{\bf k} = d_{\bf k} (0).
\end{equation}
At the next order we have
\begin{eqnarray}\label{c1}
c^{(1)}_{\bf k} &=& 
-i  \sum_{{\bf k}_1, {\bf k}_2}  
V^{\bf k}_{12} \, d^{(0)}_1 d^{(0)}_2
\Delta{ (- \omega_{{\bf k} 12}^{abb}   ) } \delta^{\bf k}_{12} ,
 \\
d^{(1)}_{\bf k} &=& \label{d1}
-2i  \sum_{{\bf k}_1, {\bf k}_2}   V^1_{2{\bf k}}
\, c^{(0)}_1 d^{(0)*}_2 
\Delta_T{ (\omega_{12 {\bf k} }^{abb} ) } \delta^1_{2{\bf k}}  ,
\end{eqnarray}
where
$$\Delta_T(x) = \int_0^T e^{ixt} \, dt = \frac {e^{ixt} - 1}{ix}.
$$
Doing one more step of the recursion, we get
\begin{eqnarray}\label{c2dot}
\dot c^{(2)}_{\bf k} &=& 
-2i  \sum_{{\bf k}_1, {\bf k}_2}  
V^{\bf k}_{12} \, d^{(0)}_1 d^{(1)}_2
e^{-i \omega_{{\bf k} 12}^{abb}  t} \delta^{\bf k}_{12} ,
 \\
\dot d^{(2)}_{\bf k} &=& \label{d2dot}
-2i  \sum_{{\bf k}_1, {\bf k}_2}   V^1_{2{\bf k}}
\, (c^{(1)}_1 d^{(0)*}_2 + c^{(0)}_1 d^{(1)*}_2) e^{i \omega_{12{\bf k} }^{abb}  t} \delta^1_{2{\bf k}}  .
\end{eqnarray}
Substituting here $c^{(1)}$ and $d^{(1)}$ from (\ref{c1}) and (\ref{d1}) and integrating over time
yields
\begin{eqnarray}\label{c2}
 c^{(2)}_{\bf k} &=&
-4  \sum_{{\bf k}_1, {\bf k}_2, {\bf k}_3, {\bf k}_4}  
V^{\bf k}_{12} 
  V^3_{4{2}}
\, d^{(0)}_1  c^{(0)}_3 d^{(0)*}_4
E(\omega_{342}^{abb}, \omega_{{\bf k} 12}^{abb} )
\delta^3_{4{2}}
 \delta^{\bf k}_{12} ,
 \\
 d^{(2)}_{\bf k} &=& \label{d2}
2 \sum_{{\bf k}_1, {\bf k}_2, {\bf k}_3, {\bf k}_4}   V^1_{2{\bf k}}
\, [
-  
V^{1}_{34} \, d^{(0)}_3 d^{(0)}_4 d^{(0)*}_2
E{ (- \omega_{{1} 34}^{abb} , -  \omega_{12{\bf k} }^{abb} ) } \delta^{1}_{34}
\nonumber \\
&&\quad \quad \quad \quad \quad \quad+ 
2   V^3_{4{2}}
\, c^{(0)}_1 c^{(0)*}_3 d^{(0)}_4 
E{ (-\omega_{34 {2} }^{abb},  -  \omega_{12{\bf k} }^{abb} ) } \delta^3_{4{2}} 
]  \delta^1_{2{\bf k}} .
\end{eqnarray}
where
\begin{equation}\label{E}
E(x,y) = 
\int_0^T \Delta_t{ (x ) }  e^{-i yt} \, dt,
\end{equation}
$$
\omega_{341}^{abb} = \omega^a_3 - \omega^b_4 -
\omega^b_{ 1}, \quad \hbox{etc}.
$$
Introduce the wave intensities
$$
I_{\bf k}  = |c_{\bf k}|^2 = \epsilon^2 |a_{\bf k}|^2,\quad
J_{\bf k}  = |d_{\bf k}|^2 = \epsilon^2 |b_{\bf k}|^2,
$$
and expand them to second order in $\epsilon$,
\begin{eqnarray}\label{wnIJ}
I_{\bf k} (T) & = & |c^{(0)}_{\bf k} + \epsilon c^{(1)}_{\bf k} + \epsilon^2 c^{(2)}_{\bf k} |^2
=|c^{(0)}_{\bf k}|^2 + \epsilon (c^{(0)*}_{\bf k} c^{(1)}_{\bf k} +c.c.) \\
\nonumber & & + \epsilon^2 |c^{(1)}_{\bf k}|^2 +  \epsilon^2 (c^{*(0)}_{\bf k}
c^{(2)}_{\bf k} +c.c.)
, \\
J_{\bf k} (T) & = & |d^{(0)}_{\bf k} + \epsilon d^{(1)}_{\bf k} + \epsilon^2 d^{(2)}_{\bf k} |^2
=|d^{(0)}_{\bf k}|^2 + \epsilon (d^{(0)*}_{\bf k} d^{(1)}_{\bf k} +c.c.) \label{wnIJ2} \\
\nonumber & & + \epsilon^2 |d^{(1)}_{\bf k}|^2 +  \epsilon^2 (d^{*(0)}_{\bf k}
d^{(2)}_{\bf k} +c.c.) \, .
\end{eqnarray}


\subsection{Phase averaging}

We consider the phases of the waves as random variables that are uncorrelated for different wave vectors. When phase-averaging the expressions (\ref{wnIJ}) and (\ref{wnIJ2}),  the linear order in $\epsilon$
drops out, as usual in Wave Turbulence. Denoting the average over those phases as $\langle \dots \rangle_\phi$, we obtain
\begin{eqnarray}
\label{c1sqphav}
\langle |c^{(1)}_{\bf k}|^2 \rangle_\phi &=& 
 2 \sum_{{\bf k}_1, {\bf k}_2}  
|V^{\bf k}_{12}|^2  \,   J^{(0)}_1 J^{(0)}_2
|\Delta_T{ (-  \omega_{{\bf k} 12}^{abb}) } |^2
\delta^{\bf k}_{12}  ,
 \\
\langle  |d^{(1)}_{\bf k}|^2 \rangle_\phi &=& \label{d1sqphab}
4  \sum_{{\bf k}_1, {\bf k}_2}   |V^1_{2{\bf k}}|^2 
\, I^{(0)}_1 J^{(0)}_2 
|\Delta_T{ (\omega_{12{\bf k} }^{abb} ) } |^2
\delta^1_{2{\bf k}} 
,
\end{eqnarray}
where $I^{(0)}_{\bf k}=|c_{\bf k}^{(0)}|^2$ and $J^{(0)}_{\bf k}=|d_{\bf k}^{(0)}|^2$.
Similarly,
\begin{eqnarray}\label{c2c0*}
\langle  c^{(2)}_{\bf k} c^{(0)*}_{\bf k} \rangle_\phi  &=&
-4  \sum_{{\bf k}_1, {\bf k}_2, {\bf k}_3, {\bf k}_4}  
V^{\bf k}_{12} 
  V^3_{4{2}}
\, \langle d^{(0)}_1  c^{(0)}_3 d^{(0)*}_4 c^{(0)*}_{\bf k} \rangle_\phi
E(\omega_{342}^{abb}, \omega_{{\bf k} 12}^{abb} )
\delta^3_{4{2}}
 \delta^{\bf k}_{12} ,
\nonumber \\
&=&
-4  \sum_{{\bf k}_1, {\bf k}_2}  
|V^{\bf k}_{12}|^2 
\, J^{(0)}_1  I^{(0)}_{\bf k} 
E(\omega_{{\bf k} 12}^{abb}, \omega_{{\bf k} 12}^{abb} )
 \delta^{\bf k}_{12} ,
\\
\langle  d^{(2)}_{\bf k} d^{(0)*}_{\bf k} \rangle_\phi  &=&
\label{d2d0*}
2 \sum_{{\bf k}_1, {\bf k}_2, {\bf k}_3, {\bf k}_4}   V^1_{2{\bf k}}
\, [
-  
V^{1}_{34} \, \langle d^{(0)}_3 d^{(0)}_4 d^{(0)*}_2 d^{(0)*}_{\bf k} \rangle_\phi
E{ (- \omega_{{1} 34}^{abb} , -  \omega_{12{\bf k} }^{abb} ) } \delta^{1}_{34}
\nonumber \\
&&\quad \quad \quad \quad \quad \quad+ 
2   V^3_{4{2}}
\, \langle c^{(0)}_1 c^{(0)*}_3 d^{(0)}_4 d^{(0)*}_{\bf k} \rangle_\phi
E{ (-\omega_{34 {2} }^{abb},  -  \omega_{12{\bf k} }^{abb} ) } \delta^3_{4{2}} 
]  \delta^1_{2{\bf k}} 
\nonumber \\
&=&
4 \sum_{{\bf k}_1, {\bf k}_2}   |V^1_{2{\bf k}}|^2
\, [
I^{(0)}_1-  
 J^{(0)}_2]  J^{(0)}_{\bf k} 
E{ (- \omega_{{1} 2{\bf k}}^{abb} , -  \omega_{12{\bf k} }^{abb} ) } 
  \delta^1_{2{\bf k}} 
.
\end{eqnarray}

We can further average over random initial amplitudes: the intensities at different wave numbers are independent random variables, for instance $\langle J^{(0)}_1 J^{(0)}_2 \rangle = \langle J^{(0)}_1\rangle \langle  J^{(0)}_2 \rangle$.

\subsection{$L\to \infty$ and  $T \to \infty$ limits. Kinetic equations.}

In wave turbulence, the limit $L\to \infty$ is taken before the limit $T \to \infty$,
 the latter being equivalent to the $\epsilon \to 0$ limit.

For the limit $L\to \infty$ we follow the standard rules to switch to a continuous description,
 $$\sum_{{\bf k}_1, {\bf k}_2} \to 
\left( \frac {L}{2\pi} \right)^{2d}
\int_{{\bf k}_1, {\bf k}_2} \, d{\bf k}_1 d{\bf k}_2 ,
$$
and replace the Kroenecker deltas with Dirac deltas:
$$ \hbox{Kroenecker-}\delta 
 \to 
\left( \frac {2\pi} {L} \right)^d \times \hbox{Dirac-}\delta.
$$
For the $T \to \infty$
limit, we take into account the following limits,
$$
|\Delta_T{ (x) } |^2 \to 2 \pi T \delta(x)
$$
and
$$
\Re [E(x,x)] \to \pi T \delta(x).
$$
Introduce the wave spectra,
$$
n^a_{\bf k} = \left( \frac {L}{2\pi} \right)^d \langle I_{\bf k} \rangle \, ,
\quad n^b_{\bf k} = \left( \frac {L}{2\pi} \right)^d \langle J_{\bf k} \rangle \, .
$$

Using $\left<I_k(T)\right>=\left<I_k(0)\right>+T \left<  {\dot I_k} \right>$, $\left<J_k(T)\right>=\left<J_k(0)\right>+T \left<  {\dot J_k} \right>$, and performing the infinite-domain and long-time limits, we arrive at the kinetic equations 
\begin{eqnarray}\label{kea}
\dot n^a_{\bf k} =4 \pi \int      |V^{\bf k}_{12}|^2 
\, (n^b_1 n^b_2 - 2 n^b_1  n^a_{\bf k})
\delta (\omega_{{\bf k} 12}^{abb} ) 
\delta^{\bf k}_{12} \, d{\bf k}_1 d{\bf k}_2 \, ,
\\\label{keb}
\dot n^b_{\bf k} =8\pi \int      |V^1_{2{\bf k}}|^2 
\, (n^a_1 n^b_2 + n^a_1  n^b_{\bf k} -n^b_2  n^b_{\bf k})
\delta (\omega_{12{\bf k} }^{abb} ) 
\delta^1_{2{\bf k}}\, d{\bf k}_1 d{\bf k}_2 \, .
\end{eqnarray}
Recall that here $\delta^{\bf k}_{12}$, etc. denote Dirac deltas rather than Kroenecker deltas.

\section{Isotropic power-law spectra in the relativistic limit}
\label{ap:kz}

\subsection{Calculation of the collision integral for power-law spectra}

\begin{figure}[h]
\begin{center}
a.\includegraphics[width=60 mm]{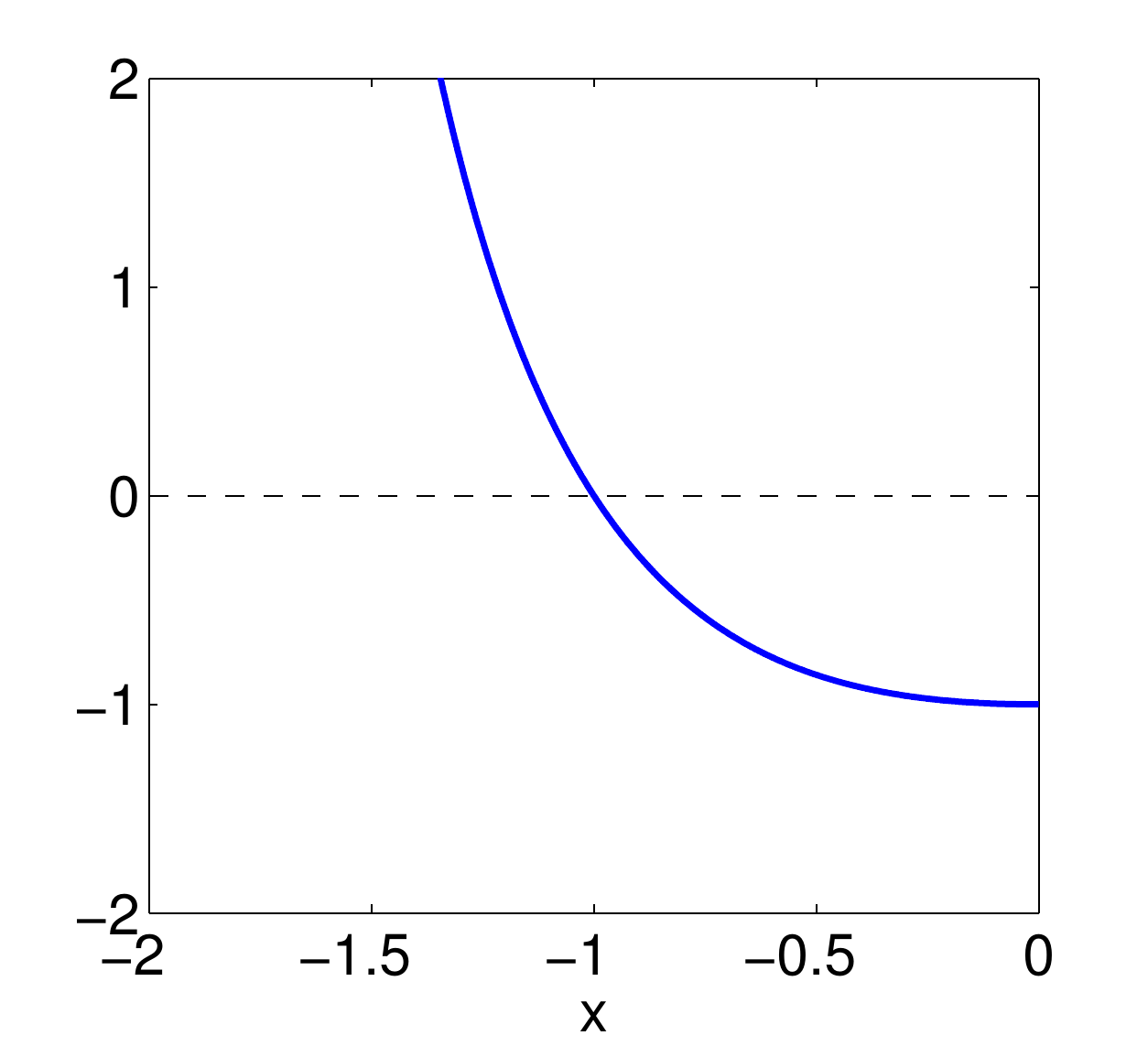} 
b.\includegraphics[width=60 mm]{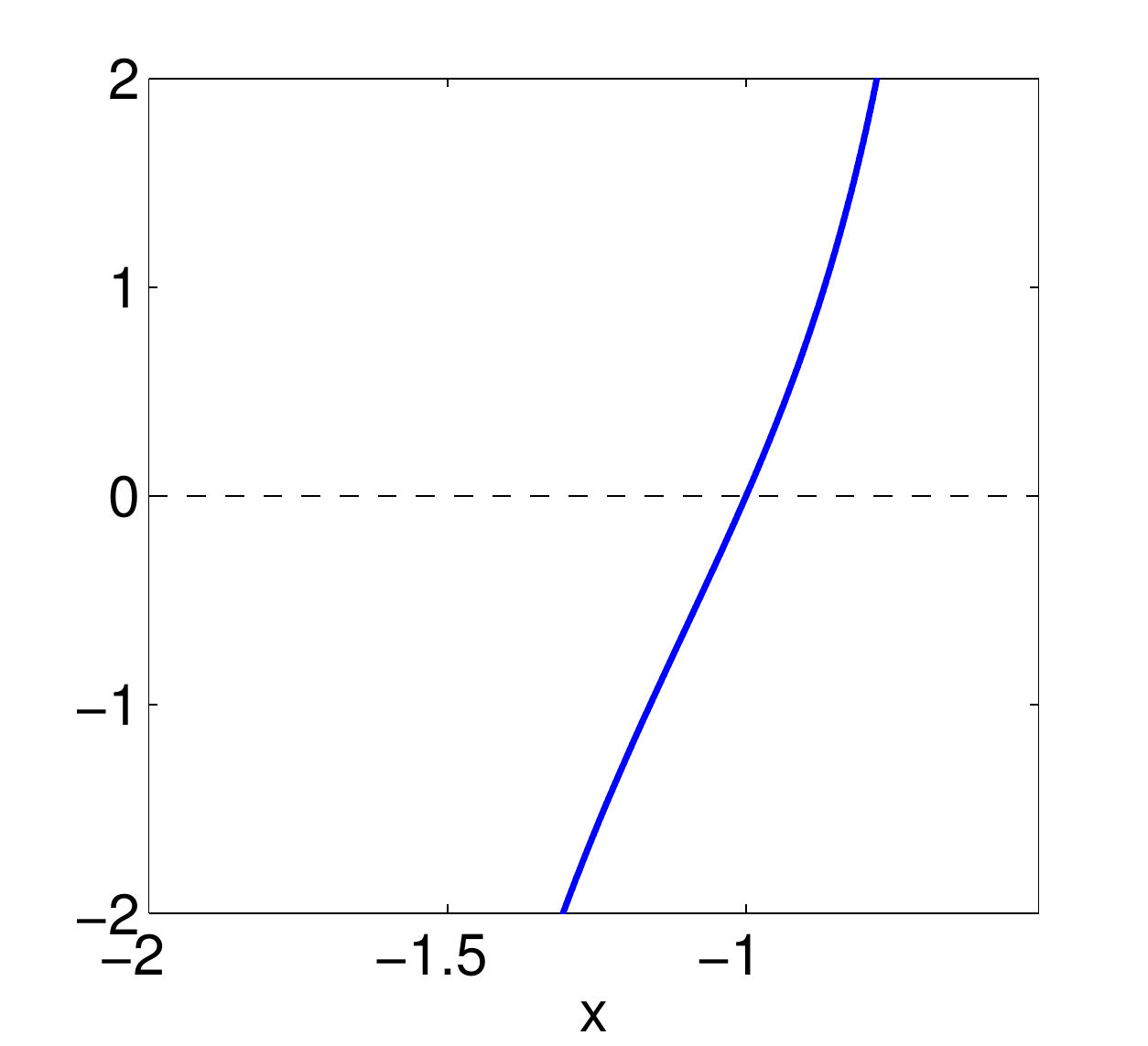} \\
c.\includegraphics[width=60 mm]{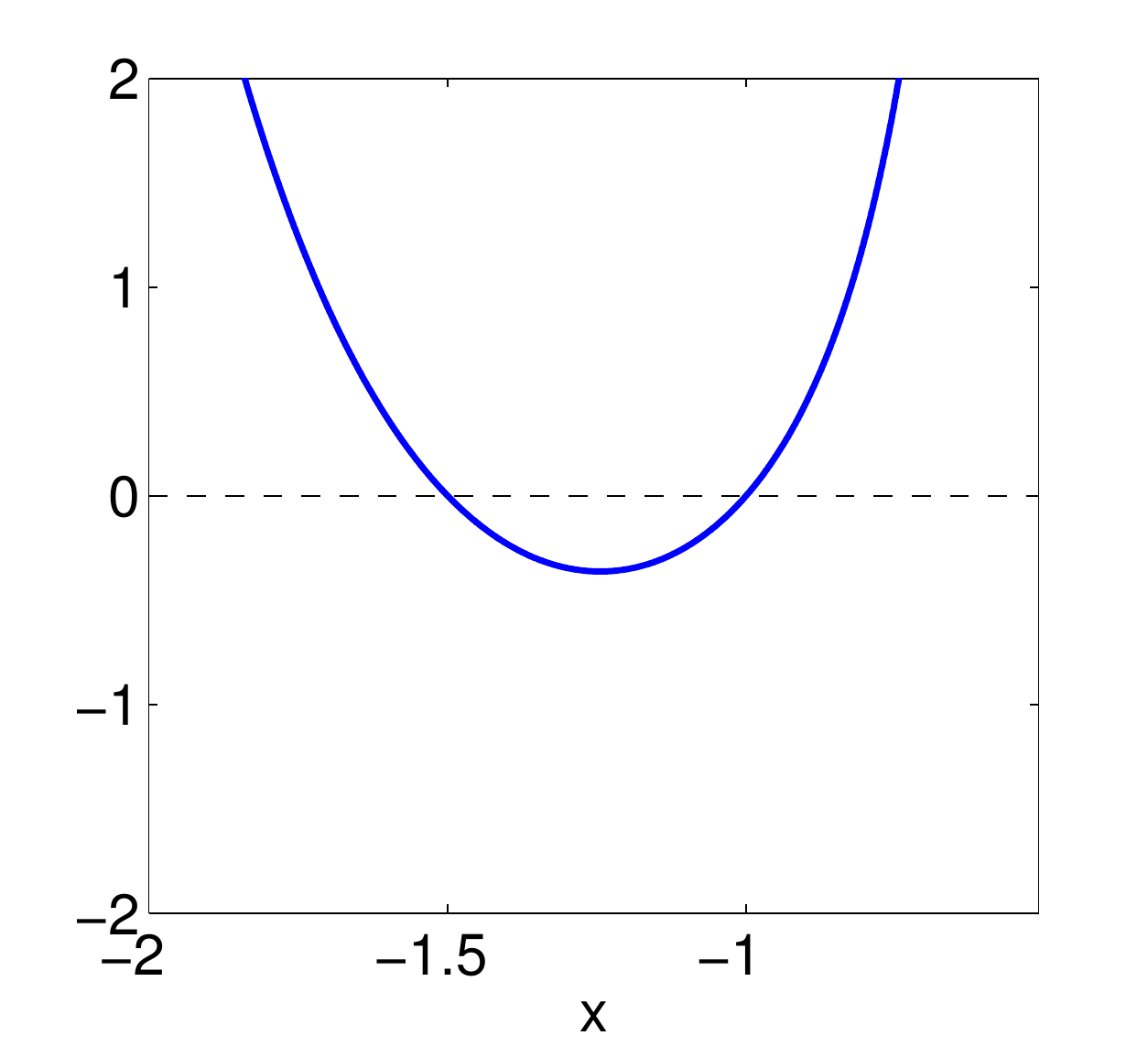}
d.\includegraphics[width=60 mm]{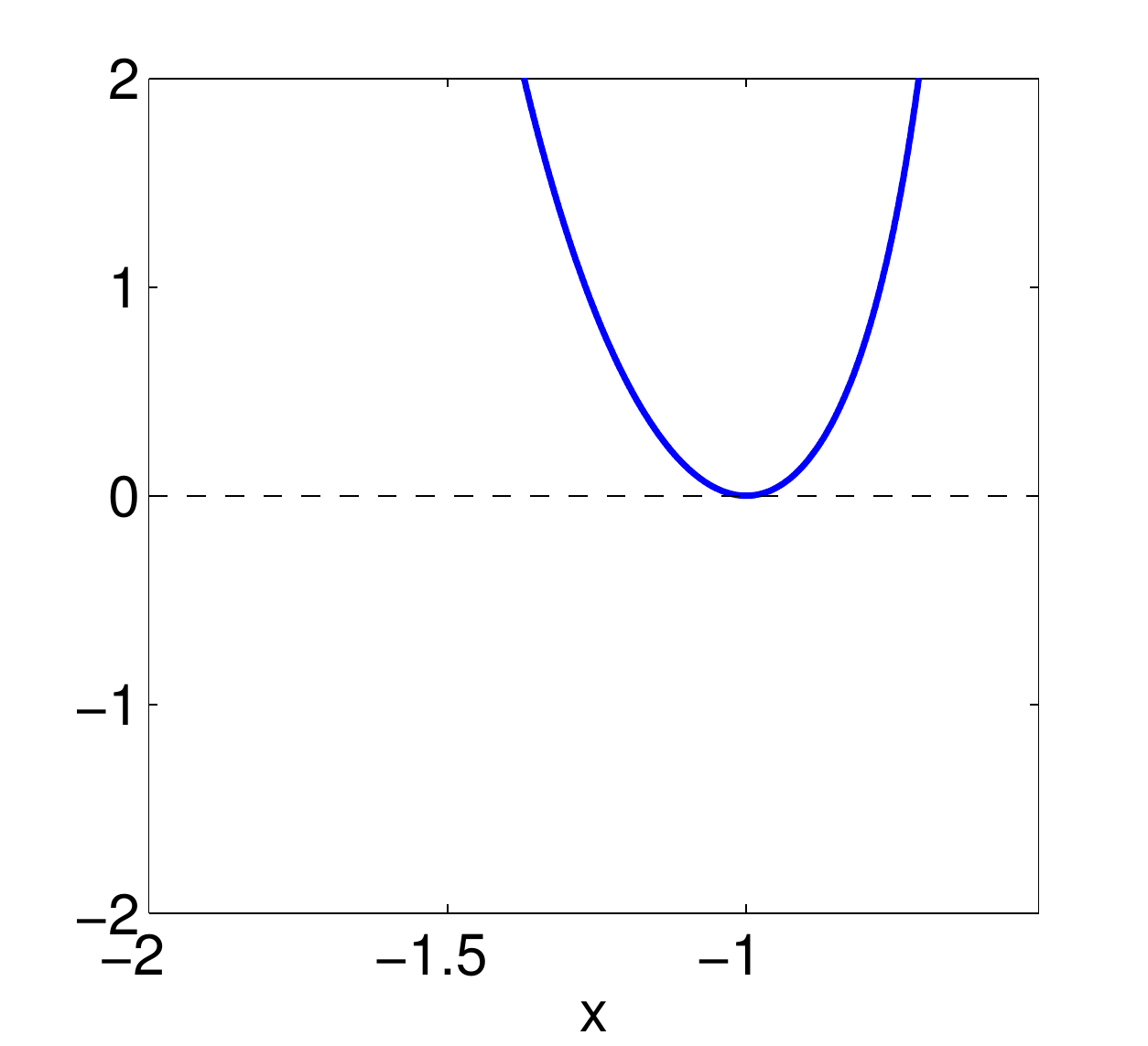}
\end{center}
\caption{Collision integrals as a function of the index $x$. a: $\frac{\dot n^a_{ k}}{2 \pi C^2 k^{2x-1}}$; b:  $\frac{\dot n^b_{ k}}{2 \pi C^2 k^{2x-1}}$; c: $\frac{\dot n^a_{ k}+\dot n^b_{ k}}{2 \pi C^2 k^{2x-1}}$; d: $\frac{2\dot n^a_{ k}+\dot n^b_{ k}}{2 \pi C^2 k^{2x-1}}$.}
\label{figApp}
\end{figure}

We consider the case $d=3$, and we assume power-law spectra of the form
 \begin{equation}
 n_k^{a;b}=C_{a;b} k^x \, . \label{powerlaw} 
  \end{equation}
 At first sight, we may be tempted to consider a more general form in which the index is different for $n_k^a$ and $n_k^b$, but that proves unfruitful: different indices would mean different degrees of homogeneity of the different contributions to the rhs of the kinetic equations and they could not sum up to zero. Secondly, to find the steady-state power-law solutions, one might like to use the so-called Zakharov transformation, as is customary in WT theory \cite{falkovich1992kolmogorov,nazarenko2011wave}. However, we will see below that in our system all integrals can be evaluated analytically for any exponent $x$, and, therefore, the Zakharov transformation is unnecessary.

Inserting   (\ref{powerlaw}) in equation (\ref{keaIr}) and integrating over $k_2$, one gets
\begin{equation}
\dot n^a_{ k} = \frac{2 \pi}{k^2}      \int_{0}^{k}  [C_b^2 k_1^x (k-k_1)^x - 2 C_a C_b k^x k_1^x] dk_1   \, .
\end{equation}
Changing variable to $u=k_1/k$, we obtain
\begin{equation}
\dot n^a_{ k} = 2 \pi k^{2x-1}   \left[  C_b^2 \int_{u=0}^{u=1} u^x (1-u)^x du - 2 C_a C_b   \int_{0}^{1} u^x du   \right] \, . \label{kepla}
\end{equation}
Performing the same steps to equation (\ref{kebIr}) yields
\begin{equation}
\dot n^b_{ k} = 4 \pi k^{2x-1}   \left[  C_a C_b \int_{1}^{\infty} \left( u^x (u-1)^x+u^x  \right) du -	C_b^2 \int_{1}^{\infty} (u-1)^x du \right] \, .\label{keplb}
\end{equation}

The convergence of these integrals depends on the value of $x$, $C_a$ and $C_b$. For $x>1/2$ the rhs of (\ref{keplb}) diverges unless $C_a=C_b$. For $x \in ]0; 1/2]$ and $x \in [-1/2; 0[$ the rhs of (\ref{keplb}) diverges. For $x=0$ it converges only if $C_b=2 C_a$, for which it vanishes. The rhs of (\ref{kepla}) also vanishes for $(x=0,C_b=2C_a)$. The corresponding solution is the equipartition of particle number $\mathcal{N}$. 
For $x\in]-2; -1/2[$ the rhs of (\ref{keplb}) converges provided $C_a=C_b$. When this condition is met the rhs of (\ref{kepla}) also converges, and the equations become
\begin{eqnarray}
\dot n^a_{ k} & = & 2\pi C^2 k^{2x-1} \left[ -\frac{2}{1+x} + \frac{\Gamma^2(1+x)}{\Gamma(2+2x)}  \right] \, , \\
\dot n^b_{ k} & = & 2\pi C^2 k^{2x-1} \left[ -\frac{2}{1+x} + 2\, \frac{\Gamma(-1-2x) \Gamma(1+x)}{\Gamma(-x)}  \right] \, ,
\end{eqnarray}
where $C=C_a=C_b$.
We plot in figure \ref{figApp} the arguments of the square brackets in the right-hand side of these two equations, i.e., we plot respectively $\frac{\dot n^a_{ k}  }{2\pi C^2 k^{2x-1}}$ and $\frac{\dot n^b_{ k}  }{2\pi C^2 k^{2x-1}}$. Both have a single zero at $x=-1$ for the range $x\in]-2; -1/2[$: this can be either the equipartition of wave energy, or the KZ spectra for a cascade of $\mathcal{N}$.

It is insightful to also plot the linear combinations $\frac{\dot n^a_{ k} +\dot n^b_{ k} }{2\pi C^2 k^{2x-1}}$ and $\frac{2 \dot n^a_{ k} +\dot n^b_{ k}}{2\pi C^2 k^{2x-1}} $. The first one indicates if the spectrum of energy is stationary, whereas the second one indicates whether the spectrum of particle number $\mathcal{N}$ is stationary. We see that $\dot n^a_{ k} +\dot n^b_{ k}$ vanishes at $x=-1$ but also at $x=-3/2$. On dimensional grounds (see e.g. \cite{nazarenko2011wave}), one can see that the latter value corresponds to the KZ spectra for an energy cascade. However, this spectrum is not a stationary solution to the full system of equations because $\dot n^a_{ k}$ and $\dot n^b_{ k}$ do not vanish independently at $x=-3/2$. 

The combination $2 \dot n^a_{ k} +\dot n^b_{ k} $ has a double zero at $x=-1$. Indeed, this index corresponds both to energy equipartition and, on dimensional grounds, to a constant flux of $\mathcal{N}$. Still it remains unclear whether it is possible to observe a stationary state with such a nonzero flux of $\mathcal{N}$: the value $x=-1$ is a single zero of $\dot n^a_{ k} +\dot n^b_{ k}$, which corresponds to energy equipartition. It is thus possible that $\dot n^a_{ k} +\dot n^b_{ k}$ is not zero when there is a flux of $\mathcal{N}$.

\subsection{KZ cascade on a condensate\label{secKZnonlocal}}

The remainders that evolve together with a strong condensate follow the kinetic equation (\ref{relKEcond}) in the relativistic limit, where the $a$- and $b$-waves have the same spectrum $n_k(T)$. Let us look for power-law solutions $n_k=C k^x$ to this equation. Substituting into (\ref{relKEcond}) and changing the integration variable to $u=k_1/k$ leads to
\begin{eqnarray}
\partial_T n_k  & = & \pi C^2 k^{2x-1}  {\cal J}(x) \, ,
\end{eqnarray}
where
\begin{eqnarray}
{\cal J}(x) & = & \int_{0}^1 u^x (1-u)^x -2 u^x \, \mathrm{d}u + 2 \int_1^{+\infty} u^x (u-1)^x+u^x-(u-1)^x \, \mathrm{d}u \\
\nonumber & = & -\frac{4}{1+x}+ \frac{2 \, \Gamma(-1-2x) \, \Gamma(1+x)}{\Gamma(-x)} + \frac{\Gamma^2(1+x)}{\Gamma(2+2x)} \, .
\end{eqnarray}
The function ${\cal J}(x)$ vanishes for $x=-1$ and $x=-3/2$, the former corresponding to energy equipartition and the latter being a KZ-type solution. Let us compute the energy flux ${\cal P}$ in this KZ state: the kinetic equation is a conservation equation for the total energy, that we can write in terms of the energy spectrum $(\omega_a+ \omega_b) 4\pi k^2 n_k = 8 \pi k^3 n_k$. Considering power-law solutions $n_k=C k^x$,
\begin{eqnarray}
\partial_T (8 \pi k^3 n_k)  & = & - \frac{d}{dk} {\cal P} = 8 \pi^2 C^2 k^{2x+2} {\cal J}(x)\, .
\end{eqnarray}
The flux ${\cal P}$ of energy going from $k<k_0$ to $k>k_0$ per unit time is obtained by integrating the right-hand-side expression from $k=0$ to $k_0$,
\begin{eqnarray}
{\cal P} = - 8 \pi^2 C^2 \frac{k_0^{2x+3}}{2x+3} {\cal J}(x)\, .\label{fluxx}
\end{eqnarray}
For the KZ exponent $x=-3/2$, this flux is independent of $k_0$, indicating a constant flux of energy cascading through the scales of the system. Indeed, both the numerator and the denominator of (\ref{fluxx}) vanish at $x=-3/2$, and Taylor expansion of ${\cal J}(x)$ to first order around this value leads to
${\cal P}  = 32 \pi^2 C^2 (\pi-4\ln 2)$. We can therefore express $C$ in terms of ${\cal P}$, which leads to the KZ solution,
\begin{eqnarray}
n_k = \frac{{\cal P}^{\frac{1}{2}} k^{-\frac{3}{2}}}{4 \pi \sqrt{2\pi-8\ln 2}} \, .
\end{eqnarray}

\section{Numerical code\label{appcode}}

Let us rewrite the complex Tachyonic Klein-Gordon equation as
\begin{equation}\label{ctkgnum}
\psi_{tt} - \Delta \psi +(-g+|\psi|^2) \psi =0 \, .
\end{equation}
We want to solve this equation in a periodic domain $[0, 2\pi ]^d$. The parameter $g>0$ allows to put the crossover wavenumber anywhere in $k$ space, i.e. the dispersion relation for a-waves is now $\omega^a_{\bf k}=\sqrt{2g+k^2}$. For better accuracy, one wants to integrate exactly the rapidly oscillating part of the solution. Here we write the equation as:
\begin{equation}\label{ctkgnum2}
\psi_{tt} = \Delta \psi +NL\, \mbox{ with } NL=(|\psi|^2-g) \psi \, . 
\end{equation}
If we drop the term $NL$, a solution in Fourier space is $\psi_{\bf k}=A_+ e^{i k t}+A_- e^{-i k t}$. Let us use the method of variation of constants: we write $\psi_{\bf k}=A_+(t) e^{i k t}+A_-(t) e^{-i k t}$ and impose the constraint $A_+'(t) e^{i k t}+A_-'(t) e^{-i k t}=0$. Inserting this decomposition into equation (\ref{ctkgnum2}) we obtain
\begin{eqnarray}
A_+' & = & \frac{i}{2k} e^{-i k t} NL_{\bf k} \, , \label{A+}\\
A_-' & = & - \frac{i}{2k} e^{i k t} NL_{\bf k} \, , \label{A-}
\end{eqnarray}
where $NL_{\bf k}$ is the Fourier amplitude of $(|\psi|^2-g) \psi$ for wavevector ${\bf k}$.
We integrate equations (\ref{A+}) and (\ref{A-}) using an Adam-Bashforth scheme with time-step $dt$:
\begin{eqnarray}
A_+(t+dt)& = & A_+(t) + \frac{i\,dt}{4 k} e^{-i k t} \left[ 3 NL_{\bf k}(t) - e^{i k dt} NL_{\bf k} (t-dt)  \right]  \, , \\
A_-(t+dt)& = & A_-(t) + \frac{i\,dt}{4 k}  e^{i k t}  \left[ -3 NL_{\bf k}(t) + e^{- i k dt} NL_{\bf k} (t-dt)  \right]  \, .
\end{eqnarray}
Let us define $\psi_{\bf k}^+(t)=A_+(t) e^{i k t}$ and $\psi_{\bf k}^-(t)=A_-(t) e^{-i k t}$:
\begin{eqnarray}
\psi_{\bf k}^+(t+dt)& = & e^{i k dt} \left[ \psi_{\bf k}^+(t) + \frac{i\,dt}{4 k}  \left(  3 NL_{\bf k}(t) - e^{i k dt} NL_{\bf k} (t-dt) \right) \right]  \, , \label{iteratepsi+}\\
\psi_{\bf k}^-(t+dt)& = & e^{-i k dt} \left[ \psi_{\bf k}^-(t) + \frac{i\,dt}{4 k}  \left(  -3 NL_{\bf k}(t) + e^{-i k dt} NL_{\bf k} (t-dt) \right) \right]  \, ,\label{iteratepsi-}
\end{eqnarray}
For nonzero $k$, the Fourier amplitude of the field is then $\psi_{\bf k}(t)=\psi_{\bf k}^+(t)+\psi_{\bf k}^-(t)$. The mode ${\bf k}={\bf 0}$ must be taken care of separately. Using a standard finite-difference approximation to the second order time-derivative, we get
\begin{equation}
\psi_{\bf 0}(t+dt)=2\psi_{\bf 0}(t)-\psi_{\bf 0}(t-dt) -dt^2 NL_{\bf 0} \, .\label{iteratepsi0}
\end{equation}
We compute the nonlinear term using a standard pseudo-spectral method with dealiasing following the $1/2$ rule.
A time step consists of the following: the nonlinear term is computed in real space. The field $\psi$ is computed as the inverse Fourier transform of $\psi_{\bf k}$ and the nonlinear term $(|\psi|^2-g) \psi$ is transformed back to Fourier space to get $NL_{\bf k}(t)$. Then equations (\ref{iteratepsi+}), (\ref{iteratepsi-}) and (\ref{iteratepsi0}) are iterated to obtain $\psi_{\bf k}$ at time $t+dt$.

\bibliography{kg}

\begin{thebibliography}{10}
\providecommand*{\bibinfo}[2]{#2}
\providecommand*{\eprint}[1]{#1}
\providecommand*{\url}[1]{#1}
\bibitem{gelman}
\bibinfo{author}{M.~Gell-Mann} and \bibinfo{author}{M.~Lévy},
  \bibinfo{journal}{Il Nuovo Cimento} \bibinfo{volume}{\textbf{16}}(4),
  \bibinfo{pages}{705} (\bibinfo{date}{1960}),
  \url{http://dx.doi.org/10.1007/BF02859738}.
\bibitem{song}
\bibinfo{author}{S.~Shu} and \bibinfo{author}{J.-R. Li},
  \bibinfo{journal}{Journal of Physics G: Nuclear and Particle Physics}
  \bibinfo{volume}{\textbf{34}}(12), \bibinfo{pages}{2727}
  (\bibinfo{date}{2007}), \url{http://stacks.iop.org/0954-3899/34/i=12/a=016}.
\bibitem{falkovich1992kolmogorov}
\bibinfo{author}{G.~Falkovich}, \bibinfo{author}{V.~E. Zakharov}, and
  \bibinfo{author}{V.~Lvov}, \bibinfo{title}{\emph{Kolmogorov spectra of
  turbulence}} (\bibinfo{publisher}{Springer}, \bibinfo{year}{1992}).
\bibitem{nazarenko2011wave}
\bibinfo{author}{S.~Nazarenko}, \bibinfo{title}{\emph{Wave turbulence}},
  \bibinfo{volume}{vol. 825} (\bibinfo{publisher}{Springer},
  \bibinfo{year}{2011}).
\bibitem{Berges:2010ez}
\bibinfo{author}{J.~Berges} and \bibinfo{author}{D.~Sexty},
  \bibinfo{journal}{Phys.Rev.} \bibinfo{volume}{\textbf{D83}},
  \bibinfo{pages}{085004} (\bibinfo{date}{2011}), \eprint{1012.5944}.
\bibitem{Gasenzer:2013era}
\bibinfo{author}{T.~Gasenzer}, \bibinfo{author}{L.~McLerran},
  \bibinfo{author}{J.~M. Pawlowski}, and \bibinfo{author}{D.~Sexty},
  \bibinfo{journal}{arXiv}  (\bibinfo{date}{2013}), \eprint{1307.5301}.
\bibitem{PhysRevLett.95.263901}
\bibinfo{author}{C.~Connaughton}, \bibinfo{author}{C.~Josserand},
  \bibinfo{author}{A.~Picozzi}, \bibinfo{author}{Y.~Pomeau}, and
  \bibinfo{author}{S.~Rica}, \bibinfo{journal}{Phys. Rev. Lett.}
  \bibinfo{volume}{\textbf{95}}, \bibinfo{pages}{263901} (\bibinfo{date}{Dec
  2005}), \url{http://link.aps.org/doi/10.1103/PhysRevLett.95.263901}.
\bibitem{zimaniy}
\bibinfo{author}{J.~Zim\'anyi}, \bibinfo{author}{G.~F\'ai}, and
  \bibinfo{author}{B.~Jakobsson}, \bibinfo{journal}{Phys. Rev. Lett.}
  \bibinfo{volume}{\textbf{43}}, \bibinfo{pages}{1705} (\bibinfo{date}{Dec
  1979}), \url{http://link.aps.org/doi/10.1103/PhysRevLett.43.1705}.
\bibitem{Bunatian:1983dq}
\bibinfo{author}{G.~Bunatian} and \bibinfo{author}{I.~Mishustin},
  \bibinfo{journal}{Nucl.Phys.} \bibinfo{volume}{\textbf{A404}},
  \bibinfo{pages}{525} (\bibinfo{date}{1983}).
\bibitem{Zakharov1985285}
\bibinfo{author}{V.~Zakharov}, \bibinfo{author}{S.~Musher}, and
  \bibinfo{author}{A.~Rubenchik}, \bibinfo{journal}{Physics Reports}
  \bibinfo{volume}{\textbf{129}}(5), \bibinfo{pages}{285 }
  (\bibinfo{date}{1985}).
\bibitem{Note1}
This definition of $T$ is common in wave turbulence. In the energy
  equipartition state, it corresponds to an energy $k_B T$ per oscillatory
  degree of freedom, where $k_B$ is a dimensionless Boltzmann constant, $k_B=(2
  \pi )^d$ in $d$ dimensions.
\bibitem{2DacousticWT}
\bibinfo{author}{V.~S. L'vov}, \bibinfo{author}{Y.~L'vov},
  \bibinfo{author}{A.~C. Newell}, and \bibinfo{author}{V.~Zakharov},
  \bibinfo{journal}{Phys. Rev. E} \bibinfo{volume}{\textbf{56}},
  \bibinfo{pages}{390} (\bibinfo{date}{Jul 1997}),
  \url{http://link.aps.org/doi/10.1103/PhysRevE.56.390}.
\bibitem{During}
\bibinfo{author}{S.~R. G.~D\"uring, A.~Picozzi}, \bibinfo{journal}{Physica D}
  \bibinfo{volume}{\textbf{238}}, \bibinfo{pages}{1524} (\bibinfo{date}{2009}).
\bibitem{Berges}
\bibinfo{author}{D.~S. J.~Berges}, \bibinfo{journal}{Phys. Rev. Lett.}
  \bibinfo{volume}{\textbf{108}}(161601) (\bibinfo{date}{2012}).
\bibitem{Lacour}
\bibinfo{author}{G.~T. P.~Lacour-Gayet}, \bibinfo{journal}{Journal de Physique}
  \bibinfo{volume}{\textbf{35}}, \bibinfo{pages}{425} (\bibinfo{date}{1974}).

\end{thebibliography}

\end{document}